\documentclass[16pt]{article}
\usepackage{epsf}
\usepackage{color}
\usepackage{framed}
\usepackage{amsmath,amsfonts,amsbsy,amssymb}
\usepackage{indentfirst}
\usepackage{mathtools}
\usepackage{arydshln} 
\usepackage{framed}

\newcommand{\nn}{\nonumber}

\def\be{\begin{equation}}
\def\ee{\end{equation}}
\def\bse{\begin{subequations}}
\def\ese{\end{subequations}}
\def\bal{\begin{align}}
\def\ealn{\end{align}}
\def\tr{\text{tr}}
\def\bs{\boldsymbol}

\begin{document}

\begin{titlepage}

\def\slash#1{{\rlap{$#1$} \thinspace/}}

\begin{flushright} 

\end{flushright} 

\vspace{0.1cm}

\begin{Large}
\begin{center}

{\bf   Generating 
 Quantum 
Matrix Geometry  \\ from \\
Gauged Quantum Mechanics
}
\end{center}
\end{Large}

\vspace{1.5cm}

\begin{center}
{\bf Kazuki Hasebe}   \\ 
\vspace{0.5cm} 
\it{
National Institute of Technology, Sendai College,  
Ayashi, Sendai, 989-3128, Japan} \\ 

\vspace{0.8cm} 

{\today} 

\end{center}

\vspace{1.5cm}

\begin{abstract}
\noindent

\baselineskip=18pt

Quantum matrix geometry is the underlying geometry of M(atrix) theory. 
Expanding upon the idea of level projection, we propose a quantum-oriented non-commutative scheme for generating the matrix geometry of the coset space  $G/H$. We employ this novel scheme to unveil  unexplored matrix geometries by utilizing gauged quantum mechanics on higher dimensional spheres. The resultant matrix geometries manifest as $\it{pure}$ quantum Nambu geometries:  Their non-commutative structures elude capture through the conventional commutator formalism of Lie algebra, necessitating the  introduction of  the  quantum Nambu algebra. This matrix geometry embodies a one-dimension-lower  quantum internal geometry  featuring nested  fuzzy structures.  While the continuum limit of this quantum geometry is represented by overlapping classical manifolds, their fuzzification cannot reproduce the original quantum geometry. We demonstrate how these quantum Nambu geometries  give rise to novel solutions in Yang-Mills matrix models, exhibiting distinct  physical properties from the known fuzzy sphere solutions. 

\end{abstract}

\end{titlepage}

\newpage 

\tableofcontents

\newpage 

\section{Introduction}

It has been almost eighty years since the inception of theoretical research on quantized space-time with Snyder's first explicit model 
\cite{Snyder-1947, Yang-1947}. This research field continues to be active, contributing to a deeper understanding of space-time.
Non-commutative geometry presents a promising mathematical framework for describing the  microscopic nature of  space-time \cite{Li-Wu-2002}. A general mathematical framework of  non-commutative geometry was set up by Connes \cite{Connes-book}. More tangible  non-commutative schemes  are those such as deformation quantization, geometric quantization and Berezin-Toeplitz quantization \cite{Ali-Englis-2005}. As these ideas are rooted in the canonical quantization method of the phase space \cite{Curtright-Zachos-2001, Zachos-Fairlie-Curtright-2006}, the corresponding non-commutative schemes are concerned with the quantization of the symplectic manifolds or Poisson manifolds. However, in the investigations of M theory, physicists encountered  even exotic non-commutative structures beyond the conventional quantization schemes, including odd dimensional fuzzy spheres  \cite{Basu-Harvey-2004, BaggerLambert2006,Gustavsson2007,BaggerLambert2007}. From  M(atrix) theory point of view \cite{Banks-Fischler-Shenker-Susskind-1996, Ishibashi-Kawai-Kitazawa-Tsuchiya-1996}, matrix geometries known as  fuzzy manifolds \cite{madore-1992, Grosse-Klimcik-Presnajder-1996, Castelino-Lee-Taylor-1997, Grosse-Reiter-1998, Guralnik&Ramgoolam2001, Alexanian-Balachandran-Immirzi-Ydri-2002, Ho-Ramgoolam-2002, Ramgoolam2002, Kimura2003, Jabbari2004, Arnlind-Bordemann-Hofer-Hoppe-Shimada-2009,Hasebe-2011,Hasebe-2012} represent fundamental  extended objects in the theory \cite{Taylor-2000, Taylor-2001}. Moreover, it has been recognized that the quantum Nambu algebra \cite{Nambu1973} 
plays  crucial roles in the formulation of M theory  (see Refs.\cite{Ho-Matsuo-2016,CurtrightZachos2003,DeBellisSS2010} as nice reviews and references therein).  
It may be evident that a new non-commutative scheme is required to address these extraordinary non-commutative spaces that extend beyond the conventional quantization methods based on the commutator formalism.\footnote{Interestingly, a cubit matrix realization is known for the quantum Nambu algebra \cite{Kawamura-2003, Kawamura-2005,  Ho-Hou-Matuso-2008}, although we do not delve into such possibilities in this paper. The deformation quantization approach to the quantum Nambu geometry is also discussed in Refs.\cite{Takhtajan-1993,Dito-Flato-Sternheimer-Takhtajan-1996,Dito-Flato-1997}.}

Associated with the developments of the higher-dimensional quantum Hall effect, the understanding of higher-dimensional non-commutative geometry has significantly advanced in the past twenty years (see \cite{Hasebe-2010, Karabali-Nair-2004} and references therein).  
We have learned that the higher dimensional non-commutative geometry  on $\mathcal{M}\simeq G/H$ can be obtained by examining  the Landau model on $\mathcal{M}$ in the non-Abelian  monopole background  
\cite{Zhang-Hu-2001,Karabali-Nair-2002,Bernevig-Hu-Toumbas-Zhang-2003,Hasebe-Kimura-2003,Hasebe-2004,Nair-Daemi-2004,Jellal-2005,Hasebe-2010-2,Balli-Behtash-Kurkcuoglu-Unal-2014,Hasebe-2014-2, Coskun-Kurkcuoglu-Toga-2017, Heckman-Tizzano-2018}. Specifically, within the lowest Landau level,  fuzzy manifolds $\mathcal{M}_F$ were successfully realized.  Nonetheless, it should be noticed that the underlying reason for the success is still missing. Furthermore, while the preceding analysis has provided a nice physical understanding of non-commutative geometries, one could argue that these analyses have not revealed unknown matrix geometries. 
Until now, substantial attention has been given to the geometry in the lowest Landau level; however, there is no logical reason for the exclusive presence of non-commutative geometry solely in this level. Indeed, it was demonstrated  that the  higher Landau levels  also give rise to  fuzzy geometries  \cite{Hasebe-2016}, which clearly shows that  level projection to $\it{any}$ Landau level generates non-commutativity.  
With regards to a two-sphere, the emergent non-commutative geometries of the higher Landau levels are the same as that of the lowest Landau level. In this sense, the geometry of higher Landau levels might not be so intriguing. Nevertheless, this does not rule out the possibility of discovering new non-commutative geometries in higher dimensional systems.
Following this idea, explorations of novel quantum matrix geometries have been conducted in various Landau models, such as  relativistic models and supersymmetric models   \cite{Hasebe-2016},  odd dimensional models \cite{Hasebe-2018} and even dimensional models \cite{Hasebe-2020, Hasebe-2021, Hasebe-2022}. It is also worthwhile to mention that quantum matrix geometries associated with the Berezin-Toeplitz quantization have been  intensively studied in recent years \cite{Ishiki-Matsumoto-Muraki-2016, Asakawa-Ishiki-Matsumoto-Matsuura-Muraki-2018, Ishiki-Matsumoto-Muraki-2018, Matsuura-Tsuchiya-2020, Nair-2020, Adachi-Ishiki-Matsumoto-Saito-2020, Adachi-Ishiki-Kanno-2022}. 

Importantly, now the higher dimensional studies  are not only relevant to theoretical interests but also to practical experiments. The idea of the synthetic dimension allows physicists to reach higher dimensional topological physics \cite{Price-et-al-2015,Price-et-al-2016,Ozawa-Price-et-al-2016}. In particular, exotic topological effects of the non-Abelian monopole in  higher dimension have already been observed  through  table top experiments very recently \cite{Ma-Jia-Bi-et-al-2023, Zhang-Chen-Zhang-et-al-2022, Ma-Bi-et-al-2021,Sugawa-et-al-2018}. It is  expected that  physical consequences arising from higher dimensional quantum  geometry will be observed in these experimental systems. 
 
In the present work,  with an appropriate interpretation of  the emergent non-commutative geometry in the Landau models, we introduce a quantum-oriented non-commutative scheme that leverages Landau models as an effective ``tool'' to generate noble quantum geometries. 
Our approach provides a concrete prescription for generating the matrix geometry of the coset manifold $\mathcal{M}\simeq G/H$. It is shown that this scheme encompasses pure quantum Nambu matrix geometry, which cannot be described by conventional non-commutative methods. We also demonstrate that these quantum Nambu matrix geometries give rise to novel classical solutions in Yang-Mills matrix models.

This paper is organized as follows. In Sec.\ref{sec:ncscheme}, we revisit the derivation of the fuzzy two-sphere from the $SO(3)$ Landau model and address the underlying reasons behind the  emergent non-commutative geometry of  the Landau models.    Sec.\ref{sec:so5lan}  presents explicit fuzzy four-sphere matrix coordinates in the $SO(5)$ Landau levels.  We investigate the matrix structures of fuzzy four-spheres  and discuss their basic properties in Sec.\ref{sec:fuzzy4}.  In Sec.\ref{sec:intfuzz}, the nested internal structures of  higher Landau level matrix geometries are exploited. We investigate the continuum limit and the classical geometry of the quantum matrix geometry  using the coherent method and the probe brane method in Sec.\ref{sec:contin}.  
In Sec.\ref{sec:ymmatrixmodel}, we demonstrate that the obtained quantum  matrix geometries realize  unexplored solutions of Yang-Mills matrix models and clarify their physical properties. Sec.\ref{sec:sum} is devoted to summary and discussions.

\section{Quantum-oriented  non-commutative scheme}\label{sec:ncscheme}

In this section, we discuss the underlying mechanism behind the emergent matrix geometry in the simple $SO(3)$ Landau model and apply this observation to propose a prescription for generating matrix geometries of $G/H$.

\subsection{Behind the scene of the emergent matrix geometry}\label{subsec:so3landaumat}

 The $SO(3)$ Landau model is a Landau model on $S^2$  and the Hamiltonian is given by 
\be
H=-\frac{1}{2M}\sum_{i=1}^3(\partial_i +iA_i)^2 |_{r=1},  \label{so3landauham}
\ee
where  $A_i$ denotes the $U(1)$ gauge field of monopole at the origin:  
\be
A_i =-\frac{I}{2r(r+x_3)} \epsilon_{ij3}x_j.
\ee
The index $I/2$ signifies the monopole charge (In the following, we assume $I$ to be a positive integer for simplicity).
While the present system is originally investigated in \cite{Wu-Yang-1976, Haldane-1983}, we will  utilize the concise notation of \cite{Hasebe-2016} in this paper.  
The eigenvalues of the Hamiltonian (\ref{so3landauham}) are obtained as 
\be
E_N=\frac{1}{2M}(I(N+\frac{1}{2})+N(N+1))~~~~(N=0,1,2,\cdots), 
\ee
and the corresponding eigenstates are given by the monopole harmonics 
\be
Y^{(N)}_{m}(\theta, \phi)=\sqrt{\frac{2N+I+1}{4\pi}} ~\mathcal{D}_{N+\frac{I}{2}} (\phi, -\theta, -\phi)_{\frac{I}{2},m}~~~(m=N+\frac{I}{2}, N+\frac{I}{2}-1, \cdots, -(N+\frac{I}{2})), \label{monoharmo}
\ee
where $\mathcal{D}$ denotes the Wigner D-function: 
\be
\mathcal{D}_l(\chi,\theta, \phi) =e^{-i\chi S_z^{(l)}}e^{-i\theta S_y^{(l)}}e^{-i\phi S_z^{(l)}}.
\ee
Here, $S^{(l)}_i$ stand for the $SU(2)$ spin matrices with  spin index $l$. 
We sandwich the coordinates on $S^2$ to derive the corresponding matrix coordinates: 
\be
( X_i^{(N)})_{mn} =\langle Y_m^{(N)} |x_i |Y_n^{(N)} \rangle \equiv \int_{S^2}d\theta d\phi \sin\theta~ {Y_m^{(N)}}^* x_i Y_n^{(N)},
\ee
where 
\be
x_1=\cos\phi\sin\theta, ~~x_2=\sin\phi\sin\theta, ~~x_3=\cos\theta. 
\ee
In the $N$th Landau level, $X_i^{[N]}$ are explicitly obtained as \cite{Hasebe-2016}
\be
X_i^{(N)} = \frac{2I}{(I+2N)(I+2N+2)}~ S_i^{(\frac{I}{2}+N)}, \label{xisiso3}
\ee
which satisfy 
\begin{subequations}
\begin{align}
&X_i^{(N)}X_i^{(N)} =\frac{I^2}{(I+2N)(I+2N+2)}\bs{1}_{I+2N+1}, \\
&[X_i^{(N)},X_j^{(N)}] =i\frac{2I}{(I+2N)(I+2N+2)}\epsilon_{ijk}X_k^{(N)}. 
\end{align}\label{xxxalgesu2}
\end{subequations}
Equation (\ref{xxxalgesu2}) represents the algebra of fuzzy two-sphere \cite{madore-1992}. Note that not only the lowest Landau level but also each of the higher Landau level matrix geometries realizes the fuzzy two-sphere matrix geometry.\footnote{For completeness, we derive the non-commutative geometry in higher Landau levels of the planar Landau model in Appendix \ref{append:planarLandaumodel}.} The physical properties of (\ref{xxxalgesu2}) as a classical solution of Yang-Mills matrix models are discussed in Appendix \ref{sec:fuzzy2}.  

\begin{figure}[tbph]
\center
\includegraphics*[width=140mm]{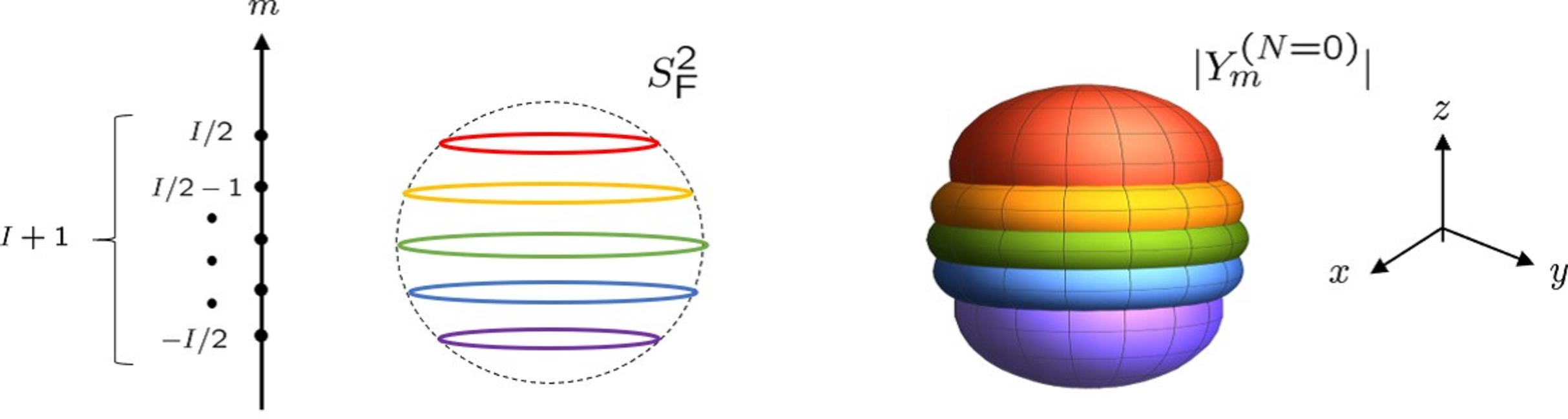}
\caption{Left: the schematic picture of the fuzzy two-sphere for $N=0$ (\ref{xisiso3}). Right: the distributions of the  magnitudes of the monopole harmonics, $|Y_m^{(N=0)}|$,  of  $m=I/2,I/2-1,\cdots ,-I/2$ for $I=4$ are depicted   as the red, orange, green, blue and violet orbitals,  respectively.  The monopole orbitals, $|Y_m^{(N=0)}|$, are localized around  the latitudes $z=2m/I$ on the two-sphere.    }
\label{s2dist.fig}
\end{figure}

We depicted the fuzzy two-sphere and the magnitudes of the monopole harmonics  in the left and the right of Fig.\ref{s2dist.fig}, respectively. One may find an apparent resemblance between  the left and the right pictures.  
The latitudes on the fuzzy two-sphere  represent the degrees of freedom of the matrix geometry, $i.e.$, the ``points'', in the fuzzy space. 
 Obviously, each point on the fuzzy space corresponds to the monopole harmonics or each state  of the $SU(2)$ irreducible representation.  Therefore, one may consider the fuzzy two-sphere to be composed of the $SU(2)$ irreducible representation. 

Reflecting the emergence of the non-commutative geometry, we can obtain the following insight. 
\begin{enumerate}
\item About the role of global symmetry  and irreducible representation: \\ 
The $SO(3)$ global symmetry of  $S^2\simeq SO(3)/SO(2)$  is naturally transformed to the $SU(2)$ symmetry on the matrix geometry side introducing the projective representation of $SO(3)$. 
In the matrix geometry,  an ``uncertainty area'' or a ``point'' corresponds to each state of the $SU(2)$ irreducible  representation. 
The irreducible representation  is ``symmetric'' in the sense that, while each state of an irreducible representation is transformed, the set of states in the irreducible representation remains unchanged under any  $SU(2)$ transformation. In the language of matrix geometry, this means that fuzzy geometry also remains unchanged under $SU(2)$ transformations, as the fuzzy two-sphere is composed of the states in the $SU(2)$ irreducible representation.     Moreover, the $SU(2)$ group is a compact group, and its irreducible representation is a finite-dimensional set with discrete quantum numbers, which aligns with the intuitive notion that a compact non-commutative space consists of finite-dimensional discrete points. In this way, while the fuzzy sphere is a discretized space, it realizes a space symmetric under   $\it{continuous}$  $SU(2)$ transformations, unlike the lattice  space, which is  symmetric only by the discrete translations   corresponding to the lattice spacing.   
This is the specific feature of the matrix geometry composed of the irreducible representation.

\item About the role of the  stabilizer group  and the gauge symmetry: \\
 The stabilizer group $SO(2)$ of $S^2\simeq SO(3)/SO(2)$ is a subgroup of $SO(3)$  that does not change a point on the classical manifold $S^2$ \cite{Nakahara-book}. A point in the classical geometry corresponds to a state of the irreducible representation on the matrix geometry side. Therefore,  the stabilizer group is considered to be  some transformation that does not change that state. The transformation that does not change physical state  is nothing but a gauge transformation.  
   To encapsulate,  the stabilizer group represents redundant symmetry of the $SO(3)$ group in the classical system when representing $S^2\simeq SO(3)/SO(2)$, and  such redundancy is naturally regarded as a gauge symmetry on the quantum mechanical side. 
    Consequently, the stabilizer group $SO(2)$ corresponds to the $U(1)\simeq SO(2)$ symmetry on the quantum mechanical side.     
    It is interesting to see that while  
  the stabilizer symmetry is an $\it{external}$ symmetry on the  classical mechanical side,  it acts as  the $\it{internal}$ symmetry on the quantum mechanical side.\footnote{This suggests that the external space and the internal space should be treated on the same footing in the matrix geometry \cite{Hasebe-2014-1}.}

\item Reinterpretation of the Landau model: \\
The above observations suggest that the matrix geometry corresponding to $S^2\simeq SO(3)/SO(2)$ is  obtained by considering a quantum system with global $SU(2)$ symmetry and $U(1)$ gauge symmetry. As we are dealing with the spatial manifold, the $U(1)$ gauge symmetry introduces the $U(1)$ 
$\textit{vector}$ potential whose field configuration  should be compatible with the $SU(2)$ global symmetry. This necessarily leads to the radially symmetric magnetic field  of the $U(1)$ monopole. Thus, the magnetic field is  just a consequence of the gauge  symmetry.   In this way, we can reproduce the original $SO(3)$ Landau system. It is important to note that the $\textit{primary}$ significance lies in the gauge symmetry itself rather than the magnetic field, although the presence of a magnetic field is commonly believed to be essential for the emergence of non-commutative geometry.

 These speculations provide a natural explanation for why the fuzzy two-sphere geometry has been successfully generated through the analyses of   the $SO(3)$ Landau model.

\end{enumerate}

\subsection{Non-commutative scheme  for generating the matrix geometry}\label{sec:idea}

With the above understanding, we now propose a  prescription for obtaining the matrix geometry of the general coset manifold, $\mathcal{M}\simeq G/H$.   We will utilize the quantum mechanics as a tool for generating matrix geometries.  
What we need to do is simply replace the $SO(3)$ in the above discussions with $G$ and $SO(2)$ with $H$.\footnote{While we will assume that $G$ is a compact group with finite dimensional irreducible representations, our discussions can also be applied to non-compact groups with discrete series of infinite dimensional irreducible representations \cite{Hasebe-2012}.}

\subsubsection{General prescription}

\begin{figure}[tbph]
\hspace{-0.3cm}
\includegraphics*[width=160mm]{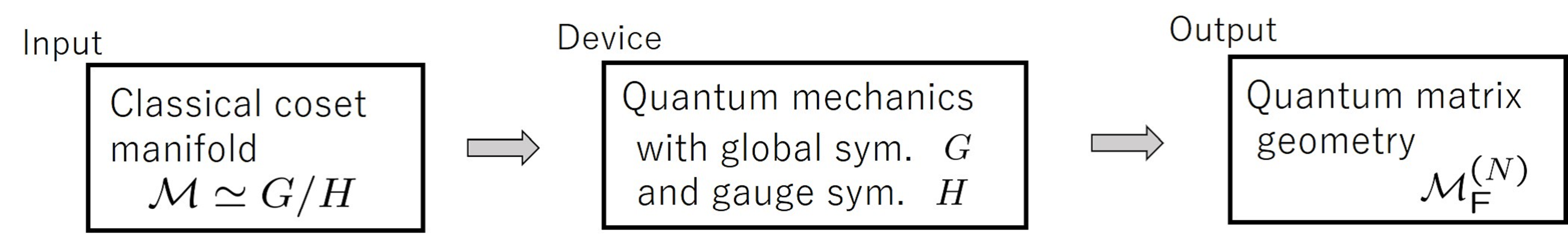}
\caption{Flow of the procedure. 
}
\label{proc.fig}
\end{figure}

\begin{enumerate}
\item  Consider quantum mechanics with gauge symmetry $H$ on base-manifold $\mathcal{M}$: 
\be
-\frac{1}{2M}\sum_{a}{D_a}^2\biggr|_{\mathcal{M}} \label{hamonm}
\ee
where $D_a=\partial_a+iA_a$ are  covariant derivatives with the gauge field $A_a$ of the gauge group $H$. The gauge field configuration has to be chosen to be compatible with the symmetry $G$ of the  base-manifold $\mathcal{M}$.
\item Solve the eigenvalue problem of the  Hamiltonian (\ref{hamonm}) to derive the degenerate eigenstates of each energy level $E_N$: 
\be
|\psi_{N, \alpha}\rangle. \label{ngeneirep}
\ee
The set of degenerate eigenstates constitute an irreducible representation of $G$.\footnote{In general, the Hamiltonian may possess  symmetries  other than $G$. In such a case,  the degenerate eigenstates constitute  states of an irreducible representation of the entire symmetry. See  Sec.\ref{sec:fuzzythree}.}  
\item
Derive the matrix elements of of the coordinates $x_a$ of $\mathcal{M}$ utilizing (\ref{ngeneirep}) to construct the matrix coordinates of $\mathcal{M}^{(N)}_{F}$: 
\be
(X_a^{(N)})_{\alpha\beta} =\langle \psi_{N,\alpha}|x_a|\psi_{N,\beta}\rangle\equiv \int_{\mathcal{M}} d\Omega ~\psi_{N,\alpha}^{\dagger} x_a  \psi_{N,\beta}, 
\ee
where $d\Omega$ is the area element of $\mathcal{M}$. 
\end{enumerate}
Notice each energy level $N$  hosts its own  matrix geometry $X_a^{[N]}$, and  distinct energy levels yield different quantum matrix coordinates  in general.  
 Consequently, multiple  quantum  geometries will be obtained from  a single classical manifold.  
The flow of this procedure is depicted in Fig.\ref{proc.fig}. 
 
\subsubsection{Advantages}

Here, we will outline the advantages of the present construction.

\begin{enumerate}
\item 
The first merit  is that we do need to  worry about mathematical inconsistency. In the present scheme,  non-commutative geometry is not postulated a priori but is what  emerges in each of the energy levels. As the original quantum system is totally  physical and the existence of mathematically consistent Hilbert space behind the quantum mechanics is founded,  there is no need to be concerned about mathematical inconsistencies.\footnote{This is inspired by  the idea of Ref.\cite{Zhang-2002}. }

\item Following the above simple prescription,  we can mechanically derive  matrix geometries for $\it{arbitrary}$ classical manifolds of the type $\mathcal{M}\simeq G/H$. Notably,  odd dimensional manifolds  are also within the realm of this scheme. Therefore, this scheme is not restricted to the symplectic manifolds unlike the  conventional quantization methods in which the quantization is basically executed by replacing the Poisson bracket with the commutator. This suggests that the present scheme is beyond the non-commutative geometry based on the canonical commutator formalism.

\item

 The present non-commutative scheme is primarily based on irreducible representations 
 of  quantum mechanics. In this sense, this  may be referred to as  a quantum-oriented scheme. The emergent matrix geometries may even encompass  pure quantum geometries that do not have their classical counterparts.  We may explore  quantum geometries that have eluded in the conventional non-commutative schemes.  
\end{enumerate}

\subsection{General Properties}\label{sec:specpro}

To examine specific properties of the  present scheme, let us consider even dimensional  spheres,  
\be
S^{2k}\simeq SO(2k+1)/SO(2k). 
\ee 

\subsubsection{Covariance}

We assume that the global symmetry $SO(2k+1)$ of $S^{2k}$ is given by  
\be
x_{a=1,2,\cdots, 2k+1} ~~\rightarrow~~R_{ab}x_b ~~~~(R_{ab}~\in~SO(2k+1)).  \label{ratrans}
\ee
The stabilizer group is defined so that the condition 
$x_a=\delta_{a, 2k+1}$ 
does not change, which is the $SO(2k)$ transformation: 
\be
x_{\mu=1,2,\cdots, 2k} ~~\rightarrow~~R_{\mu\nu}x_{\nu} ~~~~(R_{\mu\nu}~\in~SO(2k)). 
\label{rmtrans}
\ee
Transformations  (\ref{ratrans}) and (\ref{rmtrans}), respectively, correspond to  the following transformations on the quantum mechanics side: 
\be
|\psi_{\alpha}^{(i)}\rangle  ~~\rightarrow ~~|\psi_{\beta}^{(i)}\rangle U_{\beta\alpha} ~~~~~(U~\in ~Spin(2k+1)), \label{qgtrans}
\ee
and 
\be
|\psi_{\alpha}^{(i)}\rangle  ~~\rightarrow ~~g_{ij}|\psi_{\alpha}^{(j)}\rangle  ~~~~~~(g~\in ~Spin(2k)).\label{qhtrans}
\ee
Equation (\ref{qgtrans}) stands for the global transformation, and $\alpha$ denote the index of the irreducible representation of the  $Spin(2k+1)$. Similarly,  Eq.(\ref{qhtrans}) represents the gauge transformation,  and $i$ signify that of the gauge group $Spin(2k)$. 
Under these transformations, $X_a$ behave as  
\be
(X_a)_{\alpha\beta} =\sum_i \langle \psi_{\alpha}^{(i)}|x_a|\psi^{(i)}_{\beta}\rangle ~~\rightarrow~~\sum_i  U_{\alpha' \alpha}^* \langle \psi_{\alpha'}^{(i)}|x_a|\psi^{(i)}_{\beta'}\rangle U_{\beta' \beta} = (U^{\dagger}X_a U)_{\alpha\beta} =R_{ab}(X_b)_{\alpha\beta},
\ee
and 
\be
(X_a)_{\alpha\beta} =\sum_i \langle \psi_{\alpha}^{(i)}|x_a|\psi^{(i)}_{\beta}\rangle ~~\rightarrow~~\sum_{j, k} \overbrace{\sum_{i}g_{ij}^*  g_{ik}}^{=\delta_{jk}} \langle   \psi_{\alpha}^{(j)}|x_a|\psi^{(k)}_{\beta}\rangle=\sum_j \langle \psi_{\alpha}^{(j)}|x_a|\psi_{\beta}^{(j)}\rangle =(X_a)_{\alpha\beta}. 
\ee
The matrix coordinates thus transform as the $SO(2k+1)$ vector, similar to the classical   coordinates on $S^{2k}$, and they are gauge invariant.  
Generally for $\mathcal{M}\simeq G/H$, 
the matrix coordinates  are $H$ gauge invariant and  transform under  $G$ in the same way as the classical coordinates of the original  manifold $\mathcal{M}$.

\subsubsection{Beyond the commutator formalism }\label{subsec:beycomform}

 In the well known construction of the fuzzy $2k$-sphere \cite{Castelino-Lee-Taylor-1997, Grosse-Klimcik-Presnajder-1996}, the matrix coordinates are given by the totally symmetric combination of the gamma matrices ,  which satisfy the following commutation relations 
\begin{subequations}
\begin{align}
&[X_a, X_b]=4i\Sigma_{ab}, \label{1re}\\
&[X_a, \Sigma_{bc}] =-i\delta_{ab}X_c +i\delta_{ac}X_b, \label{2re}\\
&[\Sigma_{ab}, \Sigma_{cd}] =i\delta_{ac}\Sigma_{bd} -i\delta_{ad}\Sigma_{bc} +i\delta_{bd}\Sigma_{ac} -i\delta_{bc}\Sigma_{ad}. \label{3re}
\end{align}\label{xsigmatral}
\end{subequations}
The commutators of $X_a$ yield  new matrices $\Sigma_{ab}$ (\ref{1re}), which are the generators of $SO(2k+1)$. In total, $X_a$ and $\Sigma_{ab}$ together form the $SO(2k+2)$ algebra. Such a matrix geometry is known to emerge in the lowest Landau level  of the $SO(2k+1)$ Landau model \cite{Hasebe-2014-1}. The lowest Landau level matrix geometry is well described by the commutator formalism. On the other hand, for the higher Landau levels, some subtleties occur. 
The $SO(2k+1)$ angular momentum operators in the $SO(2k)$ monopole background are constructed as  \cite{Hasebe-Kimura-2003} 
\be
L_{ab} =-ix_a (\partial_b+iA_b) +ix_b (\partial_a+iA_a) +\frac{1}{r^2}F_{ab},
\ee
which satisfy the $SO(2k+1)$ algebra: 
\be
[L_{ab}, L_{cd}] =i\delta_{ac}L_{bd} -i\delta_{ad}L_{bc} +i\delta_{bd}L_{ac} -i\delta_{bc}L_{ad}. \label{so2k+1al}
\ee
Since the coordinates $x_a$ on $S^{2k}$ transform as an $SO(2k+1)$ vector, the algebra associated with the $SO(2k+1)$ transformation is represented as 
\be 
[x_a, L_{bc}] =-i\delta_{ab}x_c +i\delta_{ac}x_b. 
\label{xlab}
\ee
Let us construct matrix coordinates for a given irreducible representation of  $SO(2k+1)$, $\{\psi^{(r)}_1, \psi_2^{(r)},\cdots \psi^{(r)}_{D^{(r)}}\}$: 
\be
(X^{(r)}_a)_{\alpha\beta} \equiv \langle \psi^{(r)}_\alpha| x_a|\psi^{(r)}_\beta\rangle, ~~~~~~~~(\Sigma^{(r)}_{ab})_{\alpha\beta} 
\equiv \langle \psi^{(r)}_\alpha| L_{ab}|\psi^{(r)}_\beta\rangle. \label{definitionxasigmaab}
\ee
It is important to note that the completeness relation holds for the total set of the  irreducible representations: 
\be
\sum_r \sum_{\alpha=1}^{D^{(r)}} |\psi_{\alpha}^{(r)}\rangle \langle \psi_{\alpha}^{(r)}|=1,
\ee
but  $\it{not}$ for each individual irreducible representation:  
\be
 \sum_{\alpha=1}^{D^{(r)}} |\psi_{\alpha}^{(r)}\rangle \langle \psi_{\alpha}^{(r)}|\neq 1. \label{notunit}
\ee
Equation (\ref{notunit}) is a direct consequence of the level projection which is the heart of non-commutative  geometry  \cite{Hasebe-2016}. Due to Eq.(\ref{notunit}),   $X_a^{(r)}$ (\ref{definitionxasigmaab}) generally become non-commutative matrices, whereas  the original  coordinates $x_a$ are commutative quantities.   
From the property of the irreducible representation
\be
\langle \psi^{(r)}_\alpha| L_{ab}|\psi_\beta^{(r')}\rangle =(\Sigma_{ab}^{(r)})_{\alpha\beta} \delta_{rr'}, \label{labmat} 
\ee
one may easily reproduce the lower two equations of (\ref{xsigmatral}) using   Eqs.(\ref{so2k+1al}) and  (\ref{xlab}).  On the other hand,  unlike Eq.(\ref{labmat}), the matrix coordinates  are not completely block diagonalized, $\langle \psi^{(r)}_\alpha| x_{a}|\psi_\beta^{(r')}\rangle \neq \langle \psi^{(r)}_\alpha| x_{a}|\psi_\beta^{(r)}\rangle  \delta_{rr'}$  (see Sec.\ref{subsec:higherlanmat} for more details).  
Consequently, the first relation (\ref{1re}) turns out to be  questionable, 
\be
[X^{(r)}_a, X^{(r)}_b] \overset{?}\propto i\Sigma^{(r)}_{ab}. 
\ee
 Equation (\ref{1re}) is  not  guaranteed in general. So, 
  if the Lie algebraic geometry fails, what kind of geometry will emerge? That is the topic that we shall discuss in Secs.\ref{sec:fuzzy4} and \ref{sec:evenhigh}.  The failure of Eq.(\ref{1re}) implies that the present  scheme is beyond the realm of the conventional commutator formalism.


Here, we also mention  relationship to the Berezin-Toeplitz quantization. 
The Berezin-Toeplitz quantization is a method that maps a function to a finite dimensional matrix \cite{Schlichenmaier-2001, Nair-2020, Ali-Englis-2005}.  In this sense, the Berezin-Toeplitz quantization shares the same spirit with the present scheme. However, Berezin-Toeplitz quantization is primarily concerned with symplectic manifolds and is based on commutator formalism.    The Kernel employed in the Berezin-Toeplitz quantization corresponds to the zero-modes of the Dirac-Landau operator whose zero-modes  are essentially equivalent to the lowest Landau level eigenstates \cite{Hasebe-2020, Hasebe-2016}.
 Therefore, the Berezin-Toeplitz quantization is thus closely related to the lowest Landau level matrix geometry  and   can be viewed as a special case of the present scheme.\footnote{Recently mathematicians are also interested in higher Landau levels from the perspective of the Berezin-Toeplitz quantization  \cite{Kordyukov-2020,Charles-2020}.  } 
We will revisit this in Sec.\ref{sec:fuzzy4}.

\section{Matrix coordinates from the $SO(5)$ Landau model}\label{sec:so5lan}

 In this section, we will directly apply the present scheme to generate quantum matrix geometries for $S^4$. Using the $SO(5)$ Landau model, we will derive the complete form of matrix coordinates in arbitrary Landau levels. This section also includes a review of Ref.\cite{Hasebe-2020}.

\subsection{The $SO(5)$ Landau model}\label{subsec:higherlanmat}

Since $S^4\simeq SO(5)/SO(4)$, we need to consider a quantum mechanics on $S^4$ with $Spin(4)$ gauge degrees of freedom. For the $Spin(4)$ gauge field configuration to respect the $SO(5)$ global symmetry of $S^4$, we place a $Spin(4)$ monopole at the origin. While  the Landau model in such a $Spin(4)$ monopole background has been investigated \cite{Hasebe-2022},  we will consider a simpler system by taking one $SU(2)$ from the $Spin(4)\simeq SU(2)\otimes SU(2)$. In the following, we then consider a quantum mechanics on $S^4$ in the $SU(2)$ monopole background, which was originally introduced in Refs.\cite{Yang-1978-1, Yang-1978-2, Zhang-Hu-2001}.

Let us briefly discuss such a $SO(5)$ Landau model with a modern notation \cite{Hasebe-2020}. 
The $SO(5)$ Landau 	Hamiltonian is given by 
\be
H=-\frac{1}{2M}\sum_{a=1}^5 {D_a}^2|_{r=0} =\frac{1}{2M}\sum_{a<b}{\Lambda_{ab}}^2
\ee
where $D_a=\partial_a+iA_a$ and 
\be
\Lambda_{ab} =-ix_aD_b +ix_bD_a. 
\ee
The gauge field is chosen to be  Yang's $SU(2)$ monopole: 
\be
A_{\mu=1,2,3,4}=-\frac{1}{r(r+x_5)}\bar{\eta}_{\mu\nu}^i x_{\nu} S^{(I/2)}_i ,~~~~A_5=0,~~~ ~~(I=1,2,3,\cdots)\label{zhanghusu2gauge}
\ee
with $\bar{\eta}_{\mu\nu}^i=\epsilon_{\mu\nu i4} -\delta_{\mu i}\delta_{\nu 4}+\delta_{\nu i}\delta_{\mu 4}$. 
The $SO(5)$ Landau Hamiltonian is equal to the $SO(5)$ Casimir up to a constant. Consequently, the energy eigenvalues are specified by  two indices of the $SO(5)$ Casimir, $(p,q)_5=(N+I, N)_5$. The $SO(5)$ Landau levels are explicitly  given by 
\be
E_N =\frac{1}{2M}((N+1)I+N(N+3))~~~~~~(N=0,1,2,\cdots).
\ee
The eigenstates of each of the Landau levels form  an irreducible representation of $SO(5)$ and are referred to as the $SO(5)$ monopole harmonics in this paper\footnote{In the original  paper \cite{Yang-1978-2}, they are referred to as the $SU(2)$ monopole harmonics.}   to emphasize its $SO(5)$ covariance. 
We parametrize the  coordinates of the four-sphere with  a unit radius as      
\be
x_{\mu=1,2,3,4} =\sin\xi~y_{\mu}, ~~~x_5=\cos\xi ~~~~(\sum_{\mu=1}^4 y_\mu y_\mu=1),
\ee
where $\xi$ represents the azimuthal angle and $y_m$ denote the coordinates of  (normalized) $S^3$-hyper-latitude: 
\be
y_1=\sin\chi\sin\theta\cos\phi, ~~y_2=\sin\chi\sin\theta\sin\phi, ~~y_3=\sin\chi\cos\theta, ~~y_4=\cos\chi.       
\label{expliys}
\ee 
Normalized $SO(5)$ monopole harmonics are represented as 
\be
{\Psi}_{N; j, m_j ;k, m_k}(\xi, \chi, \theta, \phi) 
= G_{N,j,k}(\xi)\cdot \bs{Y}_{j, m_j; k, m_k}(\chi, \theta, \phi),
 \label{x5so5monopolehamonicsnorm}
\ee
where 
\begin{subequations}
\begin{align}
&G_{N,j,k}(\xi) = (-1)^{2j+1}~\sqrt{N+\frac{I}{2}+\frac{3}{2}}~ \frac{1}{\sin\xi}~ d_{N+\frac{I}{2}+1, -j+k, j+k+1} (\xi), 
\label{exdefgmfunc} \\
&\bs{Y}_{j,  m_j;~ k,  m_k}(\chi, \theta, \phi) =\sum_{m_R=-j}^j \begin{pmatrix}
 C_{j, m_R; ~\frac{I}{2}, \frac{I}{2}}^{k, m_k} ~\Phi_{j,m_j;~j, m_R}(\chi, \theta, \phi) \\
 C_{j, m_R; ~\frac{I}{2}, \frac{I}{2}-1}^{k, m_k} ~\Phi_{j,m_j;~j, m_R}(\chi, \theta, \phi)\\
\vdots \\
 C_{j, m_R; ~\frac{I}{2}, -\frac{I}{2}}^{k, m_k} ~\Phi_{j,m_j;~j, m_R}(\chi, \theta, \phi) 
\end{pmatrix}. \label{vectorlikespinsphereharmo}
\end{align}
\end{subequations}
Here, $d_{l, m, m'}(\xi)\equiv \mathcal{D}_l(0, \xi, 0)_{m, m'}$ in (\ref{exdefgmfunc}) stand for Wigner's small $D$-matrices,   $C$s in (\ref{vectorlikespinsphereharmo}) represent the Clebsch-Gordan coefficients, and $\Phi$s  in (\ref{vectorlikespinsphereharmo}) denote  the $SO(4)$ spherical harmonics  \cite{Hasebe-2018}. The $SO(5)$ monopole harmonics satisfy 
\be
\int_{S^4}d\Omega_4~ {\Psi}_{N; j, m_j ;k, m_k}^{\dagger}{\Psi}_{N'; j', m'_j ;k', m'_k} = \delta_{N,N'}\delta_{j, j'}\delta_{k, k'}\delta_{m_j, m'_j}\delta_{m_k, m'_k} \label{orthonormalconditionso5mono}
\ee
and 
\be
\sum_{n=0}^N \sum_{s=-I/2}^{I/2} \sum_{m_j =- j}^{j}\sum_{m_k =- k}^k \Psi_{N; j, m_j; k, m_k} \Psi_{N; j, m_j; k, m_k}^{\dagger} 
=\frac{(N+1)(I+N+2)(I+2N+3)}{16\pi^2}\bs{1}_{I+1},
\ee
where $d\Omega_4 \equiv d\xi d\chi d\theta d\phi ~\sin^3\xi \sin^2\chi\sin\theta $. 

\begin{figure}[tbph]
\center
\includegraphics*[width=80mm]{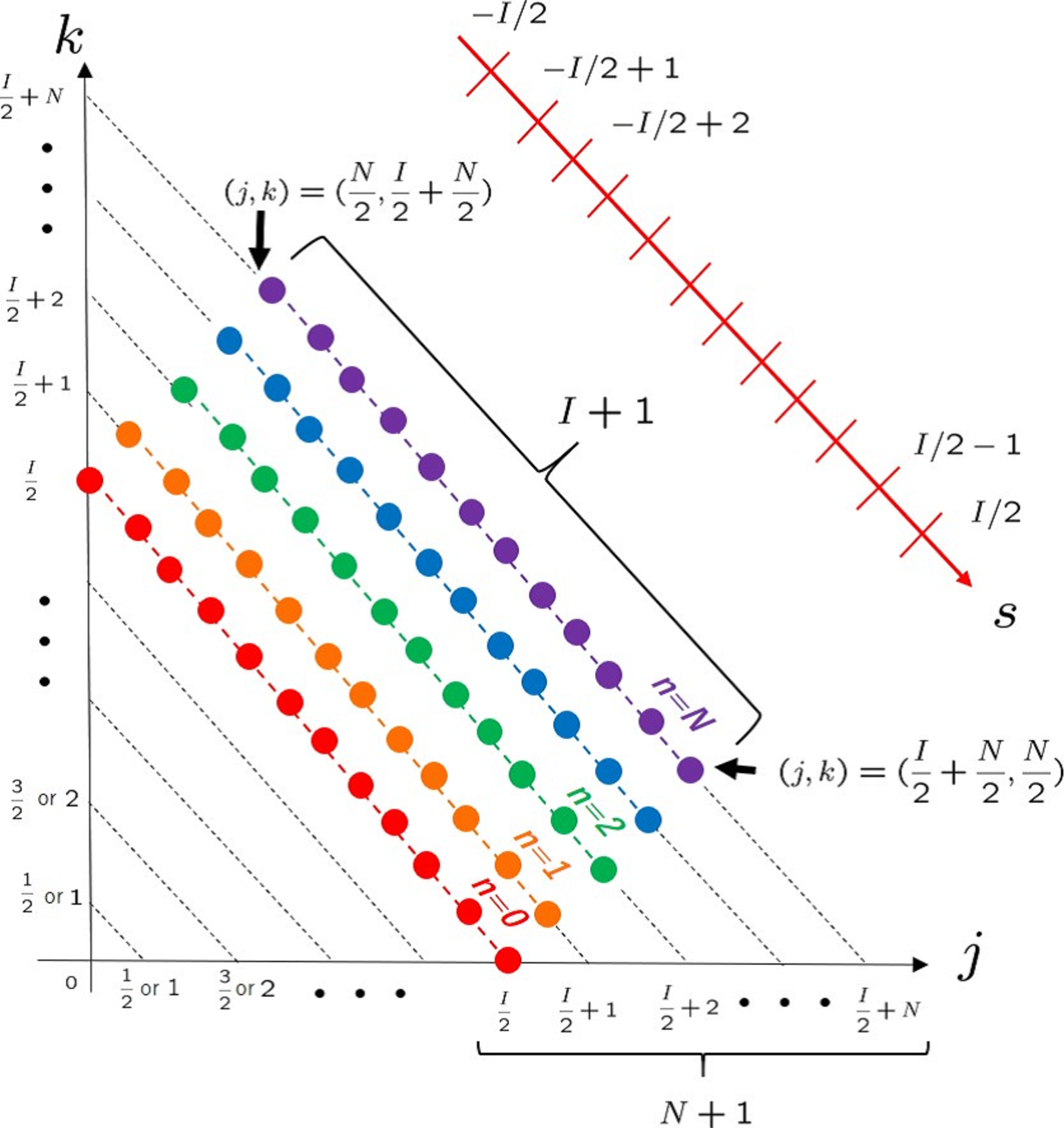}
\caption{Each of the filled circles represents an $SO(4)$ irreducible representation $(j,k)$.   The $SO(4)$ irreducible representations denoted by the filled circles amount to the $SO(5)$ irreducible representation $(p,q)_5=(N+I, N)_5$. (Taken from \cite{Hasebe-2020}.)   }
\label{so5diagram.fig}
\end{figure}

The $SO(5)$ irreducible representation $(p,q)_5=(N+I, N)_5$ is decomposed as (see Fig.\ref{so5diagram.fig})
\be
(N+I, N)_5 =\bigoplus_{n=0}^N (n) =\bigoplus_{n=0}^N \bigoplus_{s=-I/2}^{I/2}~(j,k)_4,
\ee
where 
\be
(n)\equiv \bigoplus_{s=-I/2}^{I/2}~(j,k)_4
\ee
signifies the set of the $SO(4)$ irreducible representations in the $SO(4)$ line  (the oblique line of the same color in Fig.\ref{so5diagram.fig}). The $N$th $SO(5)$ Landau level eigenstates consist of  $SO(4)$ irreducible representations on the $SO(4)$ lines with  $n=0,1,2,\cdots N$.   The $SO(4)$ bi-spin index, $(j, k)_4$, is defined as    
\be
(j, k)_4\equiv (\frac{n}{2}+\frac{I}{4} +\frac{s}{2}, ~\frac{n}{2}+\frac{I}{4} -\frac{s}{2})_4, \label{formulajandk}
\ee
with 
\be
n=0, 1, 2, 3,  \cdots, N, ~~~~~~~~~s 
 =\frac{I}{2}, ~\frac{I}{2}-1, \cdots, -\frac{I}{2}. 
\label{jminuskcond}
\ee 
The dimension of the $SO(4)$ irreducible representation (\ref{formulajandk}) is given by 
\be
d(n, I, s)=(2j+1)(2k+1) =(n+\frac{I}{2}+s+1)(n+\frac{I}{2}-s+1)=d(n, I, -s),
\ee
and that  of the $SO(4)$ line  is 
\be
d(n,I)\equiv \sum_{s=-I/2}^{I/2} d(n, I, s) =\frac{1}{6}(I+1)(I^2+(6n+5)I+6(n+1)^2).
\ee
Consequently, the $N$th Landau level degeneracy is  counted as  
\be
D(N, I) = \sum_{n=0}^N d(n, I)=\sum_{n=0}^N \sum_{s=-I/2}^{I/2} d(n, I, s)=\frac{1}{6}(N+1)(I+1)(I+N+2)(I+2N+3). \label{dnidnis}
\ee

\subsection{Matrix coordinates}\label{subsec:matrixcoor}

\begin{figure}[tbph]
\center
\hspace{-1cm}
\includegraphics*[width=150mm]{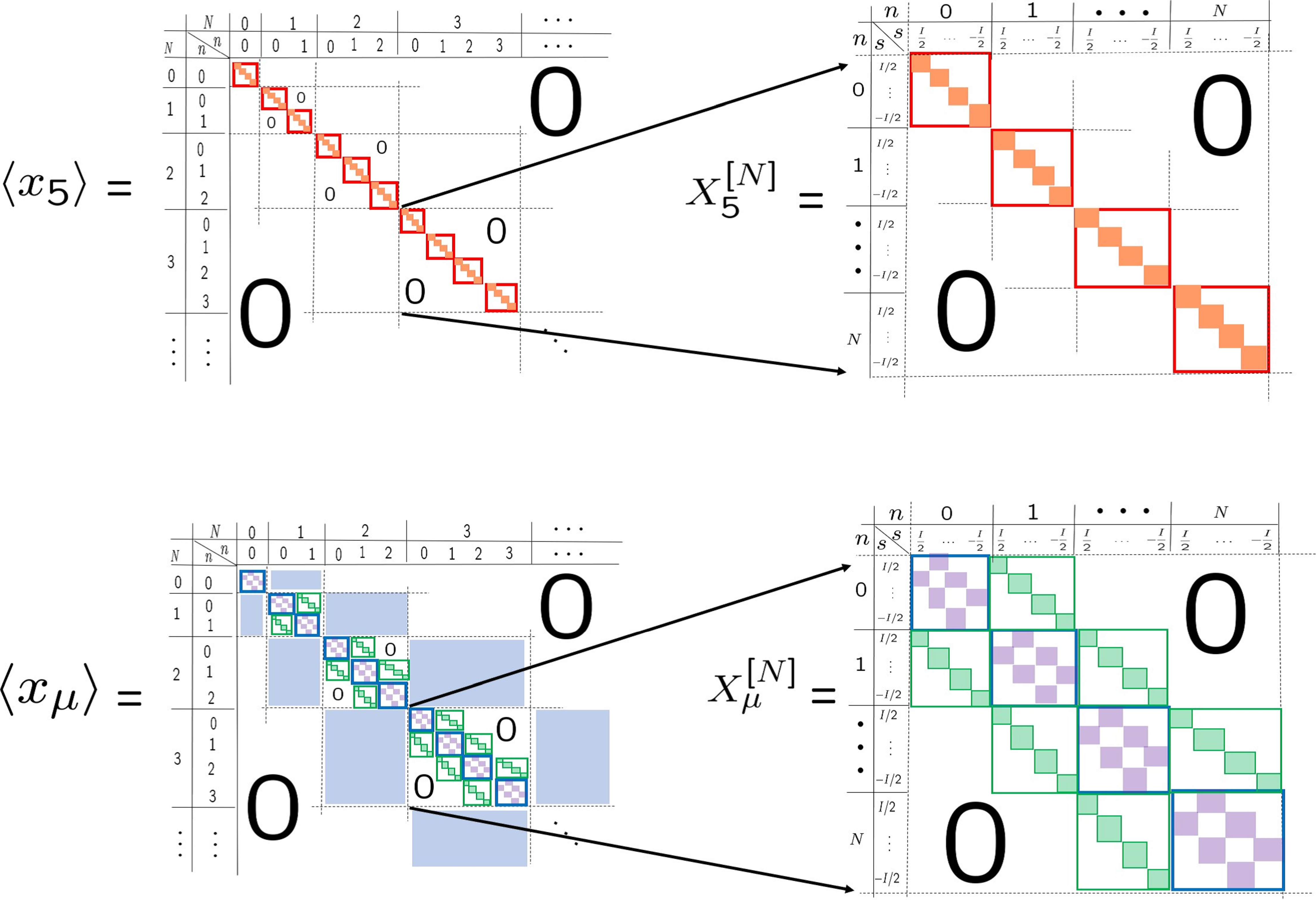}
\caption{The matrix coordinate representation of $x_a$. }
\label{xbk.fig}
\end{figure}

The matrix coordinates have non-zero components only within the same Landau level and among adjacent Landau levels \cite{Hasebe-2020}:
\begin{subequations}
\begin{align}
&\langle x_{5}\rangle \neq 0  ~\rightarrow~\Delta N=0, \label{ntransxm15}\\
&\langle x_{\mu}\rangle \neq 0  ~\rightarrow~\Delta N =0,~ \pm 1. \label{ntransxm14}
\end{align}
\end{subequations}
See the left of Fig.\ref{xbk.fig}  where  the non-zero matrix elements are denoted as the shaded color regions.   
Under the $SO(4)$ rotation around the fifth axis, $x_5$ behaves as a scalar $(j,k)=(0,0)$, while $x_{\mu}$ transform as a bi-spinor $(j,k)=(1/2, 1/2)$.  
From (\ref{formulajandk}), we can see that the $SO(4)$ selection rule implies that 
 non-zero matrix coordinates exist  only for 
\begin{subequations}
\begin{align}
&X_5^{[N]} \neq 0  ~\rightarrow~(\Delta n, \Delta s) =(0, 0), \label{transxm15}\\
&X_{\mu}^{[N]}\neq 0  ~\rightarrow~(\Delta n, \Delta s) =(\pm 1, 0), ~~(0, \pm 1). \label{transxm14}
\end{align}
\end{subequations}
Equations (\ref{transxm15})/(\ref{transxm14}) correspond to the upper/lower right of Fig.\ref{xbk.fig}.  There are two cases in which  $X_{\mu}^{[N]}$ take  finite values.  The first  is  $(\Delta n, \Delta s) =(\pm 1, 0)$ that corresponds  to the green shaded rectangles in Fig.\ref{xbk.fig}, representing  transitions between two adjacent  $SO(4)$ lines in Fig.\ref{so5diagram.fig}. The second 
$(\Delta n, \Delta s) =(0, \pm 1)$ 
corresponds to the small purple shaded rectangles in Fig.\ref{xbk.fig},  signifying transitions between two adjacent $SO(4)$ irreducible representations on  same $SO(4)$ lines in Fig.\ref{so5diagram.fig}. 
With this in mind, we will explicitly evaluate the matrix elements of $x_a$.

We can perform integrations of the azimuthal part and the $S^3$-hyper-latitude part separately. 
For instance, the ortho-normal condition (\ref{orthonormalconditionso5mono}) is evaluated as 
\be
\langle \Psi_{N'; j', m'_j; k', m'_k} |\Psi_{N; j, m_j; k, m_k}\rangle = \langle G_{N', j',k'} |G_{N, j, k} \rangle \cdot \langle Y_{j', m'_j; k', m'_k}| Y_{j, m_j; k, m_k} \rangle ,
\ee
where 
\begin{subequations}
\begin{align}
& \langle G_{N', j,k} |G_{N, j, k} \rangle=\int_0^{\pi}d\xi \sin^3\xi ~ G_{N, j, k} (\xi)^*~G_{N', j, k} (\xi)=\delta_{N, N'},\\
&\langle Y_{j', m'_j; k', m'_k}| Y_{j, m_j; k, m_k} \rangle=\int_{S^3}d\chi d\theta d\phi~\sin^2\chi~\sin\theta ~{ \bs{Y}_{j, m_j; k,  m_k}(\chi,\theta,\phi)}^{\dagger}~\bs{Y}_{j', m'_j; k', m'_k}(\chi,\theta,\phi) =\delta_{jj'}\delta_{m_j m'_j}\delta_{kk'}\delta_{m_k m'_k}. 
\end{align}
\end{subequations}
The matrix elements of $X_5^{[N]}$ are derived as\footnote{The minus sign  in (\ref{expx5mat}) is not essential but  added  for later convenience.}        
\be 
 X_5^{[N]}~~:~~-\langle \Psi_{N; j', m'_j; k', m'_k}|x_5 |\Psi_{N; j, m_j; k, m_k}\rangle 
= -\langle G_{N, j,k} |x_5|G_{N, j, k} \rangle \cdot  \delta_{j, j'}\delta_{k, k'}\delta_{m_j, m'_j}\delta_{m_k, m'_k}, \label{expx5mat}
\ee
where  
\be 
-\langle G_{N, j,k} |x_5|G_{N, j, k} \rangle 
=\frac{2n+I+2}{(2N+I+2)(2N+I+4)} \cdot 2s .  
\label{impint}
\ee 
The matrix coordinate (\ref{impint}) takes equally spaced discrete values specified by  $s=I/2, I/2-1, \cdots, -I/2$, which are regarded as the hyper-latitudes on  fuzzy four-sphere.   This structure is quite similar to that of  the fuzzy two-sphere (Fig.\ref{s2dist.fig}). However notice that   
 while the latitudes of fuzzy two-sphere are not degenerate,  the hyper-latitudes of fuzzy four-sphere are  degenerate,  resulting in an intriguing internal structure as we shall discuss in Sec.\ref{sec:intfuzz}.   
Next, we turn to  
\begin{align}
&X_{\mu}^{[N]}~~:~~\langle \Psi_{N; j', m'_j; k', m'_k}|x_{\mu} |\Psi_{N; j, m_j; k, m_k}\rangle \nn\\
&~~~~~~~~~~~~~~~~~~~
= \sum_{\sigma, \tau=+,-} \langle G_{N, j+\frac{\sigma}{2},k+\frac{\tau}{2}} |\sin \xi|G_{N, j, k} \rangle ~\langle Y_{j+\frac{\sigma}{2}, m'_j; k+\frac{\tau}{2}, m'_k}|y_{\mu} |Y_{j, m_j; k, m_k}\rangle ~ \delta_{j', j+\frac{\sigma}{2}}~\delta_{k', k+\frac{\tau}{2}}. \label{matrixelexmucom} 
\end{align}
Here, the azimuthal part is evaluated as 
\be
\langle G_{N, j+\frac{\sigma}{2},k+\frac{\tau}{2}} |\sin \xi|G_{N, j, k} \rangle = \delta_{\sigma,\tau}\langle G_{N, j+\frac{\sigma}{2},k+\frac{\sigma}{2}} |\sin \xi|G_{N, j, k} \rangle+\delta_{\sigma,-\tau}\langle G_{N, j+\frac{\sigma}{2},k-\frac{\sigma}{2}} |\sin \xi|G_{N, j, k} \rangle, \label{calnnmat}
\ee
with\footnote{ 
In the derivation of (\ref{calnnmat}), we used the formulas,  
\begin{subequations}
\begin{align}
&\int_0^{\pi} d\theta~\sin\theta~d_{l,m',n}(\theta)~\sin\theta~d_{l,m,n}(\theta) |_{m'=m\pm 1} =\frac{2n}{(2l+1)(l+1)l}\sqrt{(l\pm m+1)(l\mp m)}, \\
&\int_0^{\pi} d\theta~\sin\theta~d_{l,m,n'}(\theta)~\sin\theta~d_{l,m,n}(\theta) |_{n'=n\pm 1} =-\frac{2m}{(2l+1)(l+1)l}\sqrt{(l\pm n+1)(l\mp n)}.  
\end{align}
\end{subequations}
}
\begin{align}
&\langle G_{N, j+\frac{\sigma}{2},k+\frac{\sigma}{2}} |\sin \xi|G_{N, j, k} \rangle= -\frac{4s}{(2N+I+2)(2N+I+4)}\sqrt{(N-n+\frac{1-\sigma}{2})(N+n+I+2+\frac{1+\sigma}{2}) }, \nn\\ 
&\langle G_{N, j+\frac{\sigma}{2},k-\frac{\sigma}{2}} |\sin \xi|G_{N, j, k} \rangle= -\frac{4n+2I+4}{(2N+I+2)(2N+I+4)}\sqrt{(N+\frac{I}{2}-\sigma s +1)(N+\frac{I}{2}+\sigma s + 2) }. \label{formulasinximat}
\end{align}
The $S^3$-hyper-latitude part is
\begin{align}
&\langle Y_{j+\frac{\sigma}{2}, m'_j; ~k+\frac{\tau}{2}, m_k'}|y_{1}|Y_{j, m_j;~ k, m_k}\rangle=  \frac{\sqrt{(2j+1)(2k+1)}}{2} (-1)^{n+I+\frac{\tau}{2}} 
\begin{Bmatrix}
j+\frac{\sigma}{2} & k+\frac{\tau}{2} & \frac{I}{2} \\
k & j & \frac{1}{2}
\end{Bmatrix} \sum_{\kappa=+, -}(-1)^{\frac{\kappa-1}{2}} C_{\frac{1}{2}, \frac{\kappa}{2}; j, m_j}^{j+\frac{\sigma}{2}, m'_j}C_{\frac{1}{2}, \frac{\kappa}{2}; k, m_k}^{k+\frac{\tau}{2}, m'_k} , \nn\\
&\langle Y_{j+\frac{\sigma}{2}, m'_j; ~k+\frac{\tau}{2}, m_k'}|y_{2}|Y_{j, m_j;~ k, m_k}\rangle=-i  \frac{\sqrt{(2j+1)(2k+1)}}{2} (-1)^{n+I+\frac{\tau}{2}} 
\begin{Bmatrix}
j+\frac{\sigma}{2} & k+\frac{\tau}{2} & \frac{I}{2} \\
k & j & \frac{1}{2}
\end{Bmatrix} \sum_{\kappa=+, -} C_{\frac{1}{2}, \frac{\kappa}{2}; j, m_j}^{j+\frac{\sigma}{2}, m'_j}C_{\frac{1}{2}, \frac{\kappa}{2}; k, m_k}^{k+\frac{\tau}{2}, m'_k} , \nn\\
&\langle Y_{j+\frac{\sigma}{2}, m'_j; ~k+\frac{\tau}{2}, m_k'}|y_{3}|Y_{j, m_j;~ k, m_k}\rangle=-  \frac{\sqrt{(2j+1)(2k+1)}}{2} (-1)^{n+I+\frac{\tau}{2}} 
\begin{Bmatrix}
j+\frac{\sigma}{2} & k+\frac{\tau}{2} & \frac{I}{2} \\
k & j & \frac{1}{2}
\end{Bmatrix} \sum_{\kappa=+, -} C_{\frac{1}{2}, \frac{\kappa}{2}; j, m_j}^{j+\frac{\sigma}{2}, m'_j}C_{\frac{1}{2}, -\frac{\kappa}{2}; k, m_k}^{k+\frac{\tau}{2}, m'_k} , \nn\\
&\langle Y_{j+\frac{\sigma}{2}, m'_j; ~k+\frac{\tau}{2}, m_k'}|y_{4}|Y_{j, m_j;~ k, m_k}\rangle= i\frac{\sqrt{(2j+1)(2k+1)}}{2} (-1)^{n+I+\frac{\tau}{2}} 
\begin{Bmatrix}
j+\frac{\sigma}{2} & k+\frac{\tau}{2} & \frac{I}{2} \\
k & j & \frac{1}{2}
\end{Bmatrix} \sum_{\kappa=+, -} (-1)^{\frac{\kappa-1}{2}} C_{\frac{1}{2}, \frac{\kappa}{2}; j, m_j}^{j+\frac{\sigma}{2}, m'_j}C_{\frac{1}{2}, -\frac{\kappa}{2}; k, m_k}^{k+\frac{\tau}{2}, m'_k} , \label{compymu}
\end{align}
where $C_{\frac{1}{2}, \frac{\kappa}{2}; j, m}^{j+\frac{\tau}{2}, m'}$ denote the Clebsch-Gordan coefficients:
\be
C_{\frac{1}{2}, \frac{\kappa}{2}; j, m}^{j+\frac{\tau}{2}, m'} = \delta_{m', m+\frac{\kappa}{2}} \biggl(\delta_{\tau, 1} \sqrt{\frac{j+\kappa m +1}{2j+1}} +\kappa \delta_{\tau, -1}\sqrt{\frac{j+\kappa m}{2j+1}}\biggr). 
\ee
The formulas of Appendix D in \cite{Hasebe-2020} were utilized in the derivation of Eq.(\ref{compymu}). 
We thus derived the explicit form of the matrix coordinates in the $SO(5)$ Landau levels. 
 For a better understanding,  in Appendix \ref{app:himat}, we provide the matrix coordinates for the case of  $(N,I)=(1,1)$. 
 Note that  all of the quantities involved in  the matrix coordinate calculations, such as an integral measure and  $S^4$ coordinates, are $SO(5)$ invariant or covariant quantities. Consequently, the obtained matrix coordinates are necessarily   $SO(5)$ covariant coordinates that transform as the $SO(5)$ vector like the original $S^4$ coordinates (see Eq.(\ref{so5trxaxig})).     
 
In the case of  $I=0$, the gauge symmetry no longer exists. Therefore,  we cannot expect   fuzzy geometries  (recall that the gauge symmetry  is crucial in the present scheme).   
Indeed, when $I=0$, the energy eigenstates are given by the $SO(5)$ spherical harmonics and the  matrix coordinates become trivial: 
\be
X^{[N]}_a =0. 
\ee
The corresponding dimensions of the $SO(5)$ spherical harmonics are 
\be
D(N, I=0)=\frac{1}{6}(N+1)(N+2)(2N+3) =5, ~14, ~30, ~\cdots. \label{explicitdimes}
\ee
Therefore in these matrix dimensions,  the matrix geometries do not exist. In Ref.\cite{Castelino-Lee-Taylor-1997}, the authors argued the non-existence of five-dimensional matrix coordinates, which corresponds to the smallest dimension  in Eq.(\ref{explicitdimes}). 

\section{Pure quantum Nambu matrix geometry }\label{sec:fuzzy4}

Using the explicit matrix coordinates, we now expand concrete discussions  about the matrix geometries. 
It is  shown that the matrix coordinates satisfy 
\be
\sum_{a=1}^5 X_a^{[N]}X_a^{[N]} =c_1(N, I)~\bs{1} \label{constfuzzyfoursphere}
\ee
and 
\be
[X_a^{[N]}, X_b^{[N]}, X_c^{[N]}, X_d^{[N]}] =4! ~c_3(N, I)~ \sum_{e=1}^5\epsilon_{abcde}X_e^{[N]}, \label{nambufuzf}
\ee
where the quantum Nambu bracket denotes the totally antisymmetric combination of the four quantities inside the bracket: 
\be
[O_1, O_2, O_3, O_4] \equiv \text{sgn}(\sigma)~O_{\sigma_1}O_{\sigma_2}O_{\sigma_3}O_{\sigma_4}.
\ee
(Detail discussions about the coefficients, $c_1$ and $c_3$, will be given in Sec.\ref{subsec:higherll}.)  
Equations (\ref{constfuzzyfoursphere}) and (\ref{nambufuzf}) signify a realization of  the fuzzy four-sphere \cite{Castelino-Lee-Taylor-1997, Grosse-Klimcik-Presnajder-1996}.  
 The quantum Nambu geometry thus emerges as  the matrix geometry  in the $SO(5)$ Landau levels. 
We delve into geometric structures  hidden in the mathematics of the quantum Nambu algebra using the explicit form of the matrix coordinates.

\subsection{The lowest Landau level matrix geometry}\label{subsec:lllgeo}

\begin{figure}[tbph]
\center
\includegraphics*[width=150mm]{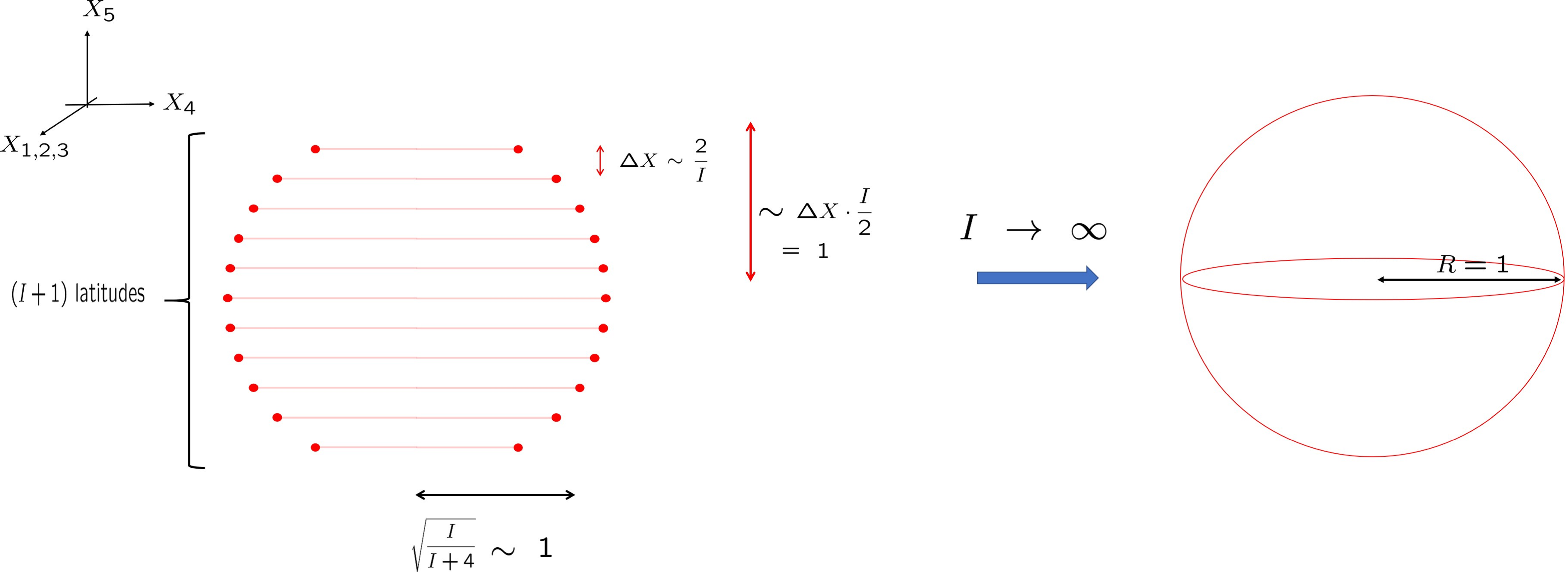}
\caption{Fuzzy four-sphere in the lowest Landau level (the left) and its continuum limit (the right).}
\label{flll.fig}
\end{figure}

 For $N=0$, Eqs.(\ref{expx5mat}) and (\ref{matrixelexmucom})  reproduce the lowest Landau level matrix coordinates  previously obtained in \cite{Hasebe-2020, Ishiki-Matsumoto-Muraki-2018}:
\be
X^{[0]}_a =\frac{1}{I+4}\Gamma_a , \label{matllso5l}
\ee
where $\Gamma^a$ represents the $I$-fold symmetric tensor product of the $SO(5)$ gamma matrices  \cite{Castelino-Lee-Taylor-1997}: 
\be
\Gamma^a =(\gamma^a \otimes 1 \otimes \cdots \otimes 1+ 1 \otimes \gamma^a \otimes  \cdots \otimes 1 + \cdots + 1 \otimes 1 \otimes  \cdots \otimes \gamma_a )_{\text{sym}} 
\ee
with
\be
\gamma_i =\begin{pmatrix}
0 & i\sigma_i \\
-i\sigma_i & 0 
\end{pmatrix}, ~~\gamma_4 =\begin{pmatrix}
0 & 1_2 \\
1_2 & 0 
\end{pmatrix},~~\gamma_5 =\begin{pmatrix}
1_2 & 0 \\
0 & -1_2 
\end{pmatrix}.
\ee
We can readily check that  $X_a^{[0]}$  satisfy (\ref{constfuzzyfoursphere}) and (\ref{nambufuzf}), 
\be
\sum_{a=1}^5 X^{[0]}_aX^{[0]}_a = \frac{I}{I+4}\bs{1}_{\frac{1}{6}(I+1)(I+2)(I+3)} \label{lllxsqua}
\ee
and 
\be
[X^{[0]}_a, X^{[0]}_b, X^{[0]}_c, X^{[0]}_d] =-(I+2)\biggl(\frac{2}{I+4}\biggr)^3 \epsilon_{abcde}X^{[0]}_e. \label{namlllaq}
\ee
The radius and the non-commutative scale are derived as 
\begin{subequations}
\begin{align}
&R=\sqrt{\frac{I}{I+4}} ~~~~(\overset{I\rightarrow\infty}{\longrightarrow}~~ 1), \\
&\Delta X = \frac{2}{I+4} ~~~~(\overset{I\rightarrow\infty}{\longrightarrow}~ 0),
\end{align}
\end{subequations}
which implies that the ordinary four-sphere with a unit radius is reproduced in the continuum limit (Fig.\ref{flll.fig}). 

 It should be emphasized that the algebra of $X_a$ can  be described within the commutator formalism, and  the quantum Nambu algebra (\ref{namlllaq}) is not indispensable for the description of the lowest Landau level matrix geometry. The matrix  coordinates 
 $X^{[0]}_a$ and $X^{[0]}_{ab}\equiv -i\frac{1}{4}[X^{[0]}_a, X^{[0]}_b]$ satisfy  a closed algebra  
\begin{align}
&[X^{[0]}_a, X^{[0]}_b] =4i\frac{1}{(I+4)^2}X^{[0]}_{a,b}, ~~[X^{[0]}_a, X^{[0]}_{bc}] = -i\delta_{ab}X^{[0]}_c+i\delta_{ac}X^{[0]}_b, \nn\\
&[X^{[0]}_{ab}, X^{[0]}_{cd}] = i\delta_{ac}X^{[0]}_{bd}-i\delta_{ad}X^{[0]}_{bc} +i\delta_{bd}X^{[0]}_{ac}-i\delta_{bc}X^{[0]}_{ad}, 
\label{su4algebralll}
\end{align}
which is the $SU(4)$  \cite{Ho-Ramgoolam-2002}. 
The quantum Nambu algebra (\ref{namlllaq}) is not exactly equivalent with the $SU(4)$ algebra  (\ref{su4algebralll}), however, they have been treated almost synonymously thus far. This is because the known matrix realization of the fuzzy four-sphere was only the fully symmetric representation that satisfies both (\ref{namlllaq}) and (\ref{su4algebralll}). 
The closed algebra (\ref{su4algebralll}) suggests that the natural symmetry in the lowest Landau level is the  $SU(4)$ rather than the original $SO(5)$. This becomes clearer in the following discussion.  
The symmetric representation can be simply realized using the Schwinger boson operators.\footnote{Historically, the Schwinger boson operators were utilized in 
the first construction of the fuzzy four-sphere \cite{Grosse-Klimcik-Presnajder-1996}}:
\be
{X}^{[0]}_a =\frac{1}{I+4}~ {\hat{\psi}_{\alpha}}^{\dagger} (\gamma_a)_{\alpha\beta} \hat{\psi}_{\beta} 
\label{schwx0}
\ee
with 
\be
[\hat{\psi}_{\alpha}, \hat{\psi}_{\beta}^{\dagger}]=\delta_{\alpha\beta}, ~~~[\hat{\psi}_{\alpha}, \hat{\psi}_{\beta}]=0. 
\ee
The boson number indicates the $SU(2)$ index of Yang's monopole : 
\be
\sum_{\alpha=1}^4 {\hat{\psi}_{\alpha}}^{\dagger}\hat{\psi}_{\alpha}=I. 
\ee
One may readily check that (\ref{schwx0})  satisfy the $SU(4)$ algebra (\ref{su4algebralll}) together with  $X_{ab}^{[0]} =-i\frac{1}{4}{\hat{\psi}_{\alpha}}^{\dagger} ([\gamma_a, \gamma_b])_{\alpha\beta} \hat{\psi}_{\beta}$. 
The fuzzy manifold constructed from the $SU(4)$ matrices of the $SU(4)$ fully symmetric representation is referred to as the fuzzy $\mathbb{C}P^3$ \cite{Alexanian-Balachandran-Immirzi-Ydri-2002}.  
Note that the dimension of the $SO(5)$ lowest Landau level $(p, q)_5=(I, 0)_5$, is  exactly equal to that of the $SU(4)$ fully symmetric representation: 
\be
D(0, I) = \frac{1}{3!}(I+1)(I+2)(I+3). 
\ee
Therefore, the fuzzy $S^4$ is equivalent to the fuzzy $\mathbb{C}P^3$ (see Ref.\cite{Abe-2004} for  discussions including matrix functions on them).    
The commutation relations of the Schwinger boson operators correspond to the canonical  quantization of the homogeneous coordinates of  
the symplectic manifold $\mathbb{C}P^3$. 
Therefore, it may be reasonable that the lowest Landau level geometry can be described within the conventional commutator formalism of the Lie algebra.  
The corresponding continuum limit is    $\mathbb{C}P^3$,  which is the coset 
\be
\mathbb{C}P^3\simeq SU(4)/U(3). 
\ee
Here, we encounter the $SU(4)$ symmetry again. It is also worth noting that $\mathbb{C}P^3$ is locally equivalent to 
\be
\mathbb{C}P^3 ~\sim~S^4\times S^2.
\ee
While the original $S^4$ itself is not  a symplectic manifold, the $S^2$-fibre twisted on $S^4$ makes the entire bundle  symplectic.

\subsection{Higher Landau level matrix geometry}\label{subsec:higherll}

From (\ref{expx5mat}), we have  
\be
\tr({X_5^{[N]}}^2) =\frac{1}{(2N+I+4)^2 (2N+I+2)^2}\sum_{n=0}^N (I+2+2n)^2 ~\sum_{s=-I/2}^{I/2} (2s)^2~d(n, I, s)~=~\frac{1}{5}c_1(N,I)D(N, I),
\ee
where 
\be
c_1(N, I)\equiv \frac{I(I+2)}{{(2N+I+4)(2N+I+2)}}. \label{neworirad}
\ee
Since all of $X^{[N]}_a$ are related by unitary transformations,  their traces are the same,  $\tr({X_1^{[N]}}^2)=\tr({X_2^{[N]}}^2) =\cdots= \tr({X_5^{[N]}}^2) =\frac{1}{5}c_1 D$.  
The ortho-normal  relation for $X_a^{}[N]$ is given by  
\be
\tr(X_a^{[N]} X_b^{[N]}) =\frac{1}{5}c_1(N, I)D(N,I)~\delta_{ab}, \label{orthonoxas}
\ee
which implies  Eq.(\ref{constfuzzyfoursphere}): 
\be
\sum_{a=1}^5 X^{[N]}_aX^{[N]}_a = c_1(N,I)\bs{1}_{D(N, I)}. \label{fuzzyfourcond1}
\ee
One can explicitly check the validity of Eq.(\ref{fuzzyfourcond1}) using Eqs.(\ref{expx5mat}) and (\ref{matrixelexmucom}).
The radius of the fuzzy four-sphere  is given by 
\be
R(N, I) \equiv \sqrt{c_1(N, I)}=\sqrt{\frac{I(I+2)}{{(2N+I+4)(2N+I+2)}}} ~\sim~\frac{I}{2N+I}. \label{orirad}
\ee
Since the matrix coordinates have two parameters, $N$ and $I$, there are two different infinity limits of the radius: 
\begin{subequations}
\begin{align}
&\lim_{I\rightarrow \infty} R(N,I) =1, \label{usinf}\\
&\lim_{N\rightarrow \infty} R(N,I) =0. \label{nonusinf}
\end{align}\label{nonusinf2}
\end{subequations}
Equation (\ref{usinf}) signifies the usual commutative limit in which the fuzzy four-sphere is reduced to the continuum four-sphere with a unit radius. On the other hand, Eq.(\ref{nonusinf})  indicates the  collapse of the fuzzy four-spheres at 
$N\rightarrow\infty$.  We will revisit this in Sec.\ref{subsec:nestfuzzy}.  
It is demonstrated that $X_a^{[N]}$  
satisfy the quantum Nambu algebra (\ref{nambufuzf}):  
\be
[X^{[N]}_a, X^{[N]}_b, X^{[N]}_c, X^{[N]}_d] =-{4!}~c_3(N,I)~ \epsilon_{abcde}X^{[N]}_e, \label{fournetrel}
\ee
where 
\be
c_3(N,I) \equiv -\frac{5}{c_1(N,I)D(N,I)}~\tr(X^{[N]}_1X^{[N]}_2X^{[N]}_3X^{[N]}_4X^{[N]}_5) . \label{definec3}
\ee
For instance,  
$\tr (X^{[N]}_1 X^{[N]}_2 X^{[N]}_3 X^{[N]}_4 X^{[N]}_5) =-\frac{2896}{7503125}, ~-\frac{217}{124416}, ~-\frac{856}{5250987}$  for  $(N,I)=(1,1),(1,2), (2,1)$.\footnote{In the lowest Landau level, we have  
\be
\tr (X^{[0]}_1 X^{[0]}_2 X^{[0]}_3 X^{[0]}_4 X^{[0]}_5) =\biggl(\frac{1}{I+4}\biggr)^5 \tr (\Gamma_1\Gamma_2\Gamma_3\Gamma_4\Gamma_5) =-\frac{1}{90(I+4)^4} I(I+1)(I+2)^2(I+3),  
\ee
and Eq.(\ref{fournetrel}) reproduces Eq.(\ref{namlllaq}).} 
The matrix coordinates of the higher Landau levels  not only satisfy the quantum Nambu algebra (\ref{fournetrel}) but also encompass all possible matrix realizations of that algebra, because the higher Landau level matrix geometries encompass all possible  irreducible representations of $SO(5)$.

It is also easy to see 
\be
[X^{[N]}_a, X^{[N]}_b] ~\not\propto~i\Sigma_{ab}^{[N]}~~~~~~~~(N\ge 1), \label{slhigherso5sig}
\ee
where $\Sigma_{ab}^{[N]}$ denote the $SO(5)$ generators in the $(N+I, I)_5$ representation. 
Equation (\ref{slhigherso5sig}) is consistent with the general discussions in Sec.\ref{subsec:beycomform}. 
While the commutators of $X_a$ do not give rise to the $SO(5)$ generators, $X_a$ themselves transform as an $SO(5)$ vector:  
\be 
[X^{[N]}_a, \Sigma_{bc}^{[N]}] =-i\delta_{ab}X^{[N]}_c +i\delta_{ac}X^{[N]}_b. \label{so5trxaxig}
\ee
The higher Landau level geometry is thus the one that adheres to the quantum Nambu algebra but $\it{not}$ the $SU(4)$ algebra in contrast to the lowest Landau level matrix geometry.  Let us recall again  that the present scheme is beyond the conventional commutator formalism.   
The quantum geometry in the higher Landau levels is thus qualitatively different to that of the lowest Landau level. 
The  algebraic structure of the higher Landau level geometry is apparent only after introducing the quantum Nambu bracket and cannot be captured by the ordinary  commutator formalism. In this sense, the  higher Landau level geometry is considered to be a $\it{pure}$ quantum Nambu geometry.

\subsection{Nested fuzzy four-sphere}\label{subsec:nestfuzzy}

\begin{figure}[tbph]
\center
\hspace{-1.0cm}
\includegraphics*[width=140mm]{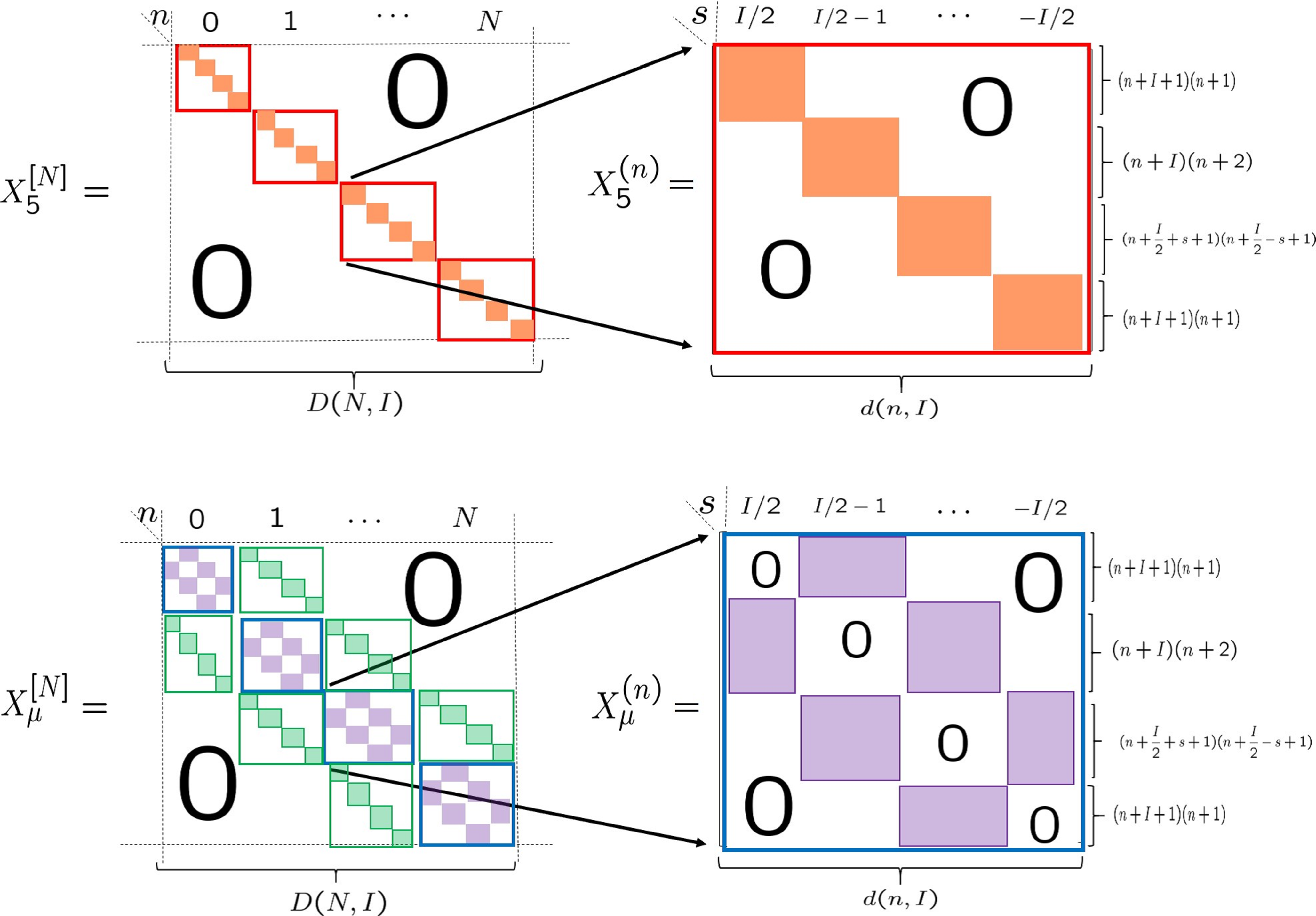}
\caption{Matrix coordinates of the $N$th Landau level. Non-zero matrix elements are denoted as  the shaded color regions. 
 }
\label{xnn.fig}
\end{figure}

Let us delve into the matrix structure of $X_a^{[N]}$. 
We represent (\ref{expx5mat}) by the following $D(N, I)\times D(N, I)$ matrix (the upper left of Fig.\ref{xnn.fig}): 
\be
X_5^{[N]} =\bigoplus_{n=0}^{N} X_5^{(n)} =\begin{pmatrix}
X_5^{(0)} & 0 & 0 & 0 & 0 \\
0 & X_5^{(1)} & 0 & 0 & 0  \\
0 & 0 & X_5^{(2)} & 0 & 0   \\
0 & 0 & 0 & \ddots   & 0    \\
0 & 0 & 0 & 0 & X_5^{(N)}    
\end{pmatrix}, \label{xfivex}
\ee
where 
\begin{align}
X_5^{(n)} &=\frac{I+2n+2}{(2N+I+4)(2N+I+2)} \bigoplus_{s=-I/2}^{I/2} ~2s~\bs{1}_{d(n,I,s)} 
\nn\\
&=\frac{I+2n+2}{(2N+I+4)(2N+I+2)} 
\begin{pmatrix}
I \bs{1}_{d(n, I, I/2)} & 0 & 0 & 0 & 0 \\
0 & (I-2) \bs{1}_{d(n, I, I/2-1)} & 0 & 0 & 0 \\
0 & 0 & (I-4) \bs{1}_{d(n, I, I/2-2)} & 0 & 0 \\
0 & 0 & 0 & \ddots & 0  \\ 
0 & 0 & 0 & 0 & -I\bs{1}_{d(n, I, -I/2)}  \\ 
\end{pmatrix} .  \label{xfive}
\end{align}
The diagonal blocks in $X_{\mu}^{[N]}$ that correspond to $X_5^{(n)}$ are denoted as $X_{\mu}^{(n)}$ (the lower right in Fig.\ref{xnn.fig}), which signify the matrix coordinates on  the $SO(4)$ line $(n)$.  
We will delve into  the matrix structure of $X_a^{(n)}$ that represents the fuzzy geometry on  the $SO(4)$ line $(n)$. The sum of the squares of $X_a^{(n)}$ is given by 
\be
\sum_{a=1}^5 X_a^{(n)}X_{a}^{(n)} =\bigoplus_{s=-I/2}^{I/2} {R^{(n,s)}}^2 \bs{1}_{d(n,I,s)}=
\begin{pmatrix}
 {R^{(n,I/2)}}^2  \bs{1}_{(n+I+1)(n+1)}  & 0 & \cdots & 0 \\
0 &   {R^{(n,I/2-1)}}^2  \bs{1}_{(n+I)(n+2)}  & 0 & 0 \\
\vdots & 0  & \ddots & 0\\
0 & 0 & 0 &   {R^{(n,-I/2)}}^2  \bs{1}_{(n+1)(n+I+1)} 
\end{pmatrix}, \label{xmuxmusummmm}
\ee
where 
\be
R^{(n,s)} \equiv \frac{I+2n+2}{(2N+I+4)(2N+I+2)}\sqrt{2(B(j,k) +  B(k,j))+(2s)^2}~=R^{(n,-s)}
\ee
with $B(j,k)$ defined by (\ref{defb}). 
Thus, $\sum_{a=1}^5 X_a^{(n)}X_{a}^{(n)}$  is not proportional to  unit matrix $\bs{1}_{d(n,I)}$ (except for the special case $I=1$)\footnote{
For $I=1$, we have only two hyper-latitudes with the same radius, and $\sum_{a=1}^5 X_a^{(n)}X_{a}^{(n)}$  is proportional to  $\bs{1}$:  
\be
\sum_{\mu=1}^a X_a^{(n)}X_{a}^{(n)} =\biggl(\frac{2n+3}{(2N+5)(2N+3)}\biggr)^2 \biggl(2\frac{(N+2)^2}{(n+2)(n+1)}+1\biggr)\bs{1}_{2d(n,1)=2(n+2)(n+1)}.
\ee
}, and so   $X_a^{(n)}$ does $\it{not}$ give rise to a complete fuzzy four-sphere geometry but  provides a  fuzzy four-sphere-like structure  referred to as the quasi-fuzzy four-sphere \cite{Hasebe-2020}.  The $I+1$ diagonal blocks  on the most right-hand side of Eq.(\ref{xmuxmusummmm})  signify the $I+1$ fuzzy hyper hyper-latitudes on the quasi-fuzzy four-sphere.   Inside the matrix coordinates $X_a^{[N]}$ (Fig.\ref{xnn.fig}), there exist  such $N+1$ quasi-fuzzy four-spheres  $X_a^{(n=0,1,2,\cdots, N)}$. The non-zero off-diagonal blocks of the $X_{\mu}^{[N]}$ (the green filled rectangles in the lower left of Fig.\ref{xnn.fig}) are  interpreted as the interactions  between the adjacent quasi-fuzzy four-spheres.  

\begin{figure}[tbph]
\center
\includegraphics*[width=160mm]{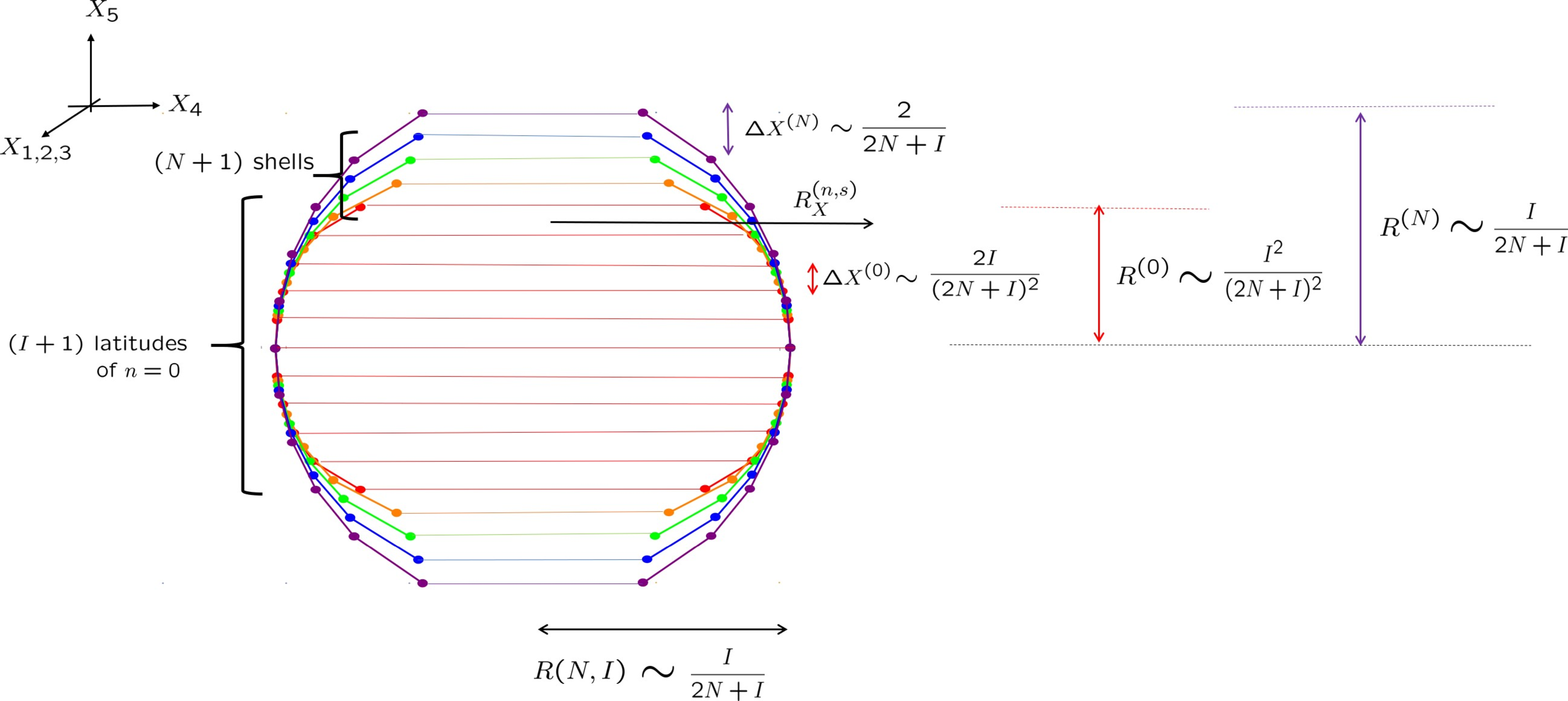}
\caption{Matrix geometry of the $N$th  higher Landau level is constituted from the $N+1$ quasi-fuzzy four-spheres (and their interactions) to exhibit a nested fuzzy  structure.  }
\label{nests4f.fig}
\end{figure}

 The nested geometry  of the $N+1$ quasi-fuzzy four-spheres is depicted in Fig.\ref{nests4f.fig}. 
One should $\it{not}$ confuse the present geometry with the nested structure made of a completely reducible representation \cite{Myers-1999}:    In the case of the completely reducible representation, the nested fuzzy structure  originates from the direct sum of the irreducible representations, while in the present, the nested fuzzy four-sphere  is constituted from a single $SO(5)$ irreducible representation and each of the quasi-fuzzy four-spheres is not made of an  $SO(5)$ irreducible representation (but rather consists of the $SO(4)$ representations on the $SO(4)$ line).\footnote{The very fuzzy fibres of  fuzzy geometry is reported in \cite{Sperling-Steinacker-2017, Steinacker-2015, Grosse-Steinacker-2004}. That structure  originates from the direct sum of  irreducible representations, and so it is more akin to \cite{Myers-1999} than the present one.   }     Consequently, each quasi-fuzzy four-sphere is  not regarded as an $SO(5)$-symmetric object. This  is  also evident from the right-hand side of (\ref{xmuxmusummmm}), which is apparently not $SO(5)$ invariant.   
The quasi-fuzzy four-spheres along with their interactions collectively form an $SO(5)$-symmetric fuzzy manifold. We would like to draw the analogy to benzene. Each   Kekul\'e structure  only respects the $C_3$ rotational symmetry, while quantum mechanical superposition of two Kekul\'e structures results in  benzene, which exhibits higher $C_6$ symmetry. Such a structure cannot be comprehended without  quantum mechanics,  and  benzene realizes a purely quantum mechanical structure with no   classical counterpart. 
In a similar sense, the nested fuzzy four-sphere can be considered  a $\it{pure}$ quantum geometry. This stems from the  present quantum-oriented scheme, which can encompass pure quantum geometries.

\begin{figure}[tbph]
\center
\hspace{-1cm}
\includegraphics*[width=140mm]{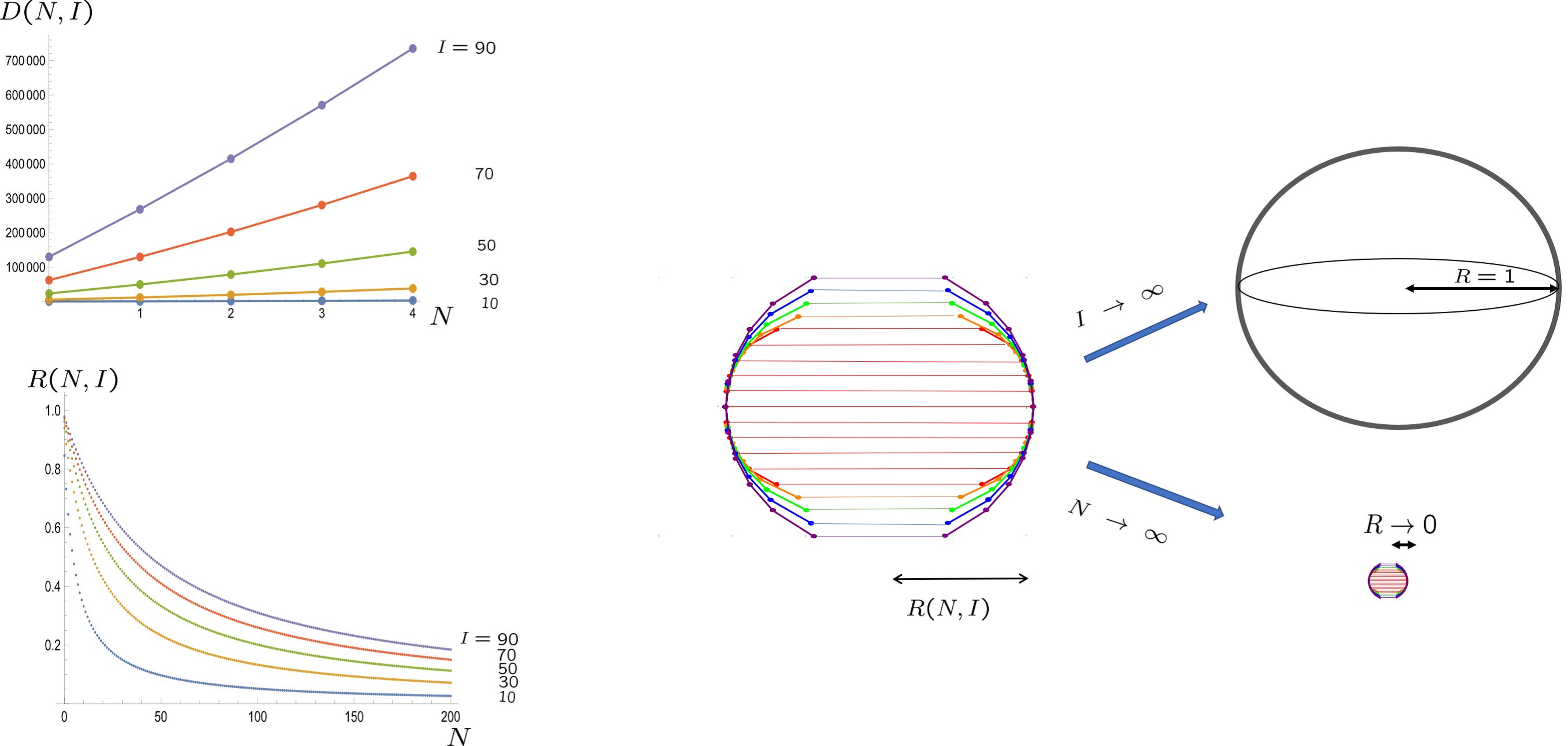}
\caption{Behaviors of the quantities of the nested fuzzy four-sphere: The degeneracies (the upper left) and the radii (the lower left),  the continuum limit $I\rightarrow \infty$ (the upper right) and the $N\rightarrow \infty$ limit (the lower right). }
\label{suml.fig}
\end{figure}

The non-commutative scale differs in  each of the quasi-fuzzy four-spheres (\ref{xfive}):  
\be
\Delta X^{(n)} =\frac{2}{(2N+I+4)(2N+I+2)}(2n+I+2), \label{comeachfuzqu}
\ee
and the ``radius'' of the quasi-fuzzy four-sphere is estimated as 
\be
R^{(n)} ~\sim~\Delta X^{(n)} \cdot \frac{I}{2} =\frac{1}{(2N+I+4)(2N+I+2)}(2n+I+2)(I+1) ~\sim~\frac{(2n+I) I}{(2N+I)^2}. \label{radnsh}
\ee
The outer quasi-fuzzy four-spheres  have  wider non-commutative scales (see Fig.\ref{nests4f.fig}). 
It can be confirmed that  the outermost quasi-fuzzy four-sphere of $n=N$   (\ref{radnsh}) exhibits the  same behavior as   the nested fuzzy four-sphere (\ref{orirad}), as anticipated.  We now provide an intuitive explanation for the previous result of the two limits (\ref{nonusinf2}). 
 In the commutative limit $I\rightarrow \infty$, while   $\Delta X^{(n)}\sim \frac{2}{I}$ (\ref{comeachfuzqu}) is reduced to zero, the number of the hyper-latitudes $I$ goes to infinity. These two contributions are compensated to realize a continuum four-sphere with a unit radius, which   
simultaneous implies that all of the $N+1$ quasi-fuzzy four-spheres are reduced to the single  four-sphere.\footnote{In the commutative limit $I\rightarrow \infty$, each point on the four-sphere is highly degenerate. Because of the $SO(5)$ symmetry, we can count this degeneracy, for instance, at the north pole $X^{(n)}_5 =1$: 
\be
I\sum_{n=0}^{N}(n+1) =I\frac{1}{2}(N+1)(N+2). 
\ee
}
 On the other hand, in the limit $N\rightarrow \infty$, while  the number of hyper-latitudes remains unchanged, the non-commutative length (\ref{comeachfuzqu}) $\Delta X^{(n)}~\sim~\frac{1}{N}$  converges to zero.  This leads to the collapse of the very nested fuzzy four-sphere,  $R^{(n)}~{\rightarrow} ~0$ (Fig.\ref{suml.fig}).

\section{Internal matrix geometry}\label{sec:intfuzz}

Fuzzy three-sphere geometry can be realized as a sub-manifold of the (unnested) fuzzy four-sphere. Here, we explore the generalization of this concept for the  nested fuzzy four-spheres.  

\subsection{Fuzzy hyper-latitudes}
 
\begin{figure}[tbph]
\center
\includegraphics*[width=160mm]{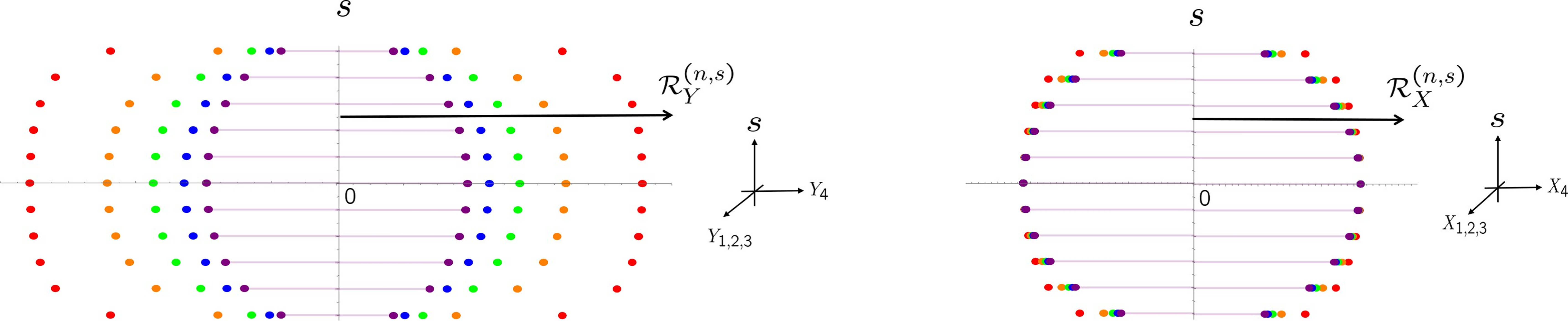}
\caption{ The distributions of  $\mathcal{R}_Y^{(n,s)}$ (the left) and  $\mathcal{R}_X^{(n,s)}$ (the right) for $I/2=5$. The $SO(4)$ line index  $n(=0, 1,2 , 3, 4)$ corresponds to red, orange, green, blue and  purple points, respectively.     }
\label{nest.fig}
\end{figure}

 The quasi-fuzzy four-sphere is constituted from the $SO(4)$ irreducible representations on the $SO(4)$ line $(n)$.  The matrix coordinates of the hyper-latitudes on the quasi-fuzzy four-sphere are readily derived from Eq.(\ref{compymu}):  
\be
(Y_{\mu}^{(n)})_{m_j', m'_k; m_j, m_k} \equiv  \langle Y_{j',m'_j, k', m'_k}| y_{\mu}|Y_{ j, m_j, k, m_k}\rangle \biggr|_{j'+k'=j+k=n+\frac{I}{2}},
\ee
which denotes a $d(n,I)\times d(n,I)$ matrix.\footnote{
The matrix $Y_{\mu}^{(n)}$ has the same matrix form as  $X_{\mu}^{(n)}$ (the lower right of Fig.\ref{xnn.fig}): 
\begin{align}
&Y_{\mu}^{(n)} =\nn\\
&\begin{pmatrix}
0 & \mathcal{Y}_{\mu}^{(+-)} (\frac{I+n-1}{2}, \frac{n+1}{2}) & 0 & 0 & 0 & 0 \\
 \mathcal{Y}_{\mu}^{(+-)} (\frac{I+n-1}{2}, \frac{n+1}{2}) ^{\dagger} & 0 & \mathcal{Y}_{\mu}^{(+-)} (\frac{I+n-2}{2}, \frac{n+2}{2}) & 0 & 0 & 0  \\ 
0 &   \mathcal{Y}_{\mu}^{(+-)} (\frac{I+n-2}{2}, \frac{n+2}{2}) ^{\dagger} & 0 & \mathcal{Y}_{\mu}^{(+-)} (\frac{I+n-3}{2}, \frac{n+3}{2}) & 0 & 0   \\
0 & 0 &    \mathcal{Y}_{\mu}^{(+-)} (\frac{I+n-3}{2}, \frac{n+3}{2}) ^{\dagger}  & 0 & \ddots  & 0 \\  
0 & 0 & 0 & \ddots   & 0 &  \mathcal{Y}_{\mu}^{(+-)} ( \frac{n}{2}, \frac{n+I}{2})  \\
0 & 0 & 0 & 0   & \mathcal{Y}_{\mu}^{(+-)} ( \frac{n}{2}, \frac{n+I}{2})^{\dagger} & 0  
\end{pmatrix}, \label{ymumatr}
\end{align}
and $A(j, k)$ and $A(k,j)$ in (\ref{radcrf3}) are given by  
\begin{subequations}
\begin{align}
&\sum_{\mu=1}^4 \mathcal{Y}_{\mu}^{(+ -)}(j,k)^{\dagger}~ \mathcal{Y}_{\mu}^{(+ -)}(j,k) =  A(j,k) ~\bs{1}_{(2j+1)(2k+1)} ,\\
&\sum_{\mu=1}^4 \mathcal{Y}_{\mu}^{(+ -)}(j-\frac{1}{2},k+\frac{1}{2})~\mathcal{Y}_{\mu}^{(+ -)}(j-\frac{1}{2},k+\frac{1}{2})^{\dagger} =  A(k,j)~ \bs{1}_{(2j+1)(2k+1)}.
\end{align}\label{fomuymusum}
\end{subequations}
} 
The sum of the squares of $Y_{\mu}^{(n)}$ is given by 
\be
\sum_{\mu=1}^4 Y_{\mu}^{(n)} Y_{\mu}^{(n)} =\bigoplus_{s=-I/2}^{I/2} {\mathcal{R}_Y^{(n,s)}}^2 \bs{1}_{d(n,I,s)}=
\begin{pmatrix}
 {\mathcal{R}_Y^{(n,\frac{I}{2})}}^2 \bs{1}_{(n+I+1)(n+1)}  & 0 & \cdots & 0 \\
0 &  {\mathcal{R}_Y^{(n,\frac{I}{2}-1)}}^2 \bs{1}_{(n+I)(n+2)} & 0 & 0 \\
0 & 0  & \ddots & 0\\
0 & 0 & 0 &   {\mathcal{R}_Y^{(n,-\frac{I}{2})}}^2 \bs{1}_{(n+I+1)(n+1)}
\end{pmatrix}, \label{sumymuymun}
\ee 
where
\begin{align}
&\mathcal{R}_Y^{(n,s)}\equiv \sqrt{ 
A(j,k) + A(k,j)} =\mathcal{R}_Y^{(n,-s)} \label{radcrf3}
\end{align}
with
\be
A(j,k)\equiv  2 
(j+1)k\begin{Bmatrix}  
j+\frac{1}{2} & k-\frac{1}{2} & \frac{I}{2} \\
k & j & \frac{1}{2}
\end{Bmatrix}^2  .
\ee
Note  that 
$A(\frac{n}{2}+\frac{I}{2}, \frac{n}{2}) 
=0.$ 
Equation (\ref{sumymuymun}) represents a block diagonal matrix, with diagonal blocks indicating the hyper-latitudes of the radius $\mathcal{R}_Y^{(n,s)}$. 
 At $I\rightarrow \infty $ and $|s| <\!< I$, we have 
\be
\mathcal{R}_Y^{(n,s)} ~\rightarrow ~1~~~~(A(j,k) ~\rightarrow ~\frac{1}{2}).
\ee
Around $s~\sim~0$,   the radii of the hyper-latitudes converge to unity, as anticipated from $\sum_{\mu=1}^4 y_{\mu}y_{\mu}=1$. 
The hyper-latitudes for $s$ as the vertical axis are depicted in the left  of Fig. \ref{nest.fig}.

We also evaluate the radii of the hyper-latitudes within the quasi-fuzzy four-sphere. 
Using Eq.(\ref{matrixelexmucom}),  
we can derive
\begin{align}
&\sum_{\mu=1}^4 X_\mu^{(n)}X_{\mu}^{(n)} 
=\bigoplus_{s=-I/2}^{I/2}{\mathcal{R}_X^{(n,s)}}^2   \bs{1}_{d(n,I,s)}=
\begin{pmatrix}
 {\mathcal{R}_X^{(n,\frac{I}{2})}}^2 \bs{1}_{(n+I+1)(n+1)}  & 0 & \cdots & 0 \\
0 &  {\mathcal{R}_X^{(n,\frac{I}{2}-1)}}^2 \bs{1}_{(n+I)(n+2)} & 0 & 0 \\
0 & 0  & \ddots & 0\\
0 & 0 & 0 &   {\mathcal{R}_X^{(n,-\frac{I}{2})}}^2 \bs{1}_{(n+I+1)(n+1)}
\end{pmatrix}, \label{xmuxmusum}
\end{align} 
where 
\be
\mathcal{R}_X^{(n,s)} \equiv  \frac{2n+I+2}{(2N+I+2)(2N+I+4)}\sqrt{  
  B(j,k)+ B(k,j) }=\mathcal{R}_X^{(n,-s)},
\ee
with 
\be
B(j,k) \equiv 4~(N+\frac{I}{2}-j+k+1)(N+\frac{I}{2}+j-k +2)~A(j,k). \label{defb}
\ee
For  the distributions of $\mathcal{R}_X^{(n,\frac{I}{2})}$, see the right of Fig.\ref{nest.fig}.    Obviously, the distribution of  points of the same color forms a quasi-fuzzy four-sphere. (The distribution of $\mathcal{R}_X^{(n,s)}$ is illustrated in Fig. \ref{nests4f.fig} with $X_5$ as the vertical axis.) 
At $I\rightarrow \infty $ and $|s| <\!< I$, we have 
\be
\mathcal{R}_X^{(n,s)} ~\rightarrow ~1~~~~~(B(j,k) ~\rightarrow ~\frac{1}{2}I^2).
\ee
When $I$ is even, $\mathcal{R}_X^{(n,s)}$ takes the maximum value at $s=j-k=0$: 
\be
\mathcal{R}_X^{(n,s=0)}=\sqrt{\frac{I(I+2)}{(2N+I+4)(2N+I+2)}} .
\ee
This quantity does not depend on  $n$, indicating that the equators of all the quasi-fuzzy four-spheres have  the same radius, which is identical to the radius of the fuzzy $S^4$  (\ref{orirad}) (see Fig.\ref{nests4f.fig} also).  

\subsection{Fuzzy three-sphere}\label{sec:fuzzythree}

\begin{figure}[tbph]
\center
\hspace{-1cm}
\includegraphics*[width=150mm]{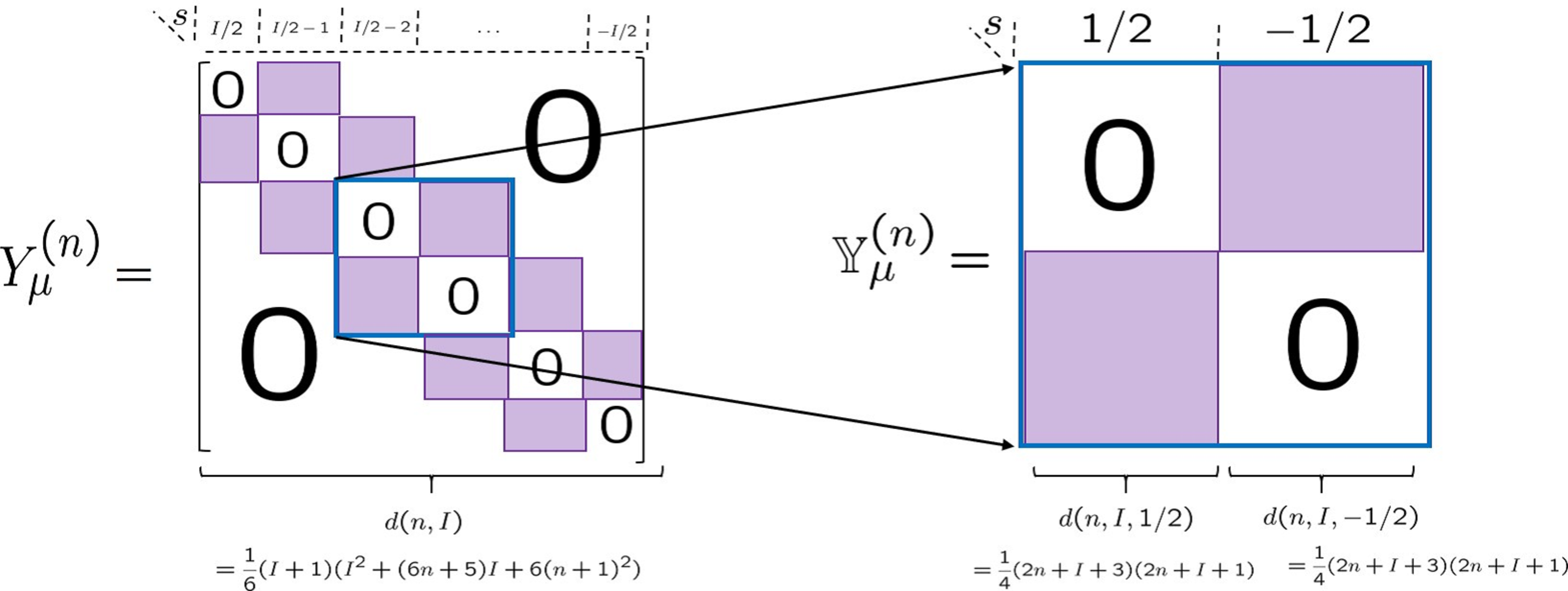}
\caption{The fuzzy three-sphere matrix coordinates $\mathbb{Y}_{\mu}^{(n)}$ from $Y_{\mu}^{(n)}$ for odd $I$. }
\label{exts3.fig}
\end{figure}

\begin{figure}[tbph]
\center
\includegraphics*[width=120mm]{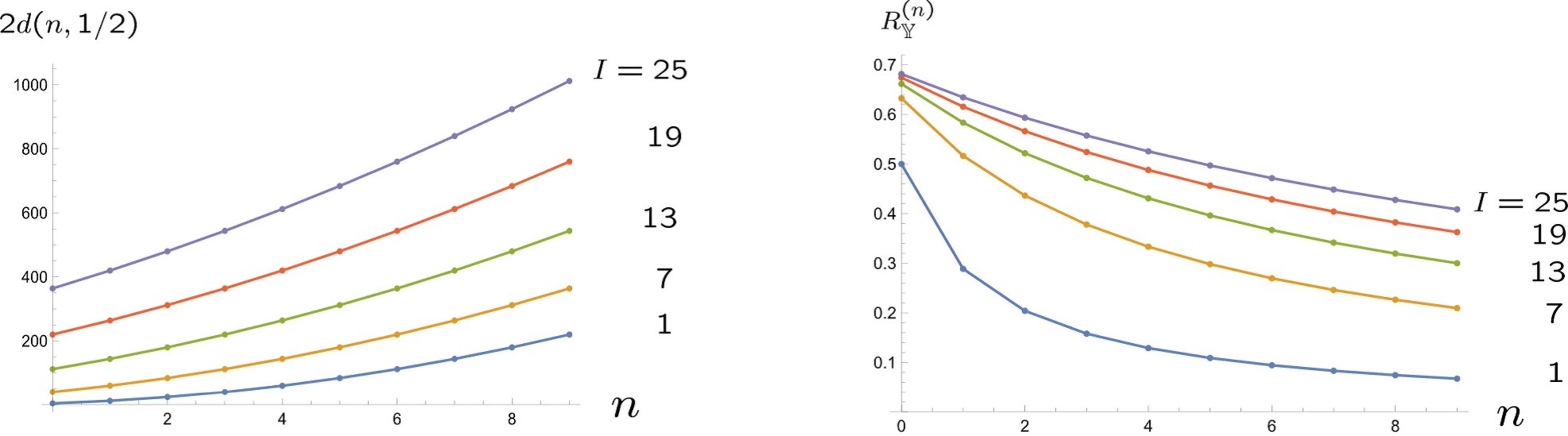}
\caption{The matrix size and the radius of the fuzzy three-sphere.  }
\label{s3pro.fig}
\end{figure}

The fuzzy three-sphere is naturally embedded within the geometry of the fuzzy four-sphere \cite{Guralnik&Ramgoolam2001, Ramgoolam2002, Jabbari2004}. This  sub-space  is composed of $SO(4)$ representations with $s=1/2 \oplus -1/2$.  In the  case of the usual (un-nested) fuzzy four-sphere, there exists only one fuzzy three-sphere around the equator of the fuzzy four-sphere. In contrast, the nested fuzzy four-sphere consists of multiple quasi-fuzzy four-spheres, each of which accommodates a  fuzzy three-sphere. Consequently, the $N$th Landau level fuzzy four-sphere hosts  $N+1$ fuzzy three spheres  around its equator.      
To extract the fuzzy three-sphere geometry, we focus on the $s=1/2\oplus -1/2$ sub-space of the matrix coordinates $Y_{\mu}^{(n)}$. 
For odd integer $I$, we can derive the fuzzy three-sphere matrix coordinates (see Fig.\ref{exts3.fig})
\be
\mathbb{Y}_{\mu}^{(n)} =\begin{pmatrix}
0 & \mathcal{Y}_{\mu}^{(+ -)} (\frac{n}{2}+\frac{I}{4}-\frac{1}{4}, \frac{n}{2}+\frac{I}{4}+\frac{1}{4}) \\
\mathcal{Y}_{\mu}^{(+ -)} (\frac{n}{2}+\frac{I}{4}-\frac{1}{4}, \frac{n}{2}+\frac{I}{4}+\frac{1}{4}) ^{\dagger} & 0 
\end{pmatrix}, \label{mathbbymu}
\ee
which  satisfy 
\be
\sum_{\mu=1}^4 \mathbb{Y}_{\mu}^{(n)}\mathbb{Y}_{\mu}^{(n)} ={{R}_{\mathbb{Y}}^{(n)}}^2 ~\bs{1}_{2d(n, I, 1/2)} \label{radfuzzs3}
\ee
where 
\be
{R}^{(n)}_{\mathbb{Y}} \equiv \sqrt{A(j,k)}|_{j=\frac{n}{2}+\frac{I}{4}-\frac{1}{4}, k =\frac{n}{2}+\frac{I}{4}+\frac{1}{4}}=\frac{I+1}{\sqrt{2(2n+I+1)(2n+I+3)}}. \label{fuzzy3rad}
\ee
Unlike the sum of the squares of $Y_{\mu}^{(n)}$ (\ref{sumymuymun}),  Eq.(\ref{radfuzzs3})
is  proportional to a unit matrix. This implies that while $Y_{\mu}^{(n)}$ themselves cannot be regarded as the coordinates of the fuzzy three-sphere, their sub-block matrices $\mathbb{Y}_{\mu}^{(n)}$ can be.  Furthermore, $\mathbb{Y}_{\mu}^{(n)}$ 
 are shown to satisfy 
\be
[\![\mathbb{Y}_{\mu}^{(n)}, \mathbb{Y}^{(n)}_{\nu}, \mathbb{Y}^{(n)}_{\rho}]\!] =8(2n+I+2)\biggl( \frac{I+1}{(2n+I+1)(2n+I+3)}\biggr)^2\epsilon_{\mu\nu\rho\sigma}\mathbb{Y}_{\sigma}^{(n)}, \label{alfuzs3}
\ee
where the ``three bracket'' $[\![\mathbb{Y}_{\mu}, \mathbb{Y}_{\nu}, \mathbb{Y}_{\rho}]\!]$ is defined as 
\be
[\![\mathbb{Y}_{\mu}^{(n)}, \mathbb{Y}_{\nu}^{(n)}, \mathbb{Y}^{(n)}_{\rho}]\!] \equiv [\mathbb{Y}^{(n)}_{\mu}, \mathbb{Y}^{(n)}_{\nu}, \mathbb{Y}^{(n)}_{\rho}, \mathit{\Gamma}_5]
\ee
with 
\be
\mathit{\Gamma}_5 \equiv \begin{pmatrix}
\bs{1}_{d(n, I, 1/2))} & 0 \\
0 & -\bs{1}_{d(n, I, -1/2)}
\end{pmatrix}. 
\ee
Equations (\ref{radfuzzs3}) and (\ref{alfuzs3}) clearly show that  $\mathbb{Y}_{\mu}^{(n)}$ realize the matrix coordinates of the fuzzy three-sphere.\footnote{With  
\be
\mathbb{Y}^{(n)}_5 \equiv \frac{I+1}{2}\frac{1}{\sqrt{2(2n+I+3)(2n+I+1)}} ~\mathit{\Gamma}_5,
\ee
$\mathbb{Y}^{(n)}_{\mu=1,2,3,4}$ satisfy the orthonormal condition:  
\be
\tr(\mathbb{Y}^{(n)}_a \mathbb{Y}^{(n)}_b) =\biggl(\frac{I+1}{4}\biggr)^2 \delta_{ab}.
\ee
Equation (\ref{alfuzs3}) is realized as a special case of  the four-algebra,
\be
[\mathbb{Y}^{(n)}_{a}, \mathbb{Y}^{(n)}_{b}, \mathbb{Y}^{(n)}_{c}, \mathbb{Y}^{(n)}_d] =-2\sqrt{2} ~(I+1)^3  \frac{2n+I+2}{\sqrt{(2n+I+3)(2n+I+1)}^5} ~\epsilon_{abcde}\mathbb{Y}^{(n)}_e. 
\ee
} Figure \ref{s3pro.fig} illustrates the behaviors of the matrix sizes and the radii (\ref{fuzzy3rad}) of fuzzy three-spheres. The qualitative features of these quantities are similar to those of the fuzzy four-sphere (Fig.\ref{suml.fig}) as the fuzzy three-spheres being embedded in of the fuzzy four-sphere. 
Note that the radius of the fuzzy three-sphere is not equal to that of the fuzzy hyper-latitude of the same $s$ (\ref{radcrf3}), $\mathcal{R}_{Y}^{(n,s=1/2)} \neq  {R}_{\mathbb{Y}}^{(n)} $, and ${R}_{\mathbb{Y}}^{(n)}$ does not converge to unity in the continuum limit, ${R}_{\mathbb{Y}}^{(n)} ~\overset{I\rightarrow \infty}{\longrightarrow} ~1/\sqrt{2} \neq 1$.

We also explain how the fuzzy three-sphere itself is obtained  within the present non-commutative framework, without referring to the geometry of the fuzzy four-sphere. Since $S^3$ can be identified with  $SO(4)/SO(3)$, the stabilizer group $SO(3)$ is interpreted as the $SU(2)$ gauge symmetry on the quantum mechanics side. Then, we consider an $SU(2)$ gauged quantum mechanics on $S^3$, known as the $SO(4)$ Landau model \cite{Hasebe-2018, Hasebe-2014-2, Nair-Daemi-2004}. In this model, $n$ represents   the Landau level index, and $s$ signifies the subband index.
The $SO(4)$ Landau model exhibits degeneracy due to the presence of the left-right $\mathbb{Z}_2$ symmetry, in addition to the global $SO(4)$ symmetry. The fuzzy three-sphere geometry $\mathbb{Y}^{(n)}_{\mu}$ emerges in  the lowest energy sub-bands with indices $s = 1/2, -1/2$, for arbitrary $n$th Landau level.
The degenerate energy eigenstates  that constitute  the fuzzy three-sphere consist of the direct sum of  irreducible representations of the global symmetry $SO(4)$, which is an irreducible representation 
 of the entire symmetry group $SO(4)\otimes \mathbb{Z}_2$.  

\section{Continuum limit and the $S^4$ geometry}\label{sec:contin}

We discuss the continuum limit and the classical geometry of the nested fuzzy four-sphere. 
While the continuum limit of the fuzzy two-sphere is the usual classical two-sphere, this is not generally the case for other fuzzy manifolds. 
For instance, the continuum limit of the unnested fuzzy $2k$-sphere yields the symplectic manifold  $SO(2k+1)/U(k)$ \cite{Ho-Ramgoolam-2002}, which is obviously distinct from $S^{2k}$.

\subsection{The second Hopf map}

The Hopf maps are a key to bridge non-commutative geometry and classical geometry \cite{Hasebe-2010}. 
The  second Hopf map 
\be
\psi_{\alpha=1,2,3,4} ~~(\psi_{\alpha}^*\psi_{\alpha}=1)~\in~S^7 ~~\rightarrow ~~x_a ={\psi_{\alpha}}^{*}(\gamma_a)_{\alpha\beta} \psi_{\beta} ~\in~S^4   ~~~(x_ax_a =(\psi_{\alpha}^{\dagger}\psi_{\alpha})^2=1)   \label{secondhopf}
\ee
provides a clear understanding of the fuzzy four-sphere geometry.  
The fuzzification is simply executed by replacing the components of the Hopf spinor with four  annihilation operators: 
\be
\psi_{\alpha} ~~\rightarrow ~~\frac{1}{\sqrt{I+4}}\hat{\psi}_{\alpha}, 
\ee
with  
\be
[\hat{\psi}_{\alpha}, {\hat{\psi}_{\beta}}^{\dagger}] =\delta_{\alpha\beta},~~[\hat{\psi}_{\alpha}, \hat{\psi}_{\beta}]=0. 
\ee
The ``quantized'' Hopf map is now given by 
\be
\hat{\psi}_{\alpha} ~~({\hat{\psi}_{\alpha}}^{\dagger}\hat{\psi}_{\alpha}=I) ~~\rightarrow ~~X_a =\frac{1}{I+4}{\hat{\psi}_{\alpha}}^\dagger(\gamma_a)_{\alpha\beta} \hat{\psi}_{\beta}, \label{quahopf}
\ee
which satisfy 
\be
X_aX_a =\frac{1}{(I+4)^2}({\hat{\psi}_{\alpha}}^{\dagger}\hat{\psi}_{\alpha})({\hat{\psi}_{\beta}}^{\dagger}\hat{\psi}_{\beta}+4)=\frac{I}{I+4}.
\ee
Notice that $X_a$ (\ref{quahopf}) coincide with the  lowest Landau level coordinate operators (\ref{schwx0}). 
The total manifold $S^7$ represents the classical manifold of the Hopf spinor and the $S^7$   modulo $U(1)$ phase is  $\mathbb{C}P^3$, which is the continuum limit of the (unnested) fuzzy four-sphere.  The second Hopf map thus presents a   relationship between the unnested fuzzy four-sphere and its continuum limit.

Also notice  that the Hopf spinor for (\ref{secondhopf}) can be chosen as 
\be
\psi=\frac{1}{\sqrt{2(1+x_5)}}\begin{pmatrix}
1+x_5 \\
0 \\
x_4-ix_3\\
x_2 -i x_1
\end{pmatrix} =\begin{pmatrix}
\cos\frac{\xi}{2} \\
0 \\
\sin\frac{\xi}{2} ~(\cos\chi-i\sin\chi\cos\theta) \\
-i\sin\frac{\xi}{2}\sin\chi \sin\theta ~e^{i\phi}
\end{pmatrix},   \label{secondhopfex}
\ee
which satisfies  
\be
\sum_{a=1}^5 x_a \gamma_a \psi =+\psi. \label{simplecoheresp}
\ee
This is the simplest version  of the  $SO(5)$ spin-coherent state equation, which plays a central role in deriving the classical geometry of the fuzzy four-sphere  in Sec.\ref{subsec:cohmet}.

\subsection{Continuum limit}\label{subsec:contlim}

To expand a  concrete discussion, let us focus on the north point of the nested fuzzy four-sphere. Since the nested fuzzy four-sphere is an $SO(5)$ symmetric object, we can choose the north pole as a reference point without loss of generality.  The north pole is represented by the index $s=I/2$, which corresponds to the $N+1$ most right edges of the oblique $SO(4)$ lines in Fig.\ref{so5diagram.fig}: 
\be
\bigoplus_{n=0}^N (j, k)|_{s=I/2}=\bigoplus_{n=0}^N ~(\frac{I}{2}+\frac{n}{2}, \frac{n}{2})_4. \label{directdumdeom} 
\ee
Since $j$ and $k$ are two independent $SU(2)$ indices,  the $(j,k)_4$ realizes a direct product of  two fuzzy spheres specified by the $SU(2)$ spins, $j$ and $k$, in the language of the fuzzy geometry. In the continuum limit $I ~{\rightarrow}~\infty$, Eq.(\ref{directdumdeom}) becomes 
\be
\bigoplus_{n=0}^N (j, k)|_{s=I/2} ~\sim~\bigoplus_{n=0}^N (\frac{I}{2}, 0), 
\ee
which suggests that the fuzzy structure of the north pole  is well approximated by the $N+1$ identical fuzzy two-spheres, each with the $SU(2)$ spin $I/2$. Since every $SO(4)$ line or quasi-fuzzy four-sphere thus accommodates a fuzzy two-sphere, each quasi-fuzzy four-sphere is described locally by  $S^4\times S^2$, or $\mathbb{C}P^3$ in the continuum.  Consequently, the nested fuzzy four-sphere is reduced to $N+1$ overlapped identical $\mathbb{C}P^3$s. 
This is also suggested by the continuum limit of the degeneracy (\ref{dnidnis})
\be
D(N, I) ~~\overset{I\rightarrow \infty}{\longrightarrow} ~~ (N+1) \cdot \frac{1}{6}I^3, \label{countdeg}
\ee
where $\frac{1}{6}I^3$ denotes the degrees of freedom of a single fuzzy $\mathbb{C}P^3$.  

It should be emphasized that  while the continuum limit of the  nested fuzzy  four-sphere is $N+1$ overlapped $\mathbb{C}P^3$s, their fuzzification does not recover the original nested fuzzy four-sphere but  just provides $N+1$ identical fuzzy $\mathbb{C}P^3s$ (or $N+1$ unnested identical fuzzy four-spheres). In other words, 
 the nested  fuzzy four-sphere geometry cannot be reproduced from its corresponding continuum  manifold. 
 This agrees with the previous observation that the nested fuzzy four-sphere is a pure quantum object.

\subsection{$S^4$ geometry}\label{subsec:classgeo}

 The coherent state method \cite{Ishiki-2015, Schneiderbauer-Steinacker-2016, Steinacker-2021}  and the probe brane method \cite{Berenstein-Dzienkowski-2012, Karczmarek-Yeh-2015, Badyn-Karczmarek-SabellaGarnier-Yeh-2015} are 
 systematic methods to obtain a classical manifold corresponding to a given matrix geometry. These two methods are related but not exactly the same \cite{Ishiki-Matsumoto-Muraki-2018, Badyn-Karczmarek-SabellaGarnier-Yeh-2015}.  Here, we derive the $S^4$ geometry from the matrix coordinates using there methods. 

\subsubsection{Coherent state method}\label{subsec:cohmet}

For  matrix coordinates $X_a$, the coherent method \cite{Ishiki-2015, Schneiderbauer-Steinacker-2016, Steinacker-2021} adopts the following matrix Hamiltonian,  
\be
H=\sum_{a=1}^5 (X^{[N]}_a -x_a \bs{1}_{D(N, I)})^2. \label{ishikiham}
\ee
We can derive classical manifold as a configuration of $x_a$  by following the three steps:   First,  we diagonalize the matrix Hamiltonian to derive the groundstate energy $E_{\text{G}}(x_a)$. Second, we examine the minimum of $E_{\text{G}}(x_a)$ as a function of $x_a$ to determine the vacuum manifold  of  $x_a$.  Last, we take the $I\rightarrow \infty$ limit of this configuration. 

The matrix Hamiltonian (\ref{ishikiham}) is rewritten as  
\be
H={X^{[N]}_a}^2 -2x_a X^{[N]}_a +r^2 \bs{1}_{D(N, I)}=\biggl(\frac{I(I+2)}{(2N+I+4)(2N+I+2)}+r^2\biggr)\bs{1}_{D(N, I)}-2x_a X^{[N]}_a,  \label{hamrewcla}
\ee
with 
\be
r\equiv \sqrt{x_ax_a}.
\ee
The cross term of Eq.(\ref{hamrewcla}) is diagonalized as 
\be
U(\xi, \chi,\theta,\phi)^{\dagger}~(x_a X^{[N]}_a )~U(\xi, \chi,\theta,\phi)=rX^{[N]}_5, \label{diagxx}
\ee
where 
\be
 U(\xi,\chi,\theta,\phi) \equiv H(\chi,\theta,\phi)^{\dagger}e^{i\xi\Sigma^{[N]}_{45}} H(\chi,\theta,\phi)~~~~~(H(\chi,\theta,\phi)\equiv e^{-i\chi\Sigma^{[N]}_{34}}e^{i\theta \Sigma^{[N]}_{31}}e^{i\phi\Sigma^{[N]}_{12}}). \label{impunitmat}
\ee
 The maximal eigenvalue  of $X_5^{[N]}$ is attained at the north pole $s=I/2$ of the outermost quasi-fuzzy four-sphere 
$n=N$ with degeneracy $d(N, I, s=I/2)$: 
\be
X^{[N]}_5\bs{e}_{\sigma}=\frac{I}{2N+I+4}\bs{e}_{\sigma}~~~~~~~(\sigma=1,2,\cdots, d(N, I, I/2)).
\ee
Here, $\bs{e}_{\sigma}$ denotes a $D(N,I)$-component unit vector with $(\bs{e}_{\sigma})_{\alpha=1,2,\cdots, D(N,I)}\equiv \delta_{\sigma\alpha}$. 
The $SO(5)$ rotation of  $\bs{e}_\sigma$ to align with the direction of $x_a$ will result in 
\be
(x_a X^{[N]}_a) \Psi_\sigma^{[N,I]}=r \frac{I}{2N+I+4} ~\Psi^{[N,I]}_\sigma, 
\label{coheeq}
\ee
where 
\be
\Psi_\sigma^{[N,I]}(\xi,\chi,\theta,\phi) \equiv U(\xi,\chi,\theta,\phi) \bs{e}_\sigma=\begin{pmatrix}
 U_{1, \sigma} \\
U_{2, \sigma} \\
\vdots \\
U_{D, \sigma}
\end{pmatrix}. 
\label{coheso5}
\ee
Equation (\ref{coheeq}) signifies a generalized $SO(5)$ spin-coherent state equation and its simplest version  $(N=0, I=1)$ corresponds to Eq.(\ref{simplecoheresp}).\footnote{From (\ref{coheeq}), we have  
\be
{\Psi_\sigma^{[N,I]}}^{\dagger}X_a^{[N]}\Psi_\sigma^{[N,I]}=r\frac{I}{2N+I+4} x_a.
\ee
This concise  form of the transformation from matrix coordinates $X_a$ to classical coordinates $x_a$ is given in  
Ref.\cite{Nair-2020}. } 
The spin-coherent states (\ref{coheso5}) 
constitute an ortho-normal set:\footnote{The $SO(5)$ spin-coherent states  are closely related to the $SO(5)$ Landau level eigenstates (\ref{x5so5monopolehamonicsnorm}), as both are realized in the unitary matrix (\ref{impunitmat}) \cite{Hasebe-2022}. See Ref.\cite{Hasebe-2023-2} for more details. }
\be
{\Psi_\sigma^{[N,I]}}(\xi,\chi,\theta,\phi)^{\dagger}~\Psi_{\tau}^{[N,I]}(\xi,\chi,\theta,\phi)=\delta_{\sigma\tau}. 
\ee
The ground state energy is then obtained as 
\be
E_{\text{G}}(r) = r^2 -2r \frac{I}{2N+I+4}+\frac{I(I+2)}{(2N+I+4)(2N+I+2)},  \label{energg}
\ee
and the corresponding eigenstates are given by (\ref{coheso5}) with degeneracy 
\be 
d(N,I, I/2) =(N+1)(N+I+1).   \label{spindegecoh}
\ee
The classical vacuum of $E_{\text{G}}(r)$ (\ref{energg}) is attained by 
\be
r=\frac{I}{2N+I+4}. \label{rdetmin}
\ee
Note that Eq.(\ref{rdetmin}) is equal to the radius of the outermost quasi-fuzzy four-sphere $n=N$ (\ref{radnsh}). From (\ref{rdetmin}), we have 
\be
\lim_{I\rightarrow\infty}r =1. 
\ee
We thus obtained the classical $S^4$ geometry $(x_ax_a=1)$ from $X_a^{[N]}$.

\subsubsection{Probe brane method}\label{subsec:probb}

The probe brane method \cite{Berenstein-Dzienkowski-2012, Karczmarek-Yeh-2015, Badyn-Karczmarek-SabellaGarnier-Yeh-2015} adopts the Dirac-operator  matrix 
\be
D(x_a) =\sum_{a=1}^5\gamma_a \otimes (X^{[N]}_a-x_a\bs{1}_{D(N, I)}). \label{probdir}
\ee
In this method, the classical manifold is obtained through the  following two steps. First, we consider the condition for the existence of the zero-modes of the Dirac-operator matrix  (\ref{probdir}). 
 For zero-modes to exist, $x_a$ must satisfy a certain condition which characterizes  a classical manifold.   Subsequently, we take $I\rightarrow \infty$ limit of the classical manifold to derive the corresponding classical geometry. 
 Due to the tensor product form of  (\ref{probdir}), it is  rather technically  intricate to derive general results unlike the case of the coherent state method.  Hence,  we conduct  numerical investigations by employing the explicit forms of $X_a^{[N]}$. The obtained numerical results imply 
\be 
\text{det}(D(x_a))|_{x_ax_a =(\frac{I}{2N+I+4})^2}=0 \label{zeromodecond2nd}
\ee 
and the  number of the zero-modes
\be  
d(N, I+1, (I+1)/2)=(N+1)(N+I+2).   \label{nozeromodes}
\ee
Equation (\ref{zeromodecond2nd}) indicates that the zero-modes exist when $x_a$ satisfy  $r =\frac{I}{2N+I+4}$, which is equal to the previous result (\ref{rdetmin}). 
 Therefore,   both the coherent state method and the probe brane method yield the identical classical geometry in the present case.  Meanwhile, the number of the zero-modes (\ref{nozeromodes}) is distinct from  that of the coherent states (\ref{spindegecoh}).

\section{Realization in  Yang-Mills matrix models }\label{sec:ymmatrixmodel}

 In this section, 
we demonstrate that the nested fuzzy four-spheres realize new classical solutions of Yang-Mills matrix models and investigate their physical properties. 
 In particular, we clarify  distinct behaviors between the lowest Landau level matrix geometry and the newly obtained higher Landau level matrix geometries.

\subsection{Basic relations}\label{sec:basf4mat}

\begin{figure}[tbph]
\center
\includegraphics*[width=160mm]{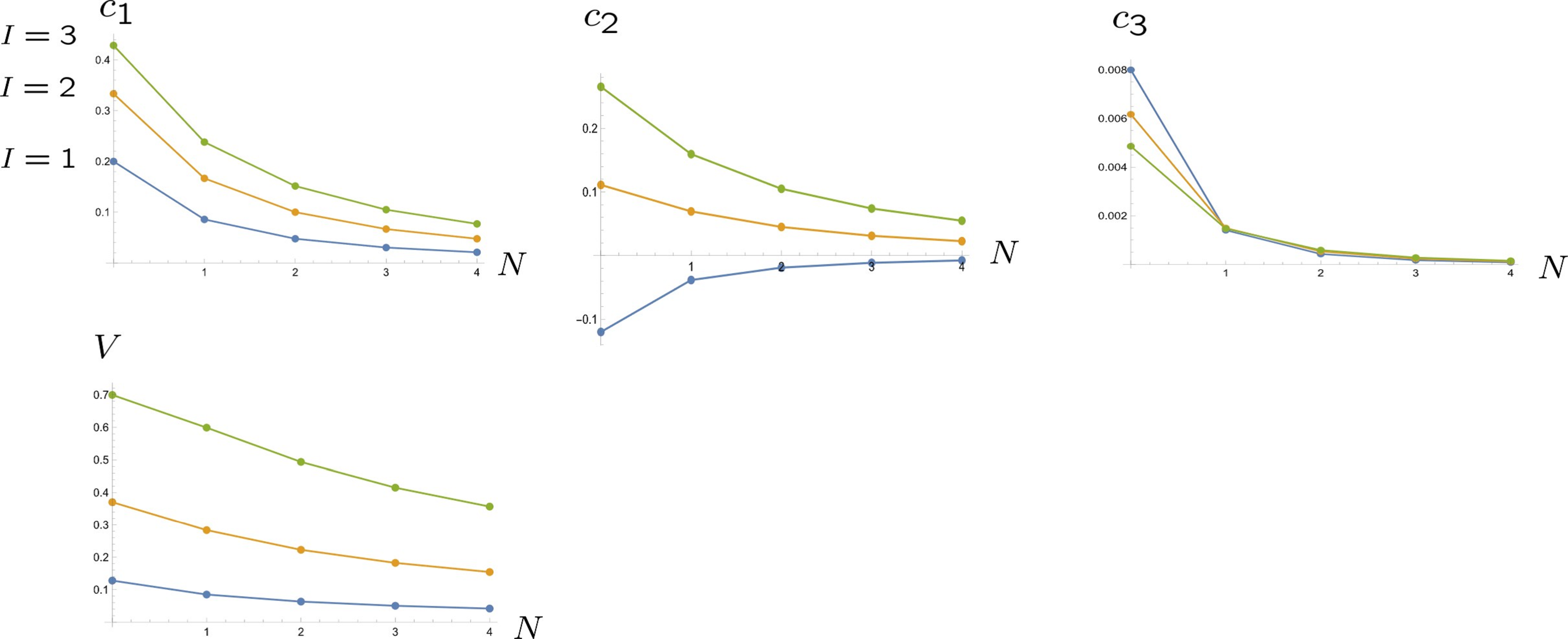}
\caption{The upper: the behaviors of $c$s. The blue, orange and green lines correspond to $I=1, 2, 3$, respectively. The lower: the behaviors of the potential (\ref{potso5}). }
\label{f4cs.fig}
\end{figure}

Using the explicit forms of $X_a^{[N]}$, we can demonstrate that $X_a^{[N]}$ satisfy 
\begin{subequations}
\begin{align}
&X_a^{[N]} X_a^{[N]} = c_1 (N, I) \bs{1}_{D(N, I)} , \label{sumfou1}\\
&X_b^{[N]} X_a^{[N]} X_b^{[N]} =c_2(N, I) X_a^{[N]}, \label{sumfou2}\\
&\epsilon_{abcde} X_b^{[N]}X_c^{[N]}X_d^{[N]}X_e^{[N]}=-4!~c_3(N, I)X_a^{[N]}, \label{sumfou3}
\end{align}\label{basalgexs}
\end{subequations}
where $c_1$ and $c_3$ are given by (\ref{neworirad}) and (\ref{definec3}), respectively. 
In principle, we can determine all of values of the $c$s through Eq.(\ref{basalgexs}). 
For instance, 
$c_2(1,1)=-\frac{47}{1225},~ c_2(1,2) =\frac{5}{72}$, $c_2(1,3) =\frac{211}{1323}$, $c_3(1,1)=\frac{181}{128625},~ c_3(1,2) =\frac{31}{20736}$, $c_3(1,3) =\frac{367}{250047}$.\footnote{In the lowest Landau level $(N=0)$, the coefficients are given by 
\be
c_1(0, I) =\frac{I}{I+4} ,~~~~c_2(0, I) =\frac{I^2 +4I-8}{(I+4)^2}, ~~~~c_3(0, I) =\frac{I+2}{3(I+4)^3}.
\ee
}   
Equations (\ref{sumfou1}) and (\ref{sumfou2}) imply 
\be
 -\frac{1}{4}([X_a^{[N]}, X_b^{[N]}]^2)=\frac{1}{2}(c_1-c_2)c_1 \bs{1}_{D(N, I)}
\ee
and the potential energy is expressed as  
\be
V(X_a^{[N]})\equiv -\frac{1}{4}\tr([X_a^{[N]}, X_b^{[N]}]^2)=\frac{1}{2}(c_1-c_2)c_1 {D(N, I)}.  \label{potso5}
\ee
The behaviors of the $c$s  and the $V$ are illustrated in Fig.\ref{f4cs.fig}. Their behaviors are similar to those of the fuzzy two-sphere (see Fig.\ref{f2cs.fig}). 

\begin{figure}[tbph]
\center
\includegraphics*[width=140mm]{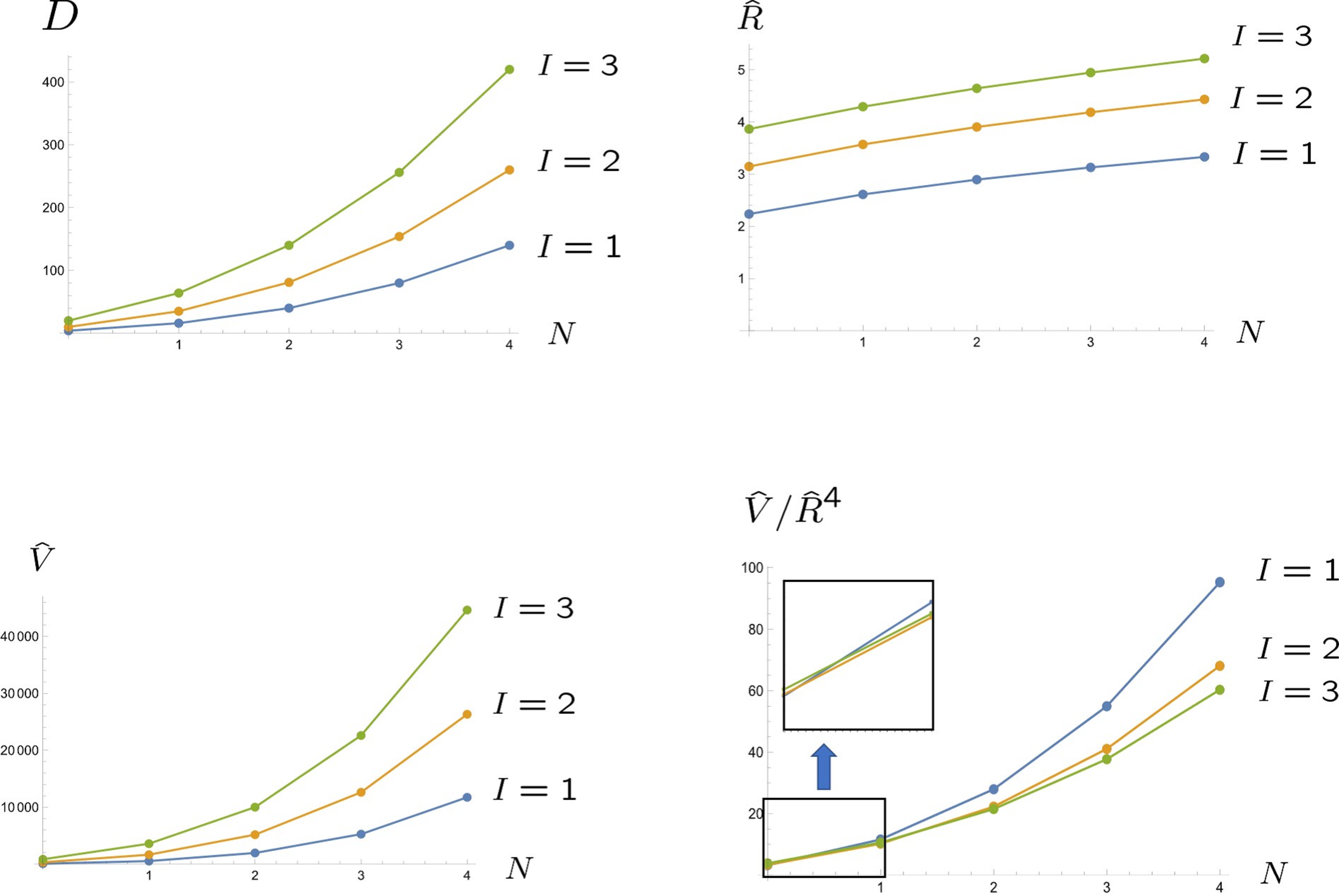}
\caption{Behaviors of the matrix size $D$ (the upper left), the radius (the upper right) and the potential energy (the lower left) and the potential energy density (the lower right).      }
\label{f4.fig}
\end{figure}

While we have utilized $X_a^{[N]}$ as  the  matrix coordinates, from an algebraic standpoint, it might be more natural to adopt "normalized" matrix coordinates that align with the quantum Nambu algebra:
\be
 [\hat{X}_a^{[N]}, \hat{X}_b^{[N]}, \hat{X}_c^{[N]}, \hat{X}_d^{[N]}]=-4! \epsilon_{abcde}\hat{X}_e^{[N]}, 
\ee
or  
\be
\hat{X}_a^{[N]} =\frac{1}{{c_3}^{1/3}} X_a^{[N]}. \label{intronomadf4co}
\ee
For $\hat{X}_a^{[N]}$, important physical quantities are given by\footnote{In the lowest Landau 
level $(N=0)$, Eq.(\ref{intronomadf4co}) is reduced to  
$\hat{X}^{[0]}_{a} =(\frac{3}{I+2})^{\frac{1}{3}} (I+4) X_a^{[0]} =(\frac{3}{I+2})^{\frac{1}{3}}\Gamma_a$  ,
and Eq.(\ref{algnormatot}) becomes 
\be
\hat{R}=I^{1/2}(I+4)^{1/2}\biggl( \frac{3}{I+2}\biggr)^{1/3}, ~~~~V
=2\biggl(\frac{3}{I+2}\biggr)^{1/3} I(I+1)(I+3)(I+4),~~~ 
\frac{V}{\hat{R}^4} =\frac{2}{3} \frac{(I+1)(I+2)(I+3)}{I(I+4)}.  
\ee
} 
\begin{subequations}
\begin{align}
&\text{radius}~~:~\hat{R} \equiv \frac{{c_1}^{1/2}}{{c_3}^{1/3}}~~~~~~~~(\hat{X}^{[N]}_a \hat{X}^{[N]}_a =\hat{R}^2 \bs{1}_{D(N, I)} ),\\
&\text{potential energy}~:~~\hat{V}=-\frac{1}{4}\tr([\hat{X}^{[N]}_a, \hat{X}^{[N]}_b]^2) =\frac{1}{2{c_3}^{4/3}}(c_1-c_2)c_1 D(N,I),\\
&\text{potential energy density}~:~~\frac{\hat{V}}{\hat{R}^4}=-\frac{1}{4}\tr([\hat{X}^{[N]}_a, \hat{X}^{[N]}_b]^2) =\frac{c_1-c_2}{2c_1} D(N,I).
\label{algnorma}
\end{align}\label{algnormatot}
\end{subequations}
The quantities in Eq.(\ref{algnormatot}) are plotted in Fig.\ref{f4.fig}.
 The radius $\hat{R}$  increases as $N$ increases (Fig.\ref{f4.fig}) unlike the original $R$ (\ref{orirad}) (Fig.\ref{suml.fig}). 
The behaviors of the quantities in Eq.(\ref{algnormatot}) are qualitative similar to those of the fuzzy two-sphere (Fig.\ref{fs2m.fig}), except for the potential energy densities  (the lower right of Fig.\ref{f4.fig}) in which 
the order of magnitudes for $I=1,2,3$ is reversed between the lowest Landau level $(N=0)$ and the higher Landau levels $(N\ge 1)$.

\subsection{As classical solutions of Yang-Mills matrix models}\label{sec:realymmat}

\subsubsection{With a mass term}\label{sec:massymfour4}

\begin{figure}[tbph]
\center
\includegraphics*[width=120mm]{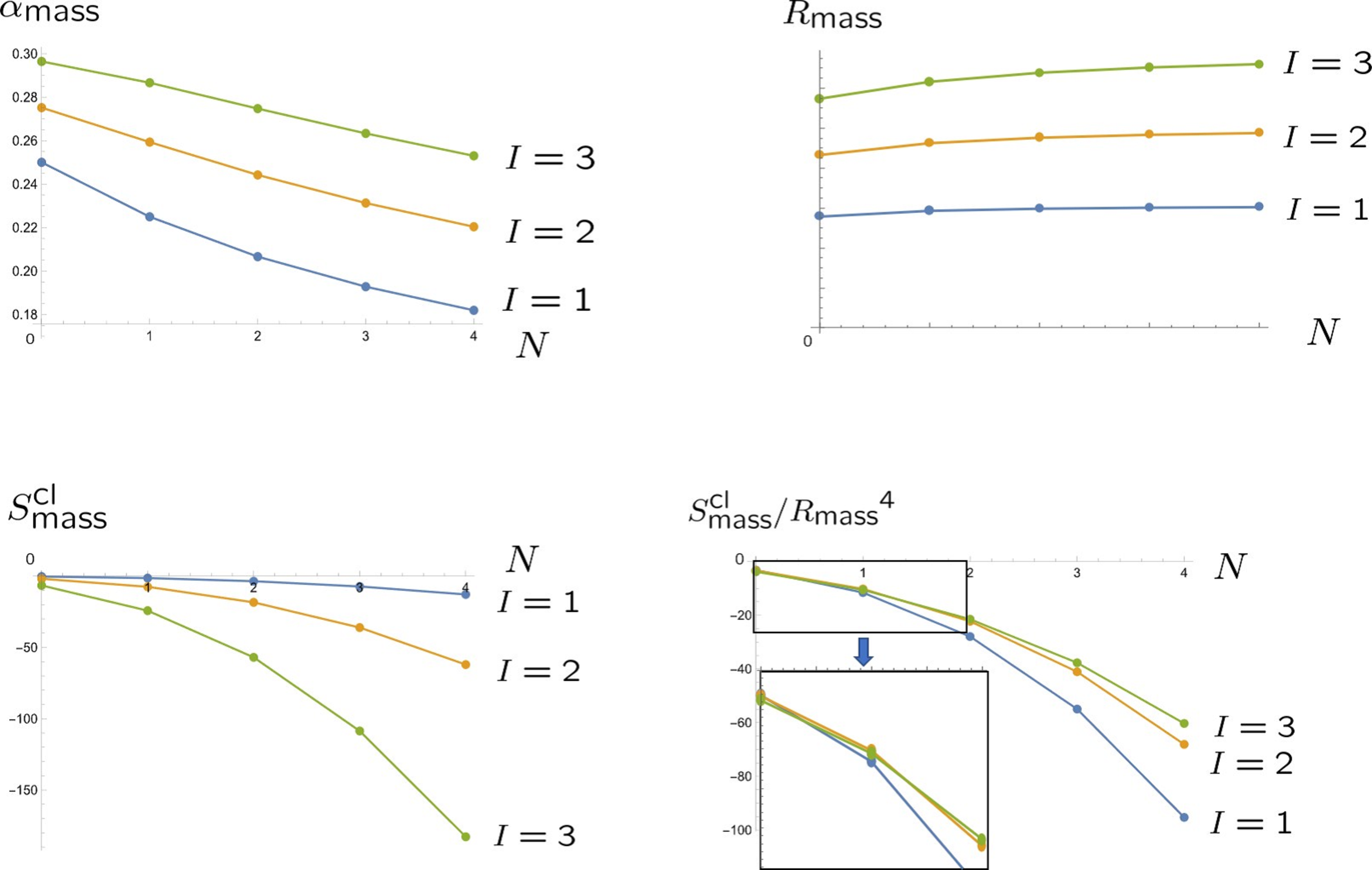}
\caption{Behaviors of Eqs.(\ref{alphmass}) and (\ref{unitmas4sum}).}
\label{massf4.fig}
\end{figure}

Let us consider Yang-Mills matrix model with a mass term \cite{Kimura2002}:  
\be
S_{\text{mass}}=-\frac{1}{4}\tr([A_a, A_b]^2) -\frac{1}{2}\rho~\tr({A_a}^2)~~~~~~(\rho >0). 
\ee
Under the scaling $A_i ~~\rightarrow~~\sqrt{\rho} A_i$,  the parameter $\rho$ turns to the overall scale factor of the action and does not have any physical effect.   
We will take $\rho=1$: 
\be
S_{\text{mass}} =-\frac{1}{4}\tr([A_a, A_b]^2) -\frac{1}{2}\tr({A_a}^2). 
\ee
The equations of motion are derived as 
\be
[[A_a, A_b], A_b] =A_a. \label{eommassive}
\ee
Using (\ref{basalgexs}), we readily see that the nested fuzzy four-spheres realize new classical solutions:\footnote{Fuzzy two-sphere and fuzzy torus are also solutions of Eq.(\ref{eommassive}) \cite{Kimura2002}.} 
\be
A_a^{\text{cl}} =\alpha_{\text{mass}}(N, I)~\hat{X}^{[N]}_a, 
\ee
where the non-commutative parameter  is given by 
\be
\alpha_{\text{mass}}(N,I)~\equiv \frac{c_3(N, I)^{1/3}}{\sqrt{2(c_1(N, I)-c_2(N,I))}}. \label{alphmass}
\ee
The non-commutative parameter $\alpha$ is  a parameter-dependent quantity unlike the case of the fuzzy two-sphere solution (see Appendix \ref{sec:matimoso3}). This  brings specific physical properties to the fuzzy four-sphere solutions. 
The physical quantities  (\ref{algnormatot}) are evaluated as\footnote{In the lowest Landau level $(N=0)$, Eq.(\ref{unitmas4sum}) is reduced to 
\be
R_{\text{mass}} =\frac{1}{4}\sqrt{I(I+4)}, ~~~~~S_{\text{mass}}^{\text{cl}} =-\frac{1}{384}I(I+1)(I+2)(I+3)(I+4),~~~~~\frac{S_{\text{mass}}^{\text{cl}}}{{R_{\text{mass}}}^4} =-\frac{2}{3}\frac{(I+1)(I+2)(I+3)}{I(I+4)}.
\ee
} 
\begin{subequations}
\begin{align}
&\text{radius}~~:~R_{\text{mass}} \equiv \sqrt{\frac{c_1}{2(c_1-c_2)}}~~~~(A_a^{\text{cl}}  A_a^{\text{cl}}  ={R_{\text{mass}}}^2 \bs{1}),\\
&\text{action}~~:~S_{\text{mass}}^{\text{cl}} =(-\frac{1}{4}+\frac{1}{2})\tr([A_a^{\text{cl}}, A_b^{\text{cl}}]^2) =\frac{1}{4}\tr([A_a^{\text{cl}}, A_b^{\text{cl}}]^2) =-\frac{1}{8}\frac{c_1}{c_1-c_2} D =-\frac{1}{4}{R_{\text{mass}}}^2 D , \\
&\text{action density}~~:~\frac{S_{\text{mass}}^{\text{cl}}}{{R_{\text{mass}}}^4} =-\frac{c_1-c_2}{2c_1}D =-\frac{1}{4 {R_{\text{mass}}}^2}D=-\frac{V}{\hat{R}^4}. \label{unitmas4s}
\end{align}\label{unitmas4sum}
\end{subequations}
The behaviors of Eqs.(\ref{alphmass}) and (\ref{unitmas4sum}) are shown in Fig.\ref{massf4.fig}. 
Similar to the case of $\hat{X}_a^{[N]}$ in Sec.\ref{sec:basf4mat}, the action densities (the lower right of Fig.\ref{massf4.fig}) exhibit qualitatively distinct behaviors  to  the fuzzy two-sphere (Fig.\ref{fs2m.fig}).  

\subsubsection{With a fifth-rank Chern-Simons term}

\begin{figure}[tbph]
\center
\includegraphics*[width=140mm]{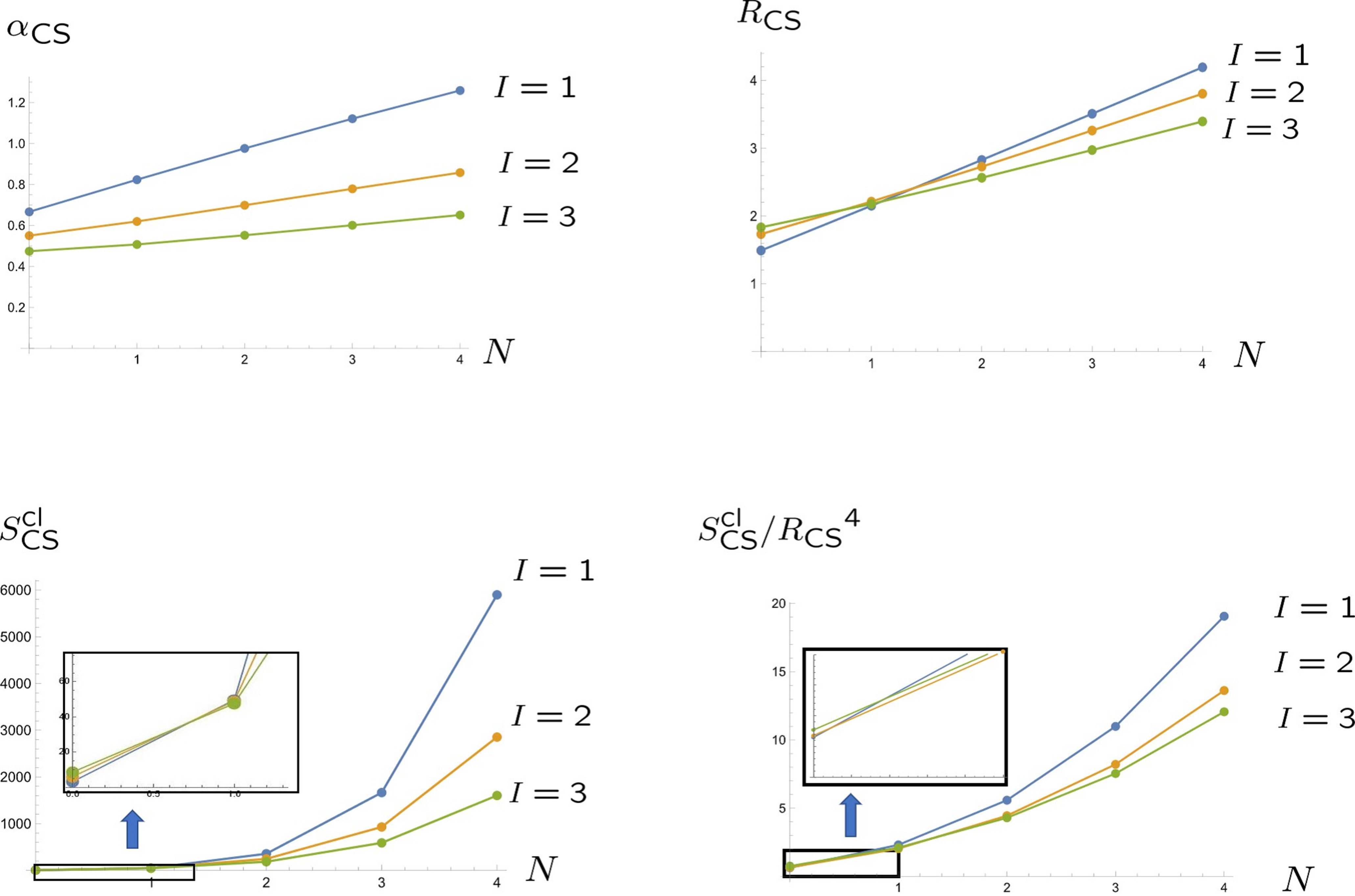}
\caption{Behaviors of the non-commutative scale (the upper left), the radius (the upper right),  the action (the lower left), and  the action density (the lower right).      }
\label{f4m.fig}
\end{figure}

We next consider the Yang-Mills matrix model with  a fifth-rank Chern-Simons term  \cite{Kimura2002}
\be
S_{\text{CS}}[X_a]=-\frac{1}{4}\text{tr}\biggl([X_a, X_b]^2 \biggr)+\frac{\lambda}{5}\epsilon_{abcde}\text{tr}\biggl(X_aX_bX_cX_dX_e\biggr). 
\ee
The coupling constant $\lambda$ can be absorbed  in the action when scaling $A_a$ as 
$A_a ~~\rightarrow~~\frac{1}{\lambda} \cdot A_a$.    
We then set $\lambda=1$ and  deal with the following action: 
\be
S_{\text{CS}}=- \frac{1}{4}\text{tr}\biggl([A_a, A_b]^2\biggr) +\frac{1}{5}\epsilon_{abcde}\text{tr}\biggl(A_aA_bA_cA_dA_e\biggr). 
\ee
The equations of motion are given by 
\be
[ [A_a, A_b], A_b]= -\epsilon_{abcde}A_bA_cA_dA_e.  
\ee
From  (\ref{basalgexs}), we easily obtain new classical solutions as   
\be
A_a^{\text{cl}} = \alpha_{\text{CS}}(N,I)~\hat{X}_a^{[N]}
\ee
with 
\be
\alpha_{\text{CS}}(N,I) \equiv \frac{1}{12}\frac{c_1(N, I)-c_2(N, I)}{{c_3(N,I)}^{2/3}},
\ee
and\footnote{In the lowest Landau level $(N=0)$, we have 
\be
\alpha=\frac{2}{3} \biggl(  \frac{3}{I+2}\biggr)^{2/3}, ~~~~
A_{a}^{\text{cl}} =\alpha \hat{X}_a^{[0]} =\frac{2}{I+2}\Gamma_a, 
\ee
which satisfies 
\be 
[A_a^{\text{cl}}, A_b^{\text{cl}}, A_c^{\text{cl}}, A_d^{\text{cl}}]=-\biggl(\frac{8}{I+2}\biggr)^2 \epsilon_{abcde}A_e^{\text{cl}}.
\ee
Equation (\ref{units4dett}) reproduces the results of Ref.\cite{Kimura2002} for $N=0$:  
\be 
R_{\text{CS}}=\frac{2\sqrt{I(I+4)}}{I+2}, ~~~~~
S_{\text{CS}}^{\text{cl}}= \frac{32}{15}\frac{I(I+1) (I+3)(I+4)}{(I+2)^3},~~~~~~~
\frac{S_{\text{CS}}^{\text{cl}}}{{R_{\text{CS}}}^4} =\frac{2}{15}\frac{(I+1)(I+2)(I+3)}{I(I+4)}. \label{units4den0} 
\ee
}  
\begin{subequations}
\begin{align}
&\text{radius}~:~~~R_{\text{CS}}\equiv  \frac{1}{12}\frac{(c_1-c_2){c_1}^{1/2}}{{c_3}}=\alpha_{\text{CS}} \hat{R}~~~~~~~(A_a^{\text{cl}} A_a^{\text{cl}} ={\alpha_N}^2  \hat{X}_a^{[N]}\hat{X}_a^{[N]}={R_{\text{CS}}}^2  \bs{1}_{D(N, I)}), \\ 
&\text{action}~:~~~S^{\text{cl}}_{\text{CS}}\equiv S_{\text{CS}}[A_a^{\text{cl}}]=-\overbrace{(\frac{1}{4}-\frac{1}{5})}^{=\frac{1}{20}} \tr([A^{\text{cl}}_a, A^{\text{cl}}_b]^2) 
=   \frac{1}{5}{\alpha_{\text{CS}}}^4 V =\frac{1}{10} \frac{(c_1-c_2)^5 c_1}{(12 c_3)^4}D(N, I),\label{clsoacva5d}\\
&\text{action density}~:~~~\frac{S_{\text{CS}}^{\text{cl}}}{{R_{\text{CS}}}^4} =\frac{1}{10} \frac{c_1-c_2}{{c_1}}D(N, I) 
=\frac{1}{5}\frac{V}{{\hat{R}_{\text{CS}}}^4} =-\frac{1}{5}\frac{S_{\text{mass}}^{\text{cl}}}{{R_{\text{mass}}}^4} . \label{units4de}
\end{align}\label{units4dett}
\end{subequations}
Figure \ref{f4m.fig} depicts the  behaviors of Eq.(\ref{units4dett}). 
There are three noteworthy points. First, the order  of $\alpha_{\text{CS}}$ 
for $I=1,2,3$ is the reverse of that of $\alpha_{\text{mass}}$ 
(the upper left in Fig.\ref{massf4.fig}). Second, the order of magnitudes of $R_{\text{CS}}$ 
for $I=1,2,3$ is reversed  between $N=0$ and $N=2$. Last,  the order of magnitudes of both $S_{\text{CS}}^{\text{cl}}$ (lower left in Fig.\ref{f4m.fig}) and $S_{\text{CS}}^{\text{cl}}/{R_{\text{CS}}}^4$  (lower right in Fig.\ref{f4m.fig}) of $N=0$ for $I=1,2,3$  is the reverse of those of $N\ge 1$. Thus, the lowest Landau level matrix geometry $(N=0)$ and the newly obtained higher Landau level matrix geometries $(N\ge 1)$  exhibit qualitatively distinct physical properties. It is rather curious that, while the matrix size $D$ is a monotonically increasing function about $I$ and the quantities such as $R_{\text{CS}}$ and $S_{\text{CS}}^{\text{cl}}$ are expected to show similar behaviors the higher Landau level matrix geometries, $i.e.$ the nested fuzzy four-spheres, do not follow this anticipation.

\section{Even higher dimensions}\label{sec:evenhigh}

We here investigate  higher Landau level matrix geometries in even higher dimensions. The associated higher form gauge field and Yang-Mills matrix model are also discussed.  

\subsection{Landau level matrix geometries}

It is known that (unnested) higher dimensional fuzzy spheres are realized as the lowest Landau level matrix geometries in  higher dimensional Landau models   \cite{Hasebe-2017, Hasebe-2014-1, Hasebe-Kimura-2003}.  Since $S^{d}\simeq SO(d+1)/SO(d)$, the corresponding gauged quantum mechanics is given by  the $SO(d+1)$ Landau model  in the $SO(d)$ non-Abelian monopole background.  
The  matrix coordinates in the lowest Landau level are given by the fully symmetric combination of the $SO(2k+1)$ gamma matrices: 
\be
X_a^{^{[N=0]}} =\frac{1}{I+2k}(\gamma_a\otimes 1 \otimes \cdots \otimes 1 +1\otimes \gamma_a \otimes \cdots \otimes 1 +\cdots +1\otimes 1 \otimes \cdots \otimes \gamma_a  )_{\text{sym}},  
\ee
which satisfy the $Spin(2k+2)$ Lie algebraic commutation relations together with the $SO(2k)$ generators. 

The higher Landau level geometries in the $SO(d+1)$ Landau model have not been investigated so far.  
Though    it is  in principle  possible to derive higher Landau level matrix geometries  by following the present non-commutative scheme,  it is rather laborious  to solve the eigenvalue problem of the higher dimensional Landau Hamiltonian.  
  Furthermore, the resulting matrix structures may be mathematically too involved to deduce  useful information about the higher dimensional non-commutative geometry.  Therefore, we will engage in a somewhat speculative yet more general discussion based on group theory.
Let us focus on  the following  $SO(2k+1)$ irreducible representation 
\be
[l_1, l_2, \cdots, l_{k}]_{SO(2k+1)} =[N+I, I, \cdots, I], \label{nthirrepchoice}
\ee
which corresponds to the $N$th Landau level eigenstates of the  $SO(2k+1)$ Landau model studied in Refs.\cite{Hasebe-2014-1, Hasebe-Kimura-2003}. From  group representation theory, the corresponding degeneracy is given by   
\be
D(N, I)_{SO(2k+1)}= \frac{2N+I+2k-1}{(2k-1)!!}\frac{(N+k-1)!}{N!(k-1)!}\frac{(I+2k-3)!!}{(I-1)!!} \frac{(N+I+2k-2)!}{(N+I+k-1)!}\prod_{l=1}^{k-2} \frac{(I+2l)!}{(I+l)!}\prod_{l=1}^{k-1}\frac{l!}{(2l)!}. 
\label{degsk+1gene} 
\ee
Meanwhile, for the one dimension lower $SO(2k)$ Landau model, the Landau levels consist of sub-bands \cite{Hasebe-2017}. The eigenstates of the $s$ band of the $n$th Landau level  constitute an $SO(2k)$  irreducible representation:
\be
[l_1, l_2, \cdots, l_{k-1},  l_{k}]_{SO(2k)} =[n+\frac{I}{2}, \frac{I}{2}, \cdots, \frac{I}{2}, s],
\ee
with degeneracy 
\begin{align}
d(n, I, s)_{SO(2k)}&=\frac{(2n+I+2k-2)^2-4s^2}{4(k-1)^2}\cdot \prod_{2\le i \le k-1} \frac{(n+I+2k-i-1)(n+i-1)}{(2k-i-1)(i-1)}\nn\\
&\cdot \prod_{2 \le i < j \le k-1 } \frac{I+2k-i-j}{2k-i-j} \cdot \prod_{2 \le i \le k-1} \frac{(I+2k-2i)^2-4s^2}{4(k-i)^2} .  
\label{degelandaunlambda}
\end{align}
There exists an  $\it{exact}$ relation  between the degeneracies (\ref{degsk+1gene}) and (\ref{degelandaunlambda}):
\be
D(N, I)_{SO(2k+1)} =\sum_{n=0}^{N} \sum_{s=-I/2}^{I/2}d(n,I,s)_{SO(2k)}.  \label{dnidnisgen}
\ee
Equation (\ref{dnidnisgen}) signifies a higher dimensional generalization of Eq.(\ref{dnidnis}) and implies that the $SO(2k+1)$ irreducible representation is constructed by   adding up the $SO(2k)$ sectors from $n=0$ to $n=N$ each of which is made of $s=I/2, \cdots, -I/2$. Since the geometric structure of the  fuzzy manifold reflects on the structure of the irreducible representation, the matrix geometry of the $N$th Landau level is expected to  exhibit $N+1$ nested fuzzy structures in arbitrary dimensions,  just like  the nested fuzzy four-sphere.  
It is also reasonable to consider that fuzzy $(2k-1)$-spheres are embedded within the nested 
fuzzy $2k$-sphere.   

\subsection{Higher form gauge field and Yang-Mills matrix model}

The lowest Landau level matrix geometry in $2k$ dimension is associated with a generalized Hopf maps \cite{Hasebe-2010} and are described by both the $SO(2k+2)$ Lie algebra and quantum Nambu $2k$ algebra.  Meanwhile,
the matrix coordinates in the higher Landau levels are not associated with the generalized Hopf map but 
are covariant under the $SO(2k+1)$ transformation like the lowest Landau level matrix coordinates. 
Therefore, the higher Landau level matrix coordinates  will not conform with the Lie algebraic description  but instead is described by the quantum Nambu algebra exclusively:
\be
[X_{a_1}, X_{a_2}, \cdots, X_{a_{2k}}] =(2k)!~ i^k~ c_3~\epsilon_{a_1 a_2 \cdots a_{2k+1}} X_{2k+1}. \label{namb2kbra}
\ee
When  one adopt more general irreducible representations beyond Eq.(\ref{nthirrepchoice}), 
the corresponding fuzzy manifold will exhibit a more exotic quantum  geometry than the nested fuzzy structure, however, due to the existence of the $SO(2k+1)$ covariance, the matrix coordinates will also adhere to the quantum Nambu algebra (\ref{namb2kbra}).  

Interestingly,  ``magnetic field'' appears on  the right-hand side of (\ref{namb2kbra}) \cite{Hasebe-2014-1}, which signifies the tensor monopole field strength : 
\be
G_{2k} = 
\frac{1}{2^{k+1}r^{2k+1}}~\epsilon_{a_1 a_2 \cdots a_{2k+1}} x_{a_{2k+1}}dx_{a_1}\wedge dx_{a_2}\wedge \cdots \wedge dx_{a_{2k}}. 
\ee
The existence of the higher form gauge field behind the quantum Nambu geometry is thus glimpsed. One may wonder where such a higher gauge symmetry comes from, where as the present quantum mechanical system only has the $SO(2k)$ gauge symmetry.   
Indeed, the tensor monopole gauge field is directly obtained from the $SO(2k)$ monopole gauge field through the Chern-Simons term \cite{Hasebe-2014-1}.

The (unnested) fuzzy $2k$-sphere realizes a solution of \cite{Kimura2003}
\be
[ [X_a, X_b], X_b]=  i^k  \epsilon_{a a_2  a_3 \cdots a_{2k+1}}X_{a_2}X_{a_3}\cdots X_{a_{2k+1}}, 
\label{eofmatcogen2k}
\ee
which is derived from the Yang-Mills matrix model with a $2k+1$ rank Chern-Simons term,
\be
S_{\text{CS}}[X_a]=-\frac{1}{4}\text{tr}([X_a, X_b]^2) -i^k\frac{1}{2k+1}\text{tr}(\epsilon_{a_1 a_2\cdots a_{2k+1}}X_{a1}X_{a_2}\cdots X_{a_{2k+1}}).
\ee
Since the equations of motion are concerned with the covariance of the matrix coordinates, it is anticipated  that the nested fuzzy $2k$-spheres realize classical solutions of Eq.(\ref{eofmatcogen2k}).
The action of for the fuzzy $2k$-sphere  solution is given by  
\be
S_{\text{CS}}(X_a=X^{\text{cl}}_a) =(-\frac{1}{4}+\frac{1}{2k+1})\text{tr}([X_a^{\text{cl}}, X_b^{\text{cl}}]^2)=\frac{2k-3}{2k+1} ~V(X^{\text{cl}}_a),
\ee
where
\be
V(X^{\text{cl}}_a) \equiv -\frac{1}{4}\text{tr}([X_a^{\text{cl}}, X^{\text{cl}}_b]^2).
\ee
While the signs of $S_{\text{CS}}(X^{\text{cl}}_a)$ and $V(X^{\text{cl}}_a)$ are opposite for $k=1$, they have the same sign for $k\ge 2$, as we have seen,  for $k=2$, in Eq. (\ref{clsoacva5d}).

\section{Summary and discussions}\label{sec:sum}

Based on the insight obtained from  the emergent fuzzy geometry in the simple $SO(3)$ Landau model,  we proposed a novel non-commutative scheme for generating the matrix geometries for arbitrary manifolds of the coset type  $G/H$. In the present approach,  manifolds need not be either symplectic  or even dimensional  unlike the conventional non-commutative schemes. We explicitly derived the matrix geometries for $S^4$ by  utilizing  the  $SO(5)$ Landau model. The emergent matrix geometries in higher Landau levels realize  pure quantum Nambu geometries in which  matrix coordinates are not closed within the canonical formalism of the Lie algebra but are described only by introducing the quantum Nambu algebra. We also demonstrated that  such pure quantum matrix geometries manifest new solutions of the Yang-Mills matrix models.    The particular features of the nested quantum geometry,  such as the internal matrix geometries, continuum limit, and classical counterpart, were clarified.   The pure Nambu matrix geometries are 
 common to the higher Landau levels of the Landau models in arbitrary dimensions.  

The conventional scheme is based on the spirit of  quantization of classical (symplectic) manifolds, $i.e.$, the replacement of the Poisson bracket with the commutator, where as the present non-commutative scheme is largely based on the mathematical structure of the Hilbert space behind quantum mechanics from the beginning.  In this sense,  the present scheme is considered to be a  quantum-oriented non-commutative scheme. That is the reason why we  obtained the pure quantum geometry. We showed this  non-commutative scheme is  practically useful in deriving novel solutions of the Matrix models.   
As Matrix model solutions, the nested matrix geometries exhibit quantitatively distinct behaviors with the unnested fuzzy four-sphere. 

The discovery of the novel quantum Nambu matrix geometries now brings various open questions, such as brane construction, relation to tachyon condensation \cite{Asakawa-Sugimoto-Terashima-2002, Terashima-2005},  realization in the Nahm equation  in higher energy physics.  
The higher form gauge field implied by the quantum Nambu algebra is closely related to the  higher Berry phase \cite{Kapustin-Spodyneiko-2020, Kitaev-talk} whose usefulness is getting appreciated in the very recent studies of strongly correlated many-body systems.   It would be intriguing  to speculate on the role of quantum Nambu geometry in condensed matter physics. We also add that
  the present scheme itself should be appropriately generalized to treat less symmetric fuzzy objects,  while  we studied highly symmetric objects in this work.   

To the best of the author's knowledge, this work is the first example of quantum matrix geometry found in the analysis of the Landau models being practically applied  to the solutions of the M(atrix) models. 
 M(atrix) theory is assumed to describe the physics at the Planck scale of $10^{19}$ Gev, while   the Landau models or the quantum Hall effect are about the  low temperature physics   at  milli-electron volt.  
It is rather amazing that same mathematics work in both physics with such a huge energy gap.

\section*{Acknowledgments}

  I would like to thank Harold Steinacker  for valuable email communications.  This work was supported by JSPS KAKENHI Grant No.~21K03542.

\appendix

\section{Groenewold-Moyal plane from planar Landau model }\label{append:planarLandaumodel}

We demonstrate a realization of the  Groenewold-Moyal plane in higher Landau levels.  
Let us consider a 2D plane subject to a constant perpendicular magnetic field: 
\be
\partial_x A_y -\partial_y A_x =B. \label{constb}
\ee
We employ the gauge-independent relation (\ref{constb}), and so all of the following results are also gauge independent.  
The covariant derivatives and the center-of-mass coordinates are respectively constructed as 
\be
D_i =\frac{\partial}{\partial x_i} +iA_i,~~~~X^{\text{CM}}_i=x_i +i\frac{1}{B} \epsilon_{ij}D_j~~~~~(i=1,2),
\ee
which satisfy  two independent commutation relations: 
\be
[D_x, D_y] =iB, ~~~[X^{\text{CM}}, Y^{\text{CM}}] =i\frac{1}{B}, ~~~~[D_i, X^{\text{CM}}_j]=0. 
\ee
We then realize two sets of  creation and annihilation operators  as
\be
a=i\frac{1}{\sqrt{2B}} (D_x-iD_y), ~~~a^{\dagger} =i\frac{1}{\sqrt{2B}}(D_x+iD_y), ~~b=\sqrt{\frac{B}{2}}~ (X^{\text{CM}}+iY^{\text{CM}}), ~~~b^{\dagger}=\sqrt{\frac{B}{2}} ~(X^{\text{CM}}-iY^{\text{CM}}),
\ee
which satisfy 
\be
[a, a^{\dagger}] =[b, b^{\dagger}]=1, ~~~~[a, b]=[a, b^{\dagger}]=0. 
\ee
The Hamiltonian of the  planar Landau model is given by 
\be
H=-\frac{1}{2M}({D_x}^2+{D_y}^2)=\frac{B}{M} (a^{\dagger}a +\frac{1}{2}). 
\ee
The corresponding energy  Landau levels and the eigenstates  are 
\be
E_N =\frac{B}{M}(N+\frac{1}{2}) ,~~~~~|N, m\rangle =\frac{1}{\sqrt{N!~m!}} {a^{\dagger}}^N {b^{\dagger}}^m|0\rangle ~~~~~~~(N, m=0,1,2,\cdots).
\ee
Using  
\be
x=X^{\text{CM}}-i\frac{1}{B} D_y =\frac{1}{\sqrt{2B}}(b+b^{\dagger}) -i\frac{1}{\sqrt{2B}}(a-a^{\dagger}), ~~~y=Y^{\text{CM}}+i\frac{1}{B} D_x =-i\frac{1}{\sqrt{2B}}(b-b^{\dagger}) +\frac{1}{\sqrt{2B}}(a+a^{\dagger}),
\ee
we readily evaluate the matrix elements of $x$ and $y$: 
\begin{align}
&\langle N, m|x|N', m'\rangle =\frac{1}{\sqrt{2B}} (\sqrt{m'}~\delta_{m, m'-1} +\sqrt{m'+1}~\delta_{m, m'+1})~\delta_{N,N'}-i\frac{1}{\sqrt{2B}} (\sqrt{N'}~\delta_{N, N'-1} -\sqrt{N'+1}~\delta_{N, N'+1})\delta_{m,m'}, \nn\\
&\langle N, m|y|N', m'\rangle =-i\frac{1}{\sqrt{2B}} (\sqrt{m'}~\delta_{m, m'-1} -\sqrt{m'+1}~\delta_{m, m'+1})~\delta_{N,N'}+\frac{1}{\sqrt{2B}} (\sqrt{N'}~\delta_{N, N'-1} +\sqrt{N'+1}~\delta_{N, N'+1})\delta_{m,m'}.
\end{align}
The intra-Landau level matrix coordinates are then obtained as  
\begin{align}
&(X^{(N)})_{mm'}\equiv \langle N, m|x|N, m'\rangle =\frac{1}{\sqrt{2B}} (\sqrt{m'}~\delta_{m, m'-1} +\sqrt{m'+1}~\delta_{m, m'+1}), \nn\\
&(Y^{(N)})_{mm'}\equiv \langle N, m|y|N, m'\rangle =-i\frac{1}{\sqrt{2B}} (\sqrt{m'}~\delta_{m, m'-1} -\sqrt{m'+1}~\delta_{m, m'+1}),
\end{align}
or
\be
X^{(N)} =\frac{1}{\sqrt{2B}}\begin{pmatrix}
0 & \sqrt{1} & 0 & 0 & 0 & 0 & \\
\sqrt{1} & 0 & \sqrt{2} & 0 & 0 & 0  \\
0 & \sqrt{2}  & 0 & \sqrt{3} & 0  & 0  \\
0 & 0 & \sqrt{3}  & 0 & \ddots  & 0  \\
0 & 0 & 0 & \ddots  & 0 & \ddots 
\end{pmatrix}, ~~~ Y^{(N)}  =i\frac{1}{\sqrt{2B}}\begin{pmatrix}
0 & -\sqrt{1} & 0 & 0 & 0 & 0 & \\
\sqrt{1} & 0 & -\sqrt{2} & 0 & 0 & 0  \\
0 & \sqrt{2}  & 0 & -\sqrt{3} & 0  & 0  \\
0 & 0 & \sqrt{3}  & 0 & \ddots  & 0  \\
0 & 0 & 0 & \ddots  & 0 & \ddots 
\end{pmatrix}. 
\label{largxy}
\ee
Notice that (\ref{largxy}) does not depend on the Landau level index $N$, and the matrix coordinates satisfy 
\be
[X^{(N)}, Y^{(N)}] =i\frac{1}{B} \bs{1}.
\ee
Obviously, the dimensionless coordinates, $\hat{X}^{(N)}={\sqrt{B}}~X^{(N)}$ and 
$\hat{Y}^{(N)}={\sqrt{B}}~Y^{(N)}$, satisfy the Heisenberg-Weyl algebra together with $\bs{1}$: 
\be
[\hat{X}^{(N)}, \hat{Y}^{(N)}] =i \bs{1}, ~~~[\hat{X}^{(N)}, \bs{1}] =[\hat{Y}^{(N)}, \bs{1}] =0.
\ee
We have thus confirmed the emergence of the Groenewold-Moyal plane in $\it{any}$ Landau level.

\section{Two-sphere matrix coordinates in Yang-Mills matrix models}\label{sec:fuzzy2}

For comparison with the fuzzy four-sphere (Sec.\ref{sec:ymmatrixmodel}), we revisit  the matrix model  analyses of the fuzzy two-sphere \cite{Iso-Kimura-Tanaka-Wakatsuki-2001, Kimura2001}, clarifying physical properties of  the matrix coordinates in the $SO(3)$ Landau model.

\subsection{Basic properties }

\begin{figure}[tbph]
\center
\includegraphics*[width=160mm]{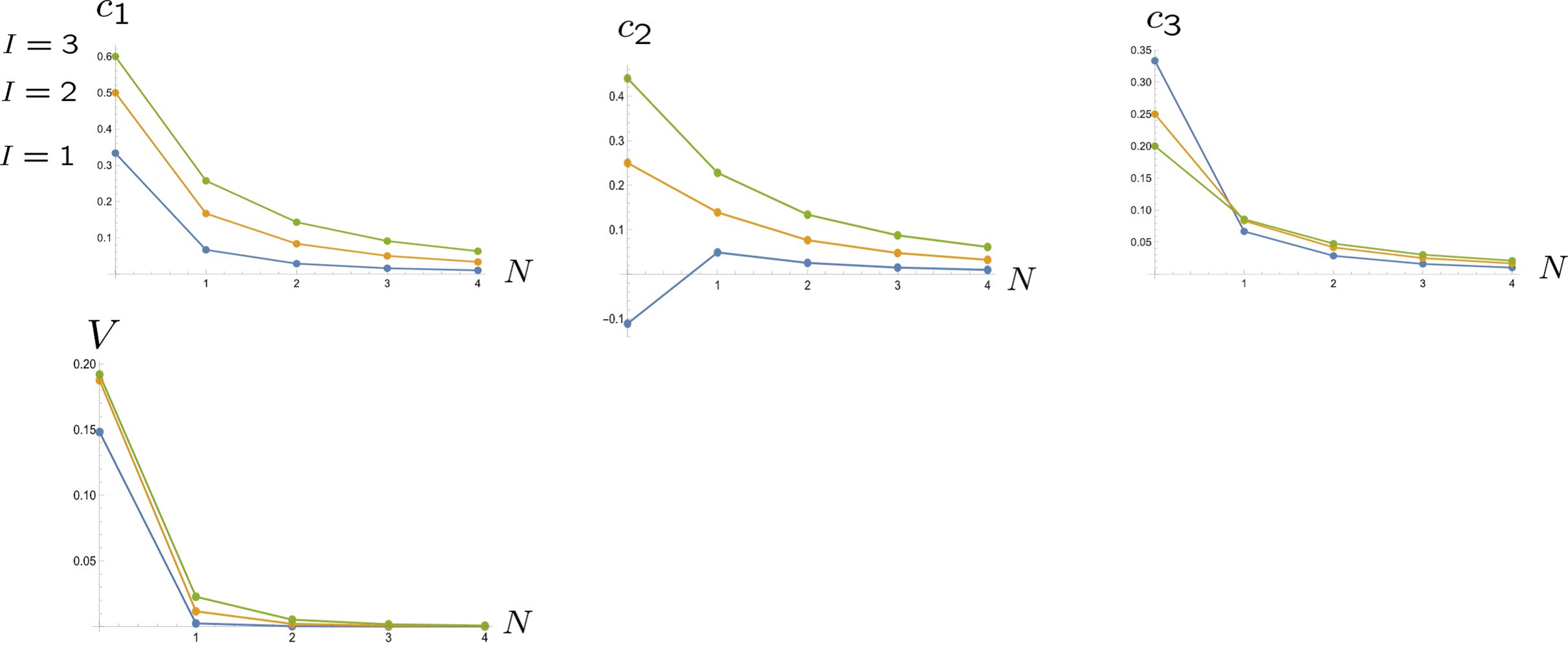}
\caption{ 
The blue, orange and green lines correspond to $I=1, 2, 3$ of Eqs.(\ref{s2cs}) and (\ref{potsu2}). 
}
\label{f2cs.fig}
\end{figure}

The  $N$th Landau level matrix coordinates 
$X_i^{(N)}$ (\ref{xisiso3}) satisfy the following relations\footnote{In the literatures on matrix models, it is common to denote the variables $D$ and $I$ as $N$ and $n$, respectively.}
\begin{subequations}
\begin{align}
&X_i^{(N)}X_i^{(N)} =c_1(N, I) \bs{1}_{D(N,I)}, \\
&X_j^{(N)}X_i^{(N)}X_j^{(N)} =c_2(N, I) X_i^{(N)}, \\
&\epsilon_{ijk}X_j^{(N)} X_k^{(N)} =2i ~c_3(N, I)~X_i^{(N)}, \label{ssrel}
\end{align}
\end{subequations}
where 
\be
D(N, I) =I+2N+1,
\ee
and 
\begin{subequations}
\begin{align}
&c_1(N, I) \equiv \frac{I^2}{{(I+2N)(I+2N+2)}}, \label{s2c1} \\
&c_2(N, I) \equiv \frac{I^2}{{(I+2N)^2(I+2N+2)^2}} ((I+2N)(I+2N+2)-4),\\
&c_3(N, I) \equiv \frac{I}{(I+2N)(I+2N+2)}.
\end{align}\label{s2cs}
\end{subequations}
These $c$s are not independent quantities; instead, they satisfy 
\be
c_1-c_2=4{c_3}^2.
\ee
From (\ref{s2cs}),  we readily have 
\be
-\frac{1}{4}[{X}_i^{(N)}, {X}_j^{(N)}]^2=\frac{1}{2}(c_2-c_2)c_1 \bs{1}_D 
\ee
and  
\be
V(N,I)=-\frac{1}{4}\tr([{X}_i^{(N)}, {X}_j^{(N)}]^2)=\frac{1}{2}(c_2-c_2)c_1 D =2\frac{I^4}{(I+2N)^3(I+2N+2)^3} (2N+I+1). \label{potsu2}
\ee
The behaviors of (\ref{s2cs}) and (\ref{potsu2})  are shown in Fig.\ref{f2cs.fig}. 
 
We introduce ``normalized'' matrix coordinates that satisfy the $SU(2)$ algebra 
\be
[\hat{X}_i^{(N)}, \hat{X}_j^{(N)}] =2i \epsilon_{ijk}\hat{X}_k^{(N)} \label{hats2com}
\ee
as 
\be
\hat{X}_i^{(N)} =\frac{1}{c_3(N, I) }{X}_i^{(N)}  =2S_i^{(\frac{I}{2}+N)}.
\ee
Notice that $\hat{X}_i^{(N)}$ depend on the $SU(2)$ index $l=N+\frac{I}{2}$ rather than $N$ and $I$, separately. Several important physical quantities are evaluated as   
\begin{subequations}
\begin{align}
&\text{radius}~~:~\hat{R}=\frac{\sqrt{c_1}}{c_3}= \sqrt{(I+2N)(I+2N+2)}, \\
&\text{potential energy}~~:~V=-\frac{1}{4}\tr([\hat{X}_i^{(N)}, \hat{X}_j^{(N)}]^2) =\frac{2}{{c_3}^2}c_1 D(N, I)=2(I+2N)(I+2N+1)(I+2N+2), \\
&\text{potential energy density}~~:~\frac{V}{\hat{R}^2} =2D(N, I)  =2(I+2N+1) .
\end{align}
\end{subequations}
The potential energy density is simply the twice  the matrix size of the fuzzy two-sphere.

\subsection{Matrix model analysis}\label{sec:matimoso3}

\begin{figure}[tbph]
\center
\includegraphics*[width=140mm]{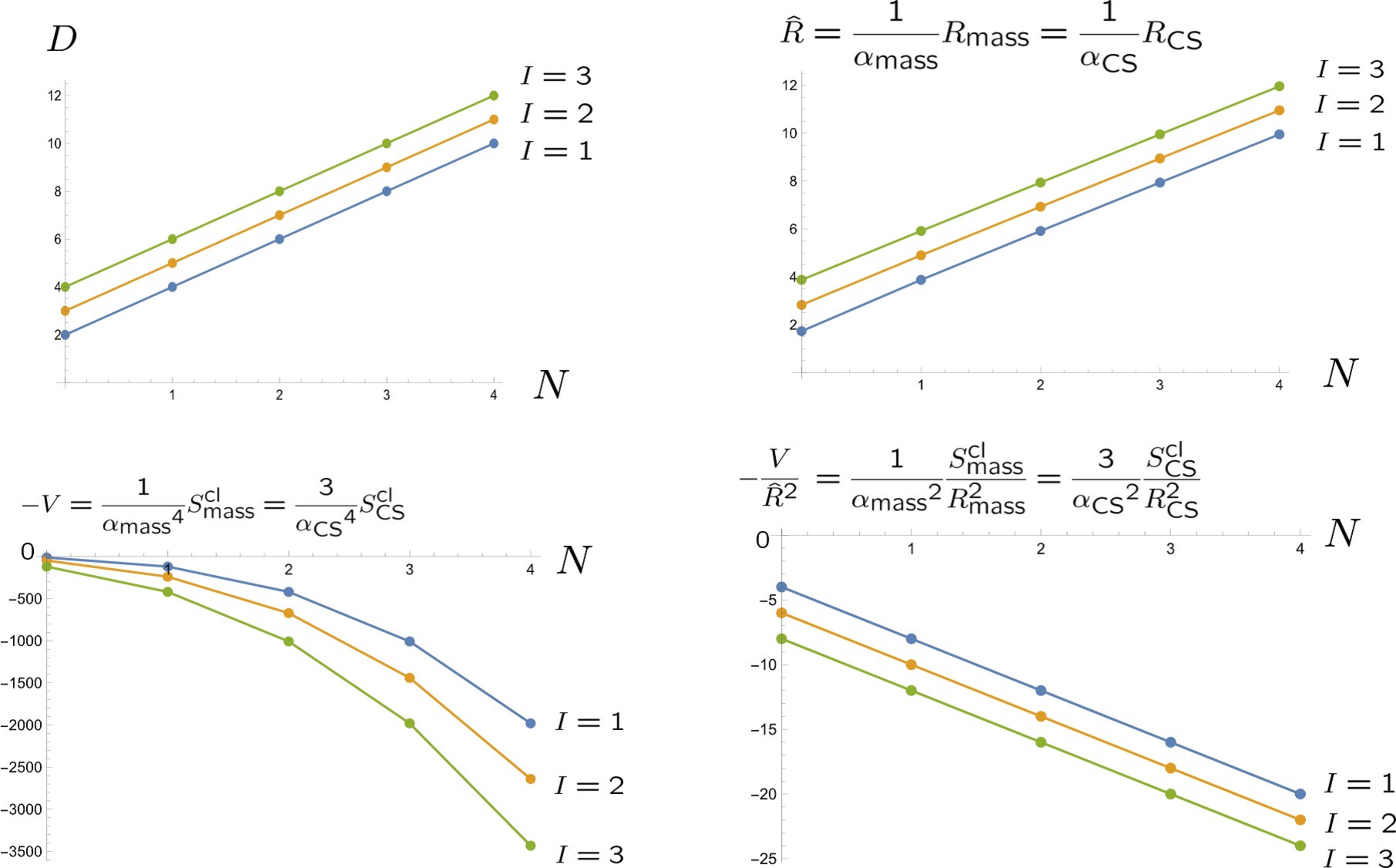}
\caption{Physical quantities of the fuzzy two-spheres. All  quantities monotonically increase or decrease as $N$ and $I$ increase.  }
\label{fs2m.fig}
\end{figure}

 Yang-Mills matrix models with a mass term and with a third-rank Chern-Simons term are given by 
\be
S_{\text{mass}} =-\frac{1}{4}\tr([A_i, A_j]^2) -\frac{1}{2}\tr({A_i}^2), ~~~~~~S_{\text{CS}}=-\frac{1}{4}\tr([A_i, A_j]^2) +i\frac{1}{3}\epsilon_{ijk}\tr(A_i A_j A_k), 
\ee
and the corresponding equations of motion are, respectively,    
\be
[[A_i, A_j], A_j] =A_i, ~~~~~[[A_i, A_j], A_j] =-i\epsilon_{ijk}A_j A_k,\label{3eommassive}
\ee
The fuzzy two-sphere is realized as a solution:  
\be
A_i^{\text{cl}} =\alpha_{\text{mass}}~\hat{X}_i^{(N)}, ~~~~~~A_i^{\text{cl}} =\alpha_{\text{CS}}~\hat{X}_i^{(N)}, \label{classmatso3ma}
\ee
with 
\be
\alpha_{\text{mass}}\equiv \frac{1}{2\sqrt{2}}, ~~~~~~~~\alpha_{\text{CS}}\equiv \frac{1}{4}.
\ee
Notice that both coefficients, $\alpha_{\text{mass}}$ and $\alpha_{\text{CS}}$, are constant unlike the case of the fuzzy four-sphere solutions (see Sec.\ref{sec:ymmatrixmodel}). 
The classical solutions 
(\ref{classmatso3ma}) have the following properties: 
\begin{subequations}
\begin{align}
&\text{radius}~~:~R_{\text{mass}} 
\equiv \alpha_{\text{mass}} \hat{R}~~~~~(A_i^{\text{cl}}  A_i^{\text{cl}}  ={R_{\text{mass}}}^2 \bs{1}),~~~~~~R_{\text{CS}} 
\equiv \alpha_{\text{CS}} \hat{R}~~~~(A_i^{\text{cl}}  A_i^{\text{cl}}  ={R_{\text{CS}}}^2 \bs{1}),\\
&\text{action}~~:~S_{\text{mass}}^{\text{cl}} =\overbrace{(-\frac{1}{4}+\frac{1}{2})}^{=1/4}\tr([A_i^{\text{cl}}, A_i^{\text{cl}}]^2) =-{\alpha_{\text{mass}}}^4 V,~~~~S_{\text{CS}}^{\text{cl}}\equiv S_{\text{CS}}[A_i^{\text{cl}}]=\overbrace{(-\frac{1}{4}+\frac{1}{3})}^{=1/12}\tr([A_i^{\text{cl}}, A_j^{\text{cl}}]^2) =- \frac{1}{3}{\alpha_{\text{CS}}}^4 V, \\
&\text{action density}~~:~\frac{S_{\text{mass}}^{\text{cl}}}{{R_{\text{mass}}}^2} =-{\alpha_{\text{mass}}}^2 \frac{V}{\hat{R}^2} =-\frac{1}{4 }D,~~~~~~~~\frac{S_{\text{CS}}^{\text{cl}}}{{R_{\text{CS}}}^2} =-\frac{1}{3}{\alpha_{\text{CS}}}^2 \frac{V}{\hat{R}^2} =-\frac{1}{24}D. 
\end{align}
\end{subequations}
See Fig.\ref{fs2m.fig} for the behaviors of these quantities. 

\section{Matrix coordinates and $SO(5)$ generators for $(N, I)=(1,1)$}\label{app:himat}

The following 16 $\times$ 16 matrices denote 
the matrix coordinates for $(N, I)=(1,1)$:  
\tiny
\begin{align}
\!\!\!\!\!\!\!\! &X^{[1]}_1=\frac{1}{70}
 \left(
\begin{array}{cc:cc:cccccc:cccccc}
0 & 0 & 0   & 9i & -\sqrt{10}i & 0 &  0          & \sqrt{5}i & 0 &   0       & 0 & 0 & 0 & 0 & 0 & 0 \\
0 & 0 & 9i  & 0  & 0           & 0 & -\sqrt{5}i & 0         & 0 & \sqrt{10}i & 0 & 0 & 0 & 0 & 0 & 0 \\
 \hdashline
 0 & -9i & 0 & 0&0 & 0 & 0 & 0 & 0 & 0 & \sqrt{10} i & 0 & 0 & 0 & -\sqrt{5} i & 0 \\
 -9i & 0 & 0 & 0&0 & 0 & 0 & 0 & 0 & 0 & 0 & \sqrt{5} i & 0 & 0 & 0 & -\sqrt{10}i  \\
 \hdashline
  \sqrt{10} i  & 0 & 0 & 0&0 & 0 & 0 & 0 & 0 & 0 & 0 & 5\sqrt{2}i & 0 & 0 & 0 & 0 \\
 0 & 0 & 0 & 0&0 & 0 & 0 & 0 & 0 & 0 & 0 & 0 & 10 i & 0 & 0 & 0 \\
  0 & \sqrt{5} i & 0 & 0&0 & 0 & 0 & 0 & 0 & 0 & 5\sqrt{2} i & 0 & 0 & 0 & 5i & 0 \\
 -\sqrt{5} i & 0 & 0 & 0&0 & 0 & 0 & 0 & 0 & 0 & 0 & 5i & 0 & 0 & 0 & 5\sqrt{2}i \\
  0 & 0 & 0 & 0&0 & 0 & 0 & 0 & 0 & 0 & 0 & 0 & 0 & 10 i & 0 & 0 \\
 0 & -\sqrt{10}i & 0 & 0&0 & 0 & 0 & 0 & 0 & 0 & 0 & 0 & 0 & 0 & 5\sqrt{2} i & 0 \\
 \hdashline
  0 & 0 & -\sqrt{10} i & 0&0 & 0 & -5\sqrt{2} i & 0 & 0 & 0 & 0 & 0 & 0 & 0 & 0 & 0 \\
 0 & 0 & 0 &  -\sqrt{5} i & -5\sqrt{2} i & 0 & 0 & -5i & 0 & 0 & 0 & 0 & 0 & 0 & 0 & 0 \\
  0 & 0 & 0 & 0&0 & -10i & 0 & 0 & 0 & 0 & 0 & 0 & 0 & 0 & 0 & 0 \\
 0 & 0 & 0 & 0&0 & 0 & 0 & 0 & -10 i & 0 & 0 & 0 & 0 & 0 & 0 & 0 \\
  0 & 0 & \sqrt{5} i & 0&0 & 0 & -5i  & 0 & 0 & -5\sqrt{2} i & 0 & 0 & 0 & 0 & 0 & 0 \\
 0 & 0 & 0 & \sqrt{10} i & 0 & 0 & 0 & -5\sqrt{2} i & 0 & 0 & 0 & 0 & 0 & 0 & 0 & 0 \\
\end{array}
\right),\nn\\
&X^{[1]}_2=\frac{1}{70}
 \left(
\begin{array}{cc:cc:cccccc:cccccc}
0 & 0 & 0   & 9 & \sqrt{10} & 0 &  0          & \sqrt{5} & 0 &   0       & 0 & 0 & 0 & 0 & 0 & 0 \\
0 & 0 & -9  & 0  & 0           & 0 & \sqrt{5} & 0         & 0 & \sqrt{10} & 0 & 0 & 0 & 0 & 0 & 0 \\
 \hdashline
 0 & -9 & 0 & 0&0 & 0 & 0 & 0 & 0 & 0 & -\sqrt{10}  & 0 & 0 & 0 & -\sqrt{5}  & 0 \\
 9 & 0 & 0 & 0&0 & 0 & 0 & 0 & 0 & 0 & 0 & -\sqrt{5}  & 0 & 0 & 0 & -\sqrt{10}  \\
 \hdashline
  \sqrt{10}   & 0 & 0 & 0&0 & 0 & 0 & 0 & 0 & 0 & 0 & 5\sqrt{2} & 0 & 0 & 0 & 0 \\
 0 & 0 & 0 & 0&0 & 0 & 0 & 0 & 0 & 0 & 0 & 0 & 10  & 0 & 0 & 0 \\
  0 & \sqrt{5}  & 0 & 0&0 & 0 & 0 & 0 & 0 & 0 & -5\sqrt{2}  & 0 & 0 & 0 & 5 & 0 \\
 \sqrt{5}  & 0 & 0 & 0&0 & 0 & 0 & 0 & 0 & 0 & 0 & -5 & 0 & 0 & 0 & 5\sqrt{2} \\
  0 & 0 & 0 & 0&0 & 0 & 0 & 0 & 0 & 0 & 0 & 0 & 0 & -10  & 0 & 0 \\
 0 & \sqrt{10} & 0 & 0&0 & 0 & 0 & 0 & 0 & 0 & 0 & 0 & 0 & 0 & -5\sqrt{2}  & 0 \\
 \hdashline
  0 & 0 & -\sqrt{10}  & 0&0 & 0 & -5\sqrt{2}  & 0 & 0 & 0 & 0 & 0 & 0 & 0 & 0 & 0 \\
 0 & 0 & 0 &  -\sqrt{5}  & 5\sqrt{2}  & 0 & 0 & -5 & 0 & 0 & 0 & 0 & 0 & 0 & 0 & 0 \\
  0 & 0 & 0 & 0&0 & 10 & 0 & 0 & 0 & 0 & 0 & 0 & 0 & 0 & 0 & 0 \\
 0 & 0 & 0 & 0&0 & 0 & 0 & 0 & -10  & 0 & 0 & 0 & 0 & 0 & 0 & 0 \\
  0 & 0 & -\sqrt{5}  & 0&0 & 0 & 5  & 0 & 0 & -5\sqrt{2}  & 0 & 0 & 0 & 0 & 0 & 0 \\
 0 & 0 & 0 & -\sqrt{10}  & 0 & 0 & 0 & 5\sqrt{2}  & 0 & 0 & 0 & 0 & 0 & 0 & 0 & 0 \\
\end{array}
\right),\nn\\
&X^{[1]}_3=\frac{1}{70}
 \left(
\begin{array}{cc:cc:cccccc:cccccc}
0 & 0 & 9i   & 0 & 0 & \sqrt{10} i &  \sqrt{5}i         & 0 & 0 &   0       & 0 & 0 & 0 & 0 & 0 & 0 \\
0 & 0 & 0  & -9i  & 0           & 0 & 0 & \sqrt{5}i        & \sqrt{10}i & 0 & 0 & 0 & 0 & 0 & 0 & 0 \\
 \hdashline
 -9i & 0 & 0 & 0&0 & 0 & 0 & 0 & 0 & 0 & 0 & -\sqrt{5}i  & 0 & -\sqrt{10}i & 0  & 0 \\
 0 & 9i & 0 & 0&0 & 0 & 0 & 0 & 0 & 0 & 0 & 0  & -\sqrt{10}i & 0 & -\sqrt{5}i & 0 \\
 \hdashline
 0   & 0 & 0 & 0&0 & 0 & 0 & 0 & 0 & 0 &   {10}i & 0 & 0 & 0 & 0 & 0 \\
 -\sqrt{10}i & 0 & 0 & 0&0 & 0 & 0 & 0 & 0 & 0 & 0 & 5\sqrt{2} i & 0  & 0 & 0 & 0 \\
  -\sqrt{5}i & 0  & 0 & 0 &0 & 0 & 0 & 0 & 0 & 0 & 0  & -5i & 0 & 5\sqrt{2}i & 0 & 0 \\
0  & - \sqrt{5} i & 0 & 0&0 & 0 & 0 & 0 & 0 & 0 & 0 & 0 &  -5\sqrt{2}i & 0 &  5i & 0 \\
  0 &  - \sqrt{10} i & 0 & 0&0 & 0 & 0 & 0 & 0 & 0 & 0 & 0 & 0 & 0  &  -5\sqrt{2}i & 0 \\
 0 & 0 & 0 & 0&0 & 0 & 0 & 0 & 0 & 0 & 0 & 0 & 0 & 0 &0  & -10i \\
 \hdashline
  0 & 0 & 0  & 0& -10i & 0 & 0  & 0 & 0 & 0 & 0 & 0 & 0 & 0 & 0 & 0 \\
 0 & 0 &   \sqrt{5}i & 0  & 0 & -5\sqrt{2}i   & 5i & 0 & 0 & 0 & 0 & 0 & 0 & 0 & 0 & 0 \\
  0 & 0 & 0 & \sqrt{10}i &0 & 0 & 0 & 5\sqrt{2}i & 0 & 0 & 0 & 0 & 0 & 0 & 0 & 0 \\
 0 & 0 & \sqrt{10}i & 0&0 & 0 & -5\sqrt{2}i & 0 & 0  & 0 & 0 & 0 & 0 & 0 & 0 & 0 \\
  0 & 0 & 0  & \sqrt{5}i & 0 &  0  & 0 & -5i & 5\sqrt{2}i & 0  & 0 & 0 & 0 & 0 & 0 & 0 \\
 0 & 0 & 0 & 0  & 0 & 0 & 0 & 0  & 0 & 10i & 0 & 0 & 0 & 0 & 0 & 0 \\
\end{array}
\right), \nn\\
&X^{[1]}_4=\frac{1}{70}
 \left(
\begin{array}{cc:cc:cccccc:cccccc}
0 & 0 & 9   & 0 & 0 & -\sqrt{10}  &  \sqrt{5}         & 0 & 0 &   0       & 0 & 0 & 0 & 0 & 0 & 0 \\
0 & 0 & 0  & 9  & 0           & 0 & 0 & -\sqrt{5}        & \sqrt{10} & 0 & 0 & 0 & 0 & 0 & 0 & 0 \\
 \hdashline
 9 & 0 & 0 & 0&0 & 0 & 0 & 0 & 0 & 0 & 0 & \sqrt{5}  & 0 & -\sqrt{10} & 0  & 0 \\
 0 & 9 & 0 & 0&0 & 0 & 0 & 0 & 0 & 0 & 0 & 0  & \sqrt{10} & 0 & -\sqrt{5} & 0 \\
 \hdashline
 0   & 0 & 0 & 0&0 & 0 & 0 & 0 & 0 & 0 &   {10} & 0 & 0 & 0 & 0 & 0 \\
 -\sqrt{10} & 0 & 0 & 0&0 & 0 & 0 & 0 & 0 & 0 & 0 & 5\sqrt{2}  & 0  & 0 & 0 & 0 \\
  \sqrt{5} & 0  & 0 & 0 &0 & 0 & 0 & 0 & 0 & 0 & 0  & 5 & 0 & 5\sqrt{2} & 0 & 0 \\
0  & - \sqrt{5}  & 0 & 0&0 & 0 & 0 & 0 & 0 & 0 & 0 & 0 &  5\sqrt{2} & 0 &  5 & 0 \\
  0 &  \sqrt{10}  & 0 & 0&0 & 0 & 0 & 0 & 0 & 0 & 0 & 0 & 0 & 0  &  5\sqrt{2} & 0 \\
 0 & 0 & 0 & 0&0 & 0 & 0 & 0 & 0 & 0 & 0 & 0 & 0 & 0 &0  & 10 \\
 \hdashline
  0 & 0 & 0  & 0& 10 & 0 & 0  & 0 & 0 & 0 & 0 & 0 & 0 & 0 & 0 & 0 \\
 0 & 0 &   \sqrt{5} & 0  & 0 & 5\sqrt{2}   & 5 & 0 & 0 & 0 & 0 & 0 & 0 & 0 & 0 & 0 \\
  0 & 0 & 0 & \sqrt{10} &0 & 0 & 0 & 5\sqrt{2} & 0 & 0 & 0 & 0 & 0 & 0 & 0 & 0 \\
 0 & 0 & -\sqrt{10} & 0&0 & 0 & 5\sqrt{2} & 0 & 0  & 0 & 0 & 0 & 0 & 0 & 0 & 0 \\
  0 & 0 & 0  & -\sqrt{5} & 0 &  0  & 0 & 5 & 5\sqrt{2} & 0  & 0 & 0 & 0 & 0 & 0 & 0 \\
 0 & 0 & 0 & 0  & 0 & 0 & 0 & 0  & 0 & 10 & 0 & 0 & 0 & 0 & 0 & 0 \\
\end{array}
\right),\nn\\
&X^{[1]}_5=\frac{1}{35}
 \left(
\begin{array}{cc:cc:cccccc:cccccc}
3 & 0 & 0   & 0 & 0 & 0  &  0        & 0 & 0 &   0       & 0 & 0 & 0 & 0 & 0 & 0 \\
0 & 3 & 0  & 0  & 0           & 0 & 0 & 0       & 0 & 0 & 0 & 0 & 0 & 0 & 0 & 0 \\
 \hdashline
 0 & 0 & -3 & 0&0 & 0 & 0 & 0 & 0 & 0 & 0 & 0 & 0 & 0 & 0  & 0 \\
 0 & 0 & 0 & -3&0 & 0 & 0 & 0 & 0 & 0 & 0 & 0  & 0 & 0 & 0 & 0 \\
 \hdashline
 0   & 0 & 0 & 0&5 & 0 & 0 & 0 & 0 & 0 &   0 & 0 & 0 & 0 & 0 & 0 \\
 0 & 0 & 0 & 0&0 &  5 & 0 & 0 & 0 & 0 & 0 & 0  & 0  & 0 & 0 & 0 \\
  0 & 0  & 0 & 0 &0 & 0 & 5  & 0 & 0 & 0 & 0  & 0 & 0 & 0 & 0 & 0 \\
0  & 0  & 0 & 0&0 & 0 & 0 &  5 & 0 & 0 & 0 & 0 & 0 & 0 &  0 & 0 \\
  0 &  0  & 0 & 0&0 & 0 & 0 & 0 & 5 & 0 & 0 & 0 & 0 & 0  &  0 & 0 \\
 0 & 0 & 0 & 0&0 & 0 & 0 & 0 & 0 & 5 & 0 & 0 & 0 & 0 &0  & 0\\
 \hdashline
  0 & 0 & 0  & 0& 0 & 0 & 0  & 0 & 0 & 0 & -5 & 0 & 0 & 0 & 0 & 0 \\
 0 & 0 &   0 & 0  & 0 & 0  & 0 & 0 & 0 & 0 & 0 & -5 & 0 & 0 & 0 & 0 \\
  0 & 0 & 0 & 0 &0 & 0 & 0 & 0 & 0 & 0 & 0 & 0 & -5 & 0 & 0 & 0 \\
 0 & 0 & 0 & 0&0 & 0 & 0 & 0 & 0  & 0 & 0 & 0 & 0 & -5 & 0 & 0 \\
  0 & 0 & 0  & 0 & 0 &  0  & 0 & 0 & 0 & 0  & 0 & 0 & 0 & 0 & -5 & 0 \\
 0 & 0 & 0 & 0  & 0 & 0 & 0 & 0  & 0 & 0 & 0 & 0 & 0 & 0 & 0 & -5 \\
\end{array}
\right).
\end{align}
\normalsize
The $SO(5)$ generators $(N, I)=(1,1)$ are 
\tiny
\begin{align}
\!\!\!\!\!\!\!\! 
&\Sigma^{[1]}_{12}=\frac{1}{2}
 \left(
\begin{array}{cc:cc:cccccc:cccccc}
1 & 0 & 0   & 0 & 0 & 0  &  0        & 0 & 0 &   0       & 0 & 0 & 0 & 0 & 0 & 0 \\
0 & -1 & 0  & 0  & 0           & 0 & 0 & 0       & 0 & 0 & 0 & 0 & 0 & 0 & 0 & 0 \\
 \hdashline
 0 & 0 & 1 & 0&0 & 0 & 0 & 0 & 0 & 0 & 0 & 0 & 0 & 0 & 0  & 0 \\
 0 & 0 & 0 & -1&0 & 0 & 0 & 0 & 0 & 0 & 0 & 0  & 0 & 0 & 0 & 0 \\
 \hdashline
 0   & 0 & 0 & 0& 3 & 0 & 0 & 0 & 0 & 0 &   0 & 0 & 0 & 0 & 0 & 0 \\
 0 & 0 & 0 & 0&0 &  1 & 0 & 0 & 0 & 0 & 0 & 0  & 0  & 0 & 0 & 0 \\
  0 & 0  & 0 & 0 &0 & 0 & 1  & 0 & 0 & 0 & 0  & 0 & 0 & 0 & 0 & 0 \\
0  & 0  & 0 & 0&0 & 0 & 0 &  -1 & 0 & 0 & 0 & 0 & 0 & 0 &  0 & 0 \\
  0 &  0  & 0 & 0&0 & 0 & 0 & 0 & -1 & 0 & 0 & 0 & 0 & 0  &  0 & 0 \\
 0 & 0 & 0 & 0&0 & 0 & 0 & 0 & 0 & -3 & 0 & 0 & 0 & 0 &0  & 0\\
 \hdashline
  0 & 0 & 0  & 0& 0 & 0 & 0  & 0 & 0 & 0 & 3 & 0 & 0 & 0 & 0 & 0 \\
 0 & 0 &   0 & 0  & 0 & 0  & 0 & 0 & 0 & 0 & 0 & 1 & 0 & 0 & 0 & 0 \\
  0 & 0 & 0 & 0 &0 & 0 & 0 & 0 & 0 & 0 & 0 & 0 & -1 & 0 & 0 & 0 \\
 0 & 0 & 0 & 0&0 & 0 & 0 & 0 & 0  & 0 & 0 & 0 & 0 & 1 & 0 & 0 \\
  0 & 0 & 0  & 0 & 0 &  0  & 0 & 0 & 0 & 0  & 0 & 0 & 0 & 0 & -1 & 0 \\
 0 & 0 & 0 & 0  & 0 & 0 & 0 & 0  & 0 & 0 & 0 & 0 & 0 & 0 & 0 & -3 \\
\end{array}
\right),\nn\\
&\Sigma^{[1]}_{13}=\frac{1}{2}
 \left(
\begin{array}{cc:cc:cccccc:cccccc}
0 & i & 0   & 0 & 0 & 0  &  0        & 0 & 0 &   0       & 0 & 0 & 0 & 0 & 0 & 0 \\
-i & 0 & 0  & 0  & 0           & 0 & 0 & 0       & 0 & 0 & 0 & 0 & 0 & 0 & 0 & 0 \\
 \hdashline
 0 & 0 & 0 & i &0 & 0 & 0 & 0 & 0 & 0 & 0 & 0 & 0 & 0 & 0  & 0 \\
 0 & 0 & -i & 0 &0 & 0 & 0 & 0 & 0 & 0 & 0 & 0  & 0 & 0 & 0 & 0 \\
 \hdashline
 0   & 0 & 0 & 0& 0 & i & \sqrt{2} i & 0 & 0 & 0 &   0 & 0 & 0 & 0 & 0 & 0 \\
 0 & 0 & 0 & 0& -i  &  0 & 0 & \sqrt{2}i & 0 & 0 & 0 & 0  & 0  & 0 & 0 & 0 \\
  0 & 0  & 0 & 0 & -\sqrt{2}i & 0 & 0  & i & \sqrt{2}i & 0 & 0  & 0 & 0 & 0 & 0 & 0 \\
0  & 0  & 0 & 0&0 &  -\sqrt{2}i & -i &  0& 0 & \sqrt{2}i & 0 & 0 & 0 & 0 &  0 & 0 \\
  0 &  0  & 0 & 0&0 & 0 & -\sqrt{2}i & 0 & 0 & i & 0 & 0 & 0 & 0  &  0 & 0 \\
 0 & 0 & 0 & 0&0 & 0 & 0 & -\sqrt{2}i & -i & 0 & 0 & 0 & 0 & 0 &0  & 0\\
 \hdashline
  0 & 0 & 0  & 0& 0 & 0 & 0  & 0 & 0 & 0 & 0 & \sqrt{2} i  & 0 &  i  & 0 & 0 \\
 0 & 0 &   0 & 0  & 0 & 0  & 0 & 0 & 0 & 0 & -\sqrt{2} i  & 0 & \sqrt{2} i  & 0 & i & 0 \\
  0 & 0 & 0 & 0 &0 & 0 & 0 & 0 & 0 & 0 & 0 & -\sqrt{2} i  & 0 & 0 & 0 & i \\
 0 & 0 & 0 & 0&0 & 0 & 0 & 0 & 0  & 0 & -i  & 0 & 0 & 0 & \sqrt{2} i  & 0 \\
  0 & 0 & 0  & 0 & 0 &  0  & 0 & 0 & 0 & 0  & 0 & -i  & 0 & -\sqrt{2} i  & 0 & \sqrt{2} i  \\
 0 & 0 & 0 & 0  & 0 & 0 & 0 & 0  & 0 & 0 & 0 & 0 & - i  & 0 & -\sqrt{2} i  & 0 \\
\end{array}
\right),\nn\\
&\Sigma^{[1]}_{14}=\frac{1}{2}
 \left(
\begin{array}{cc:cc:cccccc:cccccc}
0 & 1 & 0   & 0 & 0 & 0  &  0        & 0 & 0 &   0       & 0 & 0 & 0 & 0 & 0 & 0 \\
1 & 0 & 0  & 0  & 0           & 0 & 0 & 0       & 0 & 0 & 0 & 0 & 0 & 0 & 0 & 0 \\
 \hdashline
 0 & 0 & 0 & -1 &0 & 0 & 0 & 0 & 0 & 0 & 0 & 0 & 0 & 0 & 0  & 0 \\
 0 & 0 & -1 & 0 &0 & 0 & 0 & 0 & 0 & 0 & 0 & 0  & 0 & 0 & 0 & 0 \\
 \hdashline
 0   & 0 & 0 & 0& 0 & -1 & \sqrt{2}  & 0 & 0 & 0 &   0 & 0 & 0 & 0 & 0 & 0 \\
 0 & 0 & 0 & 0& -1  &  0 & 0 & \sqrt{2} & 0 & 0 & 0 & 0  & 0  & 0 & 0 & 0 \\
  0 & 0  & 0 & 0 & \sqrt{2} & 0 & 0  & -1 & \sqrt{2} & 0 & 0  & 0 & 0 & 0 & 0 & 0 \\
0  & 0  & 0 & 0&0 &  \sqrt{2} & -1 &  0& 0 & \sqrt{2} & 0 & 0 & 0 & 0 &  0 & 0 \\
  0 &  0  & 0 & 0&0 & 0 & \sqrt{2} & 0 & 0 & -1 & 0 & 0 & 0 & 0  &  0 & 0 \\
 0 & 0 & 0 & 0&0 & 0 & 0 & \sqrt{2} & -1 & 0 & 0 & 0 & 0 & 0 &0  & 0\\
 \hdashline
  0 & 0 & 0  & 0& 0 & 0 & 0  & 0 & 0 & 0 & 0 & -\sqrt{2}   & 0 &  1  & 0 & 0 \\
 0 & 0 &   0 & 0  & 0 & 0  & 0 & 0 & 0 & 0 & -\sqrt{2}   & 0 & -\sqrt{2}   & 0 & 1 & 0 \\
  0 & 0 & 0 & 0 &0 & 0 & 0 & 0 & 0 & 0 & 0 & -\sqrt{2}   & 0 & 0 & 0 & 1 \\
 0 & 0 & 0 & 0&0 & 0 & 0 & 0 & 0  & 0 & 1  & 0 & 0 & 0 & -\sqrt{2}   & 0 \\
  0 & 0 & 0  & 0 & 0 &  0  & 0 & 0 & 0 & 0  & 0 & 1  & 0 & -\sqrt{2}   & 0 & -\sqrt{2}   \\
 0 & 0 & 0 & 0  & 0 & 0 & 0 & 0  & 0 & 0 & 0 & 0 & 1  & 0 & -\sqrt{2}   & 0 \\
\end{array}
\right),\nn\\
&\Sigma^{[1]}_{23}=\frac{1}{2}
 \left(
\begin{array}{cc:cc:cccccc:cccccc}
0 & 1 & 0   & 0 & 0 & 0  &  0        & 0 & 0 &   0       & 0 & 0 & 0 & 0 & 0 & 0 \\
1 & 0 & 0  & 0  & 0           & 0 & 0 & 0       & 0 & 0 & 0 & 0 & 0 & 0 & 0 & 0 \\
 \hdashline
 0 & 0 & 0 & 1 &0 & 0 & 0 & 0 & 0 & 0 & 0 & 0 & 0 & 0 & 0  & 0 \\
 0 & 0 & 1 & 0 &0 & 0 & 0 & 0 & 0 & 0 & 0 & 0  & 0 & 0 & 0 & 0 \\
 \hdashline
 0   & 0 & 0 & 0& 0 & 1 & \sqrt{2}  & 0 & 0 & 0 &   0 & 0 & 0 & 0 & 0 & 0 \\
 0 & 0 & 0 & 0& 1  &  0 & 0 & \sqrt{2} & 0 & 0 & 0 & 0  & 0  & 0 & 0 & 0 \\
  0 & 0  & 0 & 0 & \sqrt{2} & 0 & 0  & 1 & \sqrt{2} & 0 & 0  & 0 & 0 & 0 & 0 & 0 \\
0  & 0  & 0 & 0&0 &  \sqrt{2} & 1 &  0& 0 & \sqrt{2} & 0 & 0 & 0 & 0 &  0 & 0 \\
  0 &  0  & 0 & 0&0 & 0 & \sqrt{2} & 0 & 0 & 1 & 0 & 0 & 0 & 0  &  0 & 0 \\
 0 & 0 & 0 & 0&0 & 0 & 0 & \sqrt{2} & 1 & 0 & 0 & 0 & 0 & 0 &0  & 0\\
 \hdashline
  0 & 0 & 0  & 0& 0 & 0 & 0  & 0 & 0 & 0 & 0 & \sqrt{2}   & 0 &  1  & 0 & 0 \\
 0 & 0 &   0 & 0  & 0 & 0  & 0 & 0 & 0 & 0 & \sqrt{2}   & 0 & \sqrt{2}   & 0 & 1 & 0 \\
  0 & 0 & 0 & 0 &0 & 0 & 0 & 0 & 0 & 0 & 0 & \sqrt{2}   & 0 & 0 & 0 & 1 \\
 0 & 0 & 0 & 0&0 & 0 & 0 & 0 & 0  & 0 & 1  & 0 & 0 & 0 & \sqrt{2}   & 0 \\
  0 & 0 & 0  & 0 & 0 &  0  & 0 & 0 & 0 & 0  & 0 & 1  & 0 & \sqrt{2}   & 0 & \sqrt{2}   \\
 0 & 0 & 0 & 0  & 0 & 0 & 0 & 0  & 0 & 0 & 0 & 0 & 1  & 0 & \sqrt{2}   & 0 \\
\end{array}
\right),\nn\\
&\Sigma^{[1]}_{24}=\frac{1}{2}
 \left(
\begin{array}{cc:cc:cccccc:cccccc}
0 & -i & 0   & 0 & 0 & 0  &  0        & 0 & 0 &   0       & 0 & 0 & 0 & 0 & 0 & 0 \\
i & 0 & 0  & 0  & 0           & 0 & 0 & 0       & 0 & 0 & 0 & 0 & 0 & 0 & 0 & 0 \\
 \hdashline
 0 & 0 & 0 & -i &0 & 0 & 0 & 0 & 0 & 0 & 0 & 0 & 0 & 0 & 0  & 0 \\
 0 & 0 & i & 0 &0 & 0 & 0 & 0 & 0 & 0 & 0 & 0  & 0 & 0 & 0 & 0 \\
 \hdashline
 0   & 0 & 0 & 0& 0 & i & -\sqrt{2} i & 0 & 0 & 0 &   0 & 0 & 0 & 0 & 0 & 0 \\
 0 & 0 & 0 & 0& -i  &  0 & 0 & -\sqrt{2}i & 0 & 0 & 0 & 0  & 0  & 0 & 0 & 0 \\
  0 & 0  & 0 & 0 & \sqrt{2}i & 0 & 0  & i & -\sqrt{2}i & 0 & 0  & 0 & 0 & 0 & 0 & 0 \\
0  & 0  & 0 & 0&0 &  \sqrt{2}i & -i &  0& 0 & -\sqrt{2}i & 0 & 0 & 0 & 0 &  0 & 0 \\
  0 &  0  & 0 & 0&0 & 0 & \sqrt{2}i & 0 & 0 & i & 0 & 0 & 0 & 0  &  0 & 0 \\
 0 & 0 & 0 & 0&0 & 0 & 0 & \sqrt{2}i & -i & 0 & 0 & 0 & 0 & 0 &0  & 0\\
 \hdashline
  0 & 0 & 0  & 0& 0 & 0 & 0  & 0 & 0 & 0 & 0 & \sqrt{2} i  & 0 &  -i  & 0 & 0 \\
 0 & 0 &   0 & 0  & 0 & 0  & 0 & 0 & 0 & 0 & -\sqrt{2} i  & 0 & \sqrt{2} i  & 0 & -i & 0 \\
  0 & 0 & 0 & 0 &0 & 0 & 0 & 0 & 0 & 0 & 0 & -\sqrt{2} i  & 0 & 0 & 0 & -i \\
 0 & 0 & 0 & 0&0 & 0 & 0 & 0 & 0  & 0 & i  & 0 & 0 & 0 & \sqrt{2} i  & 0 \\
  0 & 0 & 0  & 0 & 0 &  0  & 0 & 0 & 0 & 0  & 0 & i  & 0 & -\sqrt{2} i  & 0 & \sqrt{2} i  \\
 0 & 0 & 0 & 0  & 0 & 0 & 0 & 0  & 0 & 0 & 0 & 0 & i  & 0 & -\sqrt{2} i  & 0 \\
\end{array}
\right),
\end{align}
\normalsize
and 
\tiny
\begin{align}
\!\!\!\!\!\!\!\! 
&\Sigma^{[1]}_{34}=\frac{1}{2}
 \left(
\begin{array}{cc:cc:cccccc:cccccc}
1 & 0 & 0   & 0 & 0 & 0  &  0        & 0 & 0 &   0       & 0 & 0 & 0 & 0 & 0 & 0 \\
0 & -1 & 0  & 0  & 0           & 0 & 0 & 0       & 0 & 0 & 0 & 0 & 0 & 0 & 0 & 0 \\
 \hdashline
 0 & 0 & -1 & 0&0 & 0 & 0 & 0 & 0 & 0 & 0 & 0 & 0 & 0 & 0  & 0 \\
 0 & 0 & 0 & 1&0 & 0 & 0 & 0 & 0 & 0 & 0 & 0  & 0 & 0 & 0 & 0 \\
 \hdashline
 0   & 0 & 0 & 0& 1 & 0 & 0 & 0 & 0 & 0 &   0 & 0 & 0 & 0 & 0 & 0 \\
 0 & 0 & 0 & 0&0 &  3 & 0 & 0 & 0 & 0 & 0 & 0  & 0  & 0 & 0 & 0 \\
  0 & 0  & 0 & 0 &0 & 0 & -1  & 0 & 0 & 0 & 0  & 0 & 0 & 0 & 0 & 0 \\
0  & 0  & 0 & 0&0 & 0 & 0 &  1 & 0 & 0 & 0 & 0 & 0 & 0 &  0 & 0 \\
  0 &  0  & 0 & 0&0 & 0 & 0 & 0 & -3 & 0 & 0 & 0 & 0 & 0  &  0 & 0 \\
 0 & 0 & 0 & 0&0 & 0 & 0 & 0 & 0 & -1 & 0 & 0 & 0 & 0 &0  & 0\\
 \hdashline
  0 & 0 & 0  & 0& 0 & 0 & 0  & 0 & 0 & 0 & -1 & 0 & 0 & 0 & 0 & 0 \\
 0 & 0 &   0 & 0  & 0 & 0  & 0 & 0 & 0 & 0 & 0 & 1 & 0 & 0 & 0 & 0 \\
  0 & 0 & 0 & 0 &0 & 0 & 0 & 0 & 0 & 0 & 0 & 0 & 3 & 0 & 0 & 0 \\
 0 & 0 & 0 & 0&0 & 0 & 0 & 0 & 0  & 0 & 0 & 0 & 0 & -3 & 0 & 0 \\
  0 & 0 & 0  & 0 & 0 &  0  & 0 & 0 & 0 & 0  & 0 & 0 & 0 & 0 & -1 & 0 \\
 0 & 0 & 0 & 0  & 0 & 0 & 0 & 0  & 0 & 0 & 0 & 0 & 0 & 0 & 0 & 1 \\
\end{array}
\right),\nn\\
&\Sigma^{[1]}_{15}=\frac{1}{4}
 \left(
\begin{array}{cc:cc:cccccc:cccccc}
0 & 0 & 0   & 3 & \sqrt{10} & 0 &  0          & -\sqrt{5} & 0 &   0       & 0 & 0 & 0 & 0 & 0 & 0 \\
0 & 0 & 3  & 0  & 0           & 0 & \sqrt{5} & 0         & 0 & -\sqrt{10} & 0 & 0 & 0 & 0 & 0 & 0 \\
 \hdashline
 0 & 3 & 0 & 0&0 & 0 & 0 & 0 & 0 & 0 & \sqrt{10}  & 0 & 0 & 0 & -\sqrt{5}  & 0 \\
 3 & 0 & 0 & 0&0 & 0 & 0 & 0 & 0 & 0 & 0 & \sqrt{5}  & 0 & 0 & 0 & -\sqrt{10}  \\
 \hdashline
  \sqrt{10}   & 0 & 0 & 0&0 & 0 & 0 & 0 & 0 & 0 & 0 & \sqrt{2} & 0 & 0 & 0 & 0 \\
 0 & 0 & 0 & 0&0 & 0 & 0 & 0 & 0 & 0 & 0 & 0 & 2  & 0 & 0 & 0 \\
  0 & \sqrt{5}  & 0 & 0&0 & 0 & 0 & 0 & 0 & 0 & \sqrt{2}  & 0 & 0 & 0 & 1 & 0 \\
 -\sqrt{5}  & 0 & 0 & 0&0 & 0 & 0 & 0 & 0 & 0 & 0 & 1 & 0 & 0 & 0 & \sqrt{2} \\
  0 & 0 & 0 & 0&0 & 0 & 0 & 0 & 0 & 0 & 0 & 0 & 0 & 2  & 0 & 0 \\
 0 & -\sqrt{10} & 0 & 0&0 & 0 & 0 & 0 & 0 & 0 & 0 & 0 & 0 & 0 & \sqrt{2}  & 0 \\
 \hdashline
  0 & 0 & \sqrt{10}  & 0&0 & 0 & \sqrt{2}  & 0 & 0 & 0 & 0 & 0 & 0 & 0 & 0 & 0 \\
 0 & 0 & 0 & \sqrt{5}  & \sqrt{2}  & 0 & 0 & 1 & 0 & 0 & 0 & 0 & 0 & 0 & 0 & 0 \\
  0 & 0 & 0 & 0&0 & 2 & 0 & 0 & 0 & 0 & 0 & 0 & 0 & 0 & 0 & 0 \\
 0 & 0 & 0 & 0&0 & 0 & 0 & 0 & 2  & 0 & 0 & 0 & 0 & 0 & 0 & 0 \\
  0 & 0 & -\sqrt{5}  & 0&0 & 0 & 1  & 0 & 0 & \sqrt{2}  & 0 & 0 & 0 & 0 & 0 & 0 \\
 0 & 0 & 0 & -\sqrt{10}  & 0 & 0 & 0 & \sqrt{2}  & 0 & 0 & 0 & 0 & 0 & 0 & 0 & 0 \\
\end{array}
\right),\nn\\
&\Sigma^{[1]}_{25}=\frac{1}{4}
 \left(
\begin{array}{cc:cc:cccccc:cccccc}
0 & 0 & 0   & -3i & \sqrt{10}i & 0 &  0          & \sqrt{5}i & 0 &   0       & 0 & 0 & 0 & 0 & 0 & 0 \\
0 & 0 & 3i  & 0  & 0           & 0 & \sqrt{5}i & 0         & 0 & \sqrt{10}i & 0 & 0 & 0 & 0 & 0 & 0 \\
 \hdashline
 0 & -3i & 0 & 0&0 & 0 & 0 & 0 & 0 & 0 & \sqrt{10} i & 0 & 0 & 0 & \sqrt{5}i  & 0 \\
 3i & 0 & 0 & 0&0 & 0 & 0 & 0 & 0 & 0 & 0 & \sqrt{5}i  & 0 & 0 & 0 & \sqrt{10}i  \\
 \hdashline
  -\sqrt{10}i   & 0 & 0 & 0&0 & 0 & 0 & 0 & 0 & 0 & 0 & -\sqrt{2}i & 0 & 0 & 0 & 0 \\
 0 & 0 & 0 & 0&0 & 0 & 0 & 0 & 0 & 0 & 0 & 0 & -2i  & 0 & 0 & 0 \\
  0 & -\sqrt{5}i  & 0 & 0&0 & 0 & 0 & 0 & 0 & 0 & \sqrt{2}i  & 0 & 0 & 0 & -i & 0 \\
 -\sqrt{5}i  & 0 & 0 & 0&0 & 0 & 0 & 0 & 0 & 0 & 0 & i & 0 & 0 & 0 & -\sqrt{2}i \\
  0 & 0 & 0 & 0&0 & 0 & 0 & 0 & 0 & 0 & 0 & 0 & 0 & 2i  & 0 & 0 \\
 0 & -\sqrt{10}i & 0 & 0&0 & 0 & 0 & 0 & 0 & 0 & 0 & 0 & 0 & 0 & \sqrt{2} i & 0 \\
 \hdashline
  0 & 0 & -\sqrt{10}i  & 0&0 & 0 & -\sqrt{2}i  & 0 & 0 & 0 & 0 & 0 & 0 & 0 & 0 & 0 \\
 0 & 0 & 0 & -\sqrt{5}i  & \sqrt{2}i  & 0 & 0 & -i & 0 & 0 & 0 & 0 & 0 & 0 & 0 & 0 \\
  0 & 0 & 0 & 0&0 & 2i & 0 & 0 & 0 & 0 & 0 & 0 & 0 & 0 & 0 & 0 \\
 0 & 0 & 0 & 0&0 & 0 & 0 & 0 & -2i  & 0 & 0 & 0 & 0 & 0 & 0 & 0 \\
  0 & 0 & -\sqrt{5}i  & 0&0 & 0 & i  & 0 & 0 & -\sqrt{2}i  & 0 & 0 & 0 & 0 & 0 & 0 \\
 0 & 0 & 0 & -\sqrt{10}i  & 0 & 0 & 0 & \sqrt{2} i & 0 & 0 & 0 & 0 & 0 & 0 & 0 & 0 \\
\end{array}
\right),\nn\\
&\Sigma^{[1]}_{35}=\frac{1}{4}
 \left(
\begin{array}{cc:cc:cccccc:cccccc}
0 & 0 & 3  & 0 & 0 & -\sqrt{10}  &  -\sqrt{5}         & 0 & 0 &   0       & 0 & 0 & 0 & 0 & 0 & 0 \\
0 & 0 & 0  & -3  & 0           & 0 & 0 & -\sqrt{5}        & -\sqrt{10} & 0 & 0 & 0 & 0 & 0 & 0 & 0 \\
 \hdashline
 3 & 0 & 0 & 0&0 & 0 & 0 & 0 & 0 & 0 & 0 & -\sqrt{5}  & 0 & -\sqrt{10} & 0  & 0 \\
 0 & -3 & 0 & 0&0 & 0 & 0 & 0 & 0 & 0 & 0 & 0  & -\sqrt{10} & 0 & -\sqrt{5} & 0 \\
 \hdashline
 0   & 0 & 0 & 0&0 & 0 & 0 & 0 & 0 & 0 &   2 & 0 & 0 & 0 & 0 & 0 \\
 -\sqrt{10} & 0 & 0 & 0&0 & 0 & 0 & 0 & 0 & 0 & 0 & \sqrt{2}  & 0  & 0 & 0 & 0 \\
  -\sqrt{5} & 0  & 0 & 0 &0 & 0 & 0 & 0 & 0 & 0 & 0  & -1 & 0 & \sqrt{2} & 0 & 0 \\
0  & - \sqrt{5}  & 0 & 0&0 & 0 & 0 & 0 & 0 & 0 & 0 & 0 &  -\sqrt{2} & 0 &  1 & 0 \\
  0 &  - \sqrt{10}  & 0 & 0&0 & 0 & 0 & 0 & 0 & 0 & 0 & 0 & 0 & 0  &  -\sqrt{2} & 0 \\
 0 & 0 & 0 & 0&0 & 0 & 0 & 0 & 0 & 0 & 0 & 0 & 0 & 0 &0  & -2 \\
 \hdashline
  0 & 0 & 0  & 0& 2 & 0 & 0  & 0 & 0 & 0 & 0 & 0 & 0 & 0 & 0 & 0 \\
 0 & 0 &   -\sqrt{5} & 0  & 0 & \sqrt{2} & -1 & 0 & 0 & 0 & 0 & 0 & 0 & 0 & 0 & 0 \\
  0 & 0 & 0 & -\sqrt{10} &0 & 0 & 0 & -\sqrt{2} & 0 & 0 & 0 & 0 & 0 & 0 & 0 & 0 \\
 0 & 0 & -\sqrt{10} & 0&0 & 0 & \sqrt{2} & 0 & 0  & 0 & 0 & 0 & 0 & 0 & 0 & 0 \\
  0 & 0 & 0  & -\sqrt{5} & 0 &  0  & 0 & 1 & -\sqrt{2} & 0  & 0 & 0 & 0 & 0 & 0 & 0 \\
 0 & 0 & 0 & 0  & 0 & 0 & 0 & 0  & 0 & -2 & 0 & 0 & 0 & 0 & 0 & 0 \\
\end{array}
\right), \nn\\
&\Sigma^{[1]}_{45}=\frac{1}{4}
 \left(
\begin{array}{cc:cc:cccccc:cccccc}
0 & 0 & -3i  & 0 & 0 & -\sqrt{10}i  &  \sqrt{5}i         & 0 & 0 &   0       & 0 & 0 & 0 & 0 & 0 & 0 \\
0 & 0 & 0  & -3i  & 0           & 0 & 0 & -\sqrt{5}i        & \sqrt{10}i & 0 & 0 & 0 & 0 & 0 & 0 & 0 \\
 \hdashline
 3i & 0 & 0 & 0&0 & 0 & 0 & 0 & 0 & 0 & 0 & -\sqrt{5}i  & 0 & \sqrt{10}i & 0  & 0 \\
 0 & 3i & 0 & 0&0 & 0 & 0 & 0 & 0 & 0 & 0 & 0  & -\sqrt{10}i & 0 & \sqrt{5}i & 0 \\
 \hdashline
 0   & 0 & 0 & 0&0 & 0 & 0 & 0 & 0 & 0 &   -2i & 0 & 0 & 0 & 0 & 0 \\
 \sqrt{10}i & 0 & 0 & 0&0 & 0 & 0 & 0 & 0 & 0 & 0 & -\sqrt{2}i  & 0  & 0 & 0 & 0 \\
  -\sqrt{5}i & 0  & 0 & 0 &0 & 0 & 0 & 0 & 0 & 0 & 0  & -i & 0 & -\sqrt{2}i & 0 & 0 \\
0  &  \sqrt{5}i  & 0 & 0&0 & 0 & 0 & 0 & 0 & 0 & 0 & 0 &  -\sqrt{2}i & 0 &  -i & 0 \\
  0 &  - \sqrt{10}i  & 0 & 0&0 & 0 & 0 & 0 & 0 & 0 & 0 & 0 & 0 & 0  &  -\sqrt{2}i & 0 \\
 0 & 0 & 0 & 0&0 & 0 & 0 & 0 & 0 & 0 & 0 & 0 & 0 & 0 &0  & -2i \\
 \hdashline
  0 & 0 & 0  & 0& 2 i & 0 & 0  & 0 & 0 & 0 & 0 & 0 & 0 & 0 & 0 & 0 \\
 0 & 0 &   \sqrt{5}i & 0  & 0 & \sqrt{2}i & i & 0 & 0 & 0 & 0 & 0 & 0 & 0 & 0 & 0 \\
  0 & 0 & 0 & \sqrt{10}i &0 & 0 & 0 & \sqrt{2}i & 0 & 0 & 0 & 0 & 0 & 0 & 0 & 0 \\
 0 & 0 & -\sqrt{10}i & 0&0 & 0 & \sqrt{2}i & 0 & 0  & 0 & 0 & 0 & 0 & 0 & 0 & 0 \\
  0 & 0 & 0  & -\sqrt{5}i & 0 &  0  & 0 & i & \sqrt{2}i & 0  & 0 & 0 & 0 & 0 & 0 & 0 \\
 0 & 0 & 0 & 0  & 0 & 0 & 0 & 0  & 0 & 2i & 0 & 0 & 0 & 0 & 0 & 0 \\
\end{array}
\right).
\end{align}

\normalsize




\end{document}